\documentclass[acmlarge,screen]{acmart}

\AtBeginDocument{%
\providecommand\BibTeX{{%
\normalfont B\kern-0.5em{\scshape i\kern-0.25em b}\kern-0.8em\TeX}}}

\newcommand\blfootnote[1]{%
\begingroup
\renewcommand\thefootnote{}\footnote{#1}%
\addtocounter{footnote}{-1}%
\endgroup
}

\usepackage{enumitem}
\usepackage{tabularx}
\usepackage{subfigure}

\graphicspath{{graphics/}{.}}

\newcommand{\squishlist}{
\begin{list}{$\bullet$}
{ \setlength{\itemsep}{0pt}      \setlength{\parsep}{3pt}
\setlength{\topsep}{3pt}       \setlength{\partopsep}{0pt}
\setlength{\leftmargin}{3.5mm} \setlength{\labelwidth}{1em}
\setlength{\labelsep}{0.5em} } }

\newcommand{\squishend}{
\end{list}  }

\newcommand{\para}[1]{\paragraph{#1}}

\newcommand{\ie}{{i.e., }}
\newcommand{\eg}{{e.g., }}
\newcommand{\etal}{\textit{et al.}}
\renewcommand{\vec}[1]{\ensuremath{#1}}
\newcommand{\feat}[1]{\texttt{\detokenize{#1}}}
\newcommand{\dev}[1]{\texttt{\detokenize{#1}}}
\newcommand{\defense}[1]{\texttt{\detokenize{#1}}}
\newcommand{\app}[1]{\texttt{\detokenize{#1}}}
\newcommand{\act}[1]{\texttt{\detokenize{#1}}}
\newcommand{\ad}{\ensuremath{A}}
\newcommand{\adpair}{\ensuremath{\ad_{\text{all}}}}
\newcommand{\adpage}{\ensuremath{\ad_{\text{paging}}}}
\newcommand{\adcomm}{\ensuremath{\ad_{\text{comm}}}}

\setcopyright{none}\renewcommand\footnotetextcopyrightpermission[1]{}
\begin{document}

%{\color{red} \noindent Private draft: Please do not make public or reshare. When in doubt, please contact ludovic.barman@epfl.ch \vspace{0.5cm}}

\title[Every Byte Matters: Traffic Analysis of Bluetooth Wearable Devices]{Every Byte Matters: Traffic Analysis of Bluetooth Wearable Devices*}

\author{Ludovic Barman}
\affiliation{\institution{EPFL}}
\email{ludovic.barman@epfl.ch}

\author{Alexandre Dumur}
\affiliation{\institution{EPFL}}
\email{alexandre.dumur@epfl.ch}

\author{Apostolos Pyrgelis}
\affiliation{\institution{EPFL}}
\email{apostolos.pyrgelis@epfl.ch}

\author{Jean-Pierre Hubaux}
\affiliation{\institution{EPFL}}
\email{jean-pierre.hubaux@epfl.ch}
\renewcommand{\shortauthors}{Barman \etal}

\blfootnote{$^*$Presented at ACM Ubicomp 2021 and published in the Proceedings of the ACM on Interactive, Mobile, Wearable and Ubiquitous Technologies (IMWUT), Vol.~5, No.~2, Article 54, June 2021. \url{https://doi.org/10.1145/3463512}}
\makeatletter
\let\@authorsaddresses\@empty
\makeatother

\begin{abstract}
Wearable devices such as smartwatches, fitness trackers, and blood-pressure monitors  process, store, and communicate sensitive and personal information related to the health, life-style, habits and interests of the wearer.
This data is typically synchronized with a companion app running on a smartphone over a Bluetooth (Classic or Low Energy) connection.
In this work, we investigate what can be inferred from the metadata (such as the packet timings and sizes) of encrypted Bluetooth communications between a wearable device and its connected smartphone.
We show that a passive eavesdropper can use \emph{traffic-analysis attacks} to accurately recognize (a) communicating devices, even without having access to the MAC address, (b) human actions (\eg monitoring heart rate, exercising) performed on wearable devices ranging from fitness trackers to smartwatches, (c) the mere opening of specific applications on a Wear OS smartwatch (\eg the opening of a medical app, which can immediately reveal a condition of the wearer), (d) fine-grained actions (\eg recording an insulin injection) within a specific application that helps diabetic users to monitor their condition, and (e) the profile and habits of the wearer by continuously monitoring her traffic over an extended period.
We run traffic-analysis attacks by collecting a dataset of Bluetooth communications concerning a diverse set of wearable devices, by designing features based on packet sizes and timings, and by using machine learning to classify the encrypted traffic to actions performed by the wearer.
Then, we explore standard defense strategies against traffic-analysis attacks such as padding, delaying packets, or injecting dummy traffic.
We show that these defenses do not provide sufficient protection against our attacks and introduce significant costs.
Overall, our research highlights the need to rethink how applications exchange sensitive information over Bluetooth, to minimize unnecessary data exchanges, and to research and design new defenses against traffic-analysis tailored to the wearable setting.
\end{abstract}



\maketitle
\settopmatter{printacmref=false}\setcopyright{none}\renewcommand\footnotetextcopyrightpermission[1]{}\pagestyle{plain}\thispagestyle{empty}

\begin{figure}
\centering
\includegraphics[width=\linewidth]{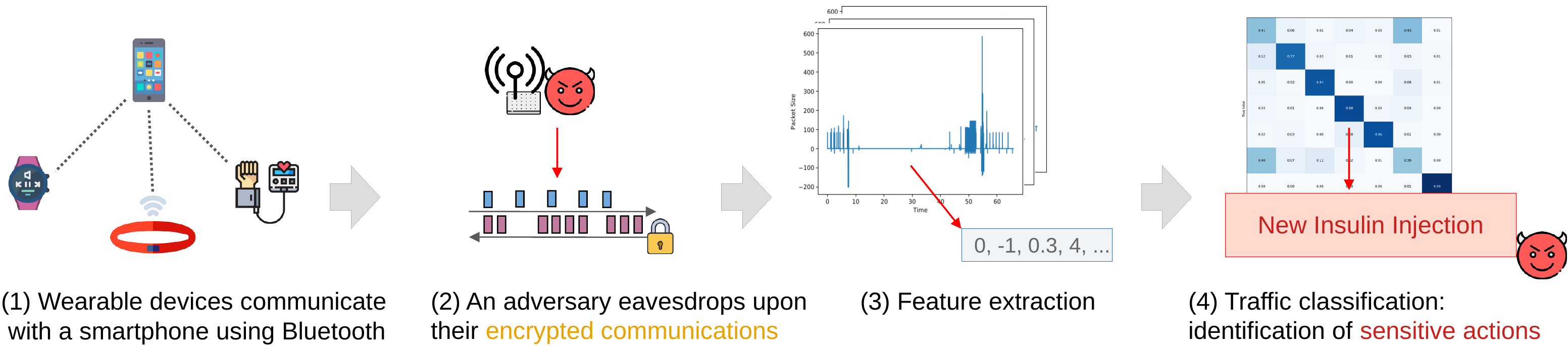}

\caption{Traffic-analysis attack on the encrypted traffic of Bluetooth wearable devices.
After an offline training phase (not depicted here), the passive eavesdropper can recognize the action to record an insulin injection despite the use of encryption.}
\label{fig:methodology}

\end{figure}

\section{Introduction}
\label{sec:introduction}

With the rising interest in personalized health care and ``quantified self", wearable Bluetooth devices are becoming pervasive in our societies.
To improve aspects of their daily lives, users increasingly\footnote{In 2019, $46$ million wearable devices were sold. This number is expected to rise to $158$ millions by 2022~\citep{statista_wearables}.} use smartwatches, medical and fitness-related devices such as activity trackers, step counters, blood-pressure monitors and sleep trackers.
These devices process, store, and transmit personal and sensitive data linked to the wearer's identity and health status:
fine-grained and long-term activity levels, heart rates, arrhythmia alerts, medication and sleep schedules, etc.
Such personal information should be protected from third parties, as it can be used to build profiles, to identify and track users (\eg for advertising purposes), or can be sold to insurance companies for the quantification of users' medical risks.
Medical wearable devices that are FDA-approved\footnote{U.S. Food and Drug Administration. This institution notably edicts rules for medical devices sold in the U.S.} are already subject to such a requirement: they ``should have appropriate protections in place that prevent sensitive information from being read by unauthorized parties either in storage or in transmission''~\citep{fda_approval}.

For enhanced functionalities, most wearable devices communicate with a smartphone via Bluetooth.
These devices can use encryption at the Bluetooth link-layer or  application-layer.
In this work, we show that these encrypted communications leak information about their contents via their communication patterns (\eg distribution of packet sizes and inter-arrival timings).
We consider a local passive eavesdropper who attempts to infer sensitive information from the communications.
This adversary could be a nosy neighbor, an advertiser in a shopping mall attempting to infer the shoppers' habits, an employer~\citep{rise_employee_health_tracking}, or a more nefarious adversary collecting data for sale to insurance companies.
These threats against privacy are concrete:
For over a decade, advertisers have been using Bluetooth, Wi-Fi, and cellular networks to track consumers in stores~\citep{google_track_bluetooth, tlf_tracking, carrier_selling_location}, and to link their profiles to online advertisement databases~\citep{liveramp_onboarding}; these advertisers could use Bluetooth traffic patterns to learn new information or to better profile users.
%The more recent Bluetooth-based \emph{proximity advertising} is employed to track users and display local targeted advertisements in transportation systems, airports and supermarkets~\citep{the_drum_london_cabs, proxbook, stores_secret_surveillance}.
%These technologies highlight the willingness of some advertisers to use Bluetooth as an advertisement and potentially profiling mechanism.

We infer sensitive information from the encrypted communications of wearable devices by using \emph{traffic-analysis attacks}~\citep{danezis07introducing}, a technique that exploits the communication patterns (\eg packet sizes and timings) of encrypted traffic.
These attacks have been successfully demonstrated in diverse settings: to recognize web pages on Tor traffic~\citep{panchenko2011website, dyer2012peek, wang13improved, wang16realistically},
to fingerprint devices~\cite{pang07finger80211} or to infer the activities performed in a user's smart home~\citep{srinivasan2008protecting,acar2018peek} and to recognize user activities and applications used on a smartphone (\eg sending an e-mail or browsing a web page)~\citep{conti2015analyzing,saltaformaggio2016eavesdropping,zhang11inferring,taylor2016appscanner,taylor2017robust}.
The focus of recent related works has been on inferring user activities in the IoT and smart home setting~\citep{trimananda2020packet,acar2018peek,apthorpe2017spying,apthorpe2019keeping,alshehri2020attacking}, eavesdropping on WLAN or Internet traffic (or both).
Very few related works consider the communications of Bluetooth wearable devices at the Personal Area Network (PAN) scale, despite the sensitive nature of the information exchanged.
Das \etal~\citep{das2016uncovering} demonstrate how the bitrate of some fitness trackers is correlated with the user's gait; however, their analysis is restricted to $6$ BLE devices, and they do not attempt to recognize devices, applications or users actions.
To the best of our knowledge, ours is the first work that performs in-depth device, application and user-action identification on the Bluetooth communications of wearable devices.

To perform traffic analysis on wearable devices, we build a Bluetooth (Classic and Low Energy) traffic-collection framework.
We set up a testbed consisting of a diverse set of wearable devices (smartwatches, fitness trackers, blood-pressure monitors, etc.) connected to a smartphone, and we use a Bluetooth sniffer to capture the traffic.
To generate realistic traffic traces, we manually use the wearable devices in the intended way: \eg in the case of a fitness monitor, by repeatedly performing a short running exercise until we obtain a sufficient number of samples.
%This is in contrast with some previous works (in the context of Wi-Fi traffic) that use UI fuzzing to quickly provide many traffic samples generated by smartphones~\citep{conti2015analyzing,taylor2017robust,taylor2016appscanner}.
%To have a realistic dataset, we manually perform real actions, and we obtain a dataset of labeled (encrypted) Bluetooth Classic and Low Energy traces.
We obtain a dataset of labeled (encrypted) Bluetooth Classic and Low Energy traces; we then design features based on packet sizes and timings, and we apply machine-learning classification techniques for identifying devices, applications, and user actions despite the encryption.

Our experimental results show that an eavesdropper can exploit encrypted communication patterns to passively identify devices, even in the absence of Bluetooth addresses or friendly names.
We show that the timings of encrypted communications allow to identify specific models/versions of a device, and hence defeats the Bluetooth address-randomization protection employed by the Bluetooth Low Energy protocol.
Moreover, we demonstrate that the adversary, for instance a nosy neighbor, can use our traffic-analysis attacks to accurately recognize user actions (\eg recording the heart rate, beginning a workout or receiving an SMS) performed on different wearable devices such as smartwatches, step counters, and fitness trackers.
We then focus on smartwatches and show that an eavesdropper, for instance an unscrupulous advertiser, can recognize the mere opening of specific applications, which can be immediately sensitive (\eg in the case of a medical app) or used to build a profile (\eg based on religious or political apps).
%Previous works that also achieve application identification on smartphones by performing traffic-analysis rely on flow separation based on IP addresses~\citep{conti2015analyzing,taylor2016appscanner,taylor2017robust}, which does not apply for Bluetooth.\todo{Is this important point?are we doing something to avoid this issues in Bluetooth?}
Furthermore, we show that our methodology generalizes well: The model trained on a smartphone/wearable device pair performs accurately on a different device pair, indicating its cost-effectiveness.
We also highlight how a targeted adversary can recognize sensitive user activities within a specific application; we accurately recognize the action of recording an insulin injection in a diabetes-helper application (Figure~\ref{fig:methodology}).
Finally, we show that an attacker can infer a wearer's habits and build her profile when eavesdropping her Bluetooth traffic over a long time-period.

% first defense: data minimization by changing a bit how the app works

% defense are not perfect but they can be a first step that developpers can reason about when creating an application

% data developpers should be careful about the shape of data they exchange, even before thinking of encryption

Finally, we focus on countermeasures against the presented traffic-analysis attacks.
%To this end, we first review the most common defenses against traffic-analysis attacks in the literature and identify the most prominent approaches.
We evaluate how standard approaches against traffic analysis (\ie padding or delaying messages, injecting dummy traffic) perform against our attacks. % and experimentally evaluate their effectiveness and costs.
Our results show that these defenses provide insufficient protection: though they yield a drop in the adversary's accuracy, none of them reduces it to that of random guessing.
%while these defenses are by no means optimal,
Furthermore, their costs are high (from tens to hundreds of kilobytes of additional exchanged data) for wearable device communications where energy consumption is crucial.
We also observe that the effectiveness of these defenses varies greatly with the adversarial task: although the ``right'' defense drops the attacker's accuracy, non-adapted defenses have almost no effect yet still incur high costs.
This suggests that a ``one-size fits all'' defense for preventing device fingerprinting, as well as application and action fingerprinting, is unlikely to exist at a reasonable cost.
%In particular, we observe that per-packet padding incur highly variable costs (from hundreds of kilobytes to megabytes), while delaying packets to the next second has consistent costs and higher effectiveness, but a limited applicability.
%Injecting dummy traffic achieves good performance at reasonable cost ($\approx 50$KB) but requires the distribution of packet sizes to be known in advance.
Our defense evaluation highlights the need to rethink how wearable devices exchange sensitive data over Bluetooth communications and prompts for the design of novel defenses.
For example, for applications that can support it, \emph{data minimization} is a valid strategy: data that is not exchanged cannot be fingerprinted, and we observe in our experiments that low-volume exchanges are naturally protected from traffic analysis.
Moreover, bulk transfers (\eg synchronizing a step counter every day at midnight) also hamper the task of the adversary.

Overall, the purpose of our work is to raise awareness, notably among device manufacturers and application developers, of the limited confidentiality on the Bluetooth link with today's applications and devices.
For instance, manufacturers might be willing to implement mitigations directly to the wearable devices' firmware, hence it is urgent that the research community provides them with acceptable solutions that will protect the next generation of wearable devices.
Application developers might want to carefully consider the information their application can leak through its communication patterns and to implement an application-level defense (for instance, pad all communications to a fixed size).
We hope that our research results will be a starting point for further research on the communication metadata of Bluetooth wearable devices.
To facilitate future research on the topic, we open-source our dataset for research purposes: it consists of Bluetooth traces of wearable devices used to perform multiple actions and that have been, for most devices, manually recorded over months.
Although our experiments focus only on Bluetooth communication metadata, our dataset contains all captured data (\eg link-layer information, pairings) that might be of independent interest.
%Finally, we also open-source the framework developed for the automated recording of traffic generated by open-source compatible devices (\ie Wear OS smartwatches).

\noindent In summary, the contributions of this work are as follows:
\begin{itemize}
\item We show traffic-analysis attacks that, based on the encrypted traffic of wearable devices, recognize:
\begin{enumerate}
\item communicating devices, up to the model number of the same device;
\item user activities and the opening of specific applications;
\item fine-grained sensitive actions (\ie recording an insulin injection) within an application;
\item actions recorded over a long time-period, which can be used to build a profile.
\end{enumerate}
\item We experimentally evaluate standard defenses against traffic analysis and suggest possible strategies;
\item We make available a large dataset of Bluetooth traffic captures for future research.
\end{itemize}

The remainder of the paper is organized as follows: in §\ref{sec:bt_background}, we introduce background information about the Bluetooth protocol used by wearable devices for their communications.
In §\ref{sec:system_model}, we describe the system and threat model considered in our work, and in §\ref{sec:methodology}, we present the methodology employed for our dataset collection and the traffic-analysis attacks.
Then, in §\ref{sec:dev_id} and §\ref{sec:act_id}, we demonstrate the results of our traffic-analysis attacks that identify devices, actions and applications, from the encrypted Bluetooth traffic of wearable devices.
In §\ref{sec:protections}, we evaluate the performance of standard defenses against our attacks.
We discuss the contributions of the paper and its limitations in §\ref{sec:discussion} and §\ref{sec:limitations}, the related work in §\ref{sec:related-work}, and then conclude in §\ref{sec:conclusion}.

\section{Bluetooth Background}
\label{sec:bt_background}

There exist two flavors of Bluetooth: Bluetooth Classic, referred to as ``BR/EDR'' for Basic Rate/Enhanced Data Rate, and Bluetooth Low Energy (BLE).
The former is used for data-intensive or latency-sensitive scenarios (\eg audio streaming), whereas the latter is used for low-power or low-throughput scenarios.
Most wearable devices we collected for our testbed use Bluetooth Low Energy, except for smartwatches that typically use Bluetooth Classic.
Both Bluetooth specifications~\citep{bluetooth2016bluetooth} are produced by the Bluetooth Special Interest Group (SIG), and their latest version is 5.2.% (introduced in December 2019) .

\para{Data Exchange.} In both Bluetooth Classic and BLE, data exchange occurs after a pairing process between one or more slaves (the wearable device) and a master (\eg the smartphone).
We remark that a smartwatch can be a master for another wearable device.
Initially, to discover each other, devices broadcast and listen on \emph{advertising channels} in Bluetooth Low-Energy, or on channels determined by a predefined hopping sequence in Bluetooth Classic.
Then, the pairing process assigns the slave to the master's \emph{Piconet}, and both devices communicate using the non-advertising channels.
%to be precise, I think Classic has no dedicated advertising channels ,only a hopping pattern derived from the general inquiry address
%Bluetooth Classic has $80$ channels, while Bluetooth LE has $37$ data channels and $3$ advertising channels.
%Often, devices stop advertising when they are part of an active connection.
Unlike Bluetooth Classic, BLE also supports data exchange without establishing a link-layer connection: devices simply broadcast short information to its surroundings on the advertising channel.

To communicate in a Piconet, all devices use the same frequency hopping pattern (derived from the master's MAC address and clock).
The channel is divided into time-slots; odd time-slots are for the slaves, and even time-slots for the master.
When the connection is Asynchronous Connectionless (ACL), devices communicate opportunistically during unreserved time-slots.
In the case of Synchronous Connection-Oriented (SCO) connections, devices communicate at predetermined time-slots without acknowledgment.
%This is used in latency-sensitive data such as audio.

\para{Security.} The security properties are similar between Bluetooth Classic and BLE\@.
Devices first establish and authenticate a long-term key through a \emph{pairing} process.
In this process, both devices also derive a short- or long-term key.
``Legacy'' pairings are generally insecure by today's standards~\citep{wong2005repairing}: they rely on a short PIN and can be easily brute-forced.
Secure Simple Pairing (SSP) is a more recent pairing protocol based on elliptic curve cryptography~\citep{phan2012analyzing}.
It is available since Bluetooth 2.1.

To authenticate the long-term key, four authorization mechanisms exist:
\begin{itemize}[noitemsep,topsep=2pt,parsep=0pt]
\item JustWorks, which uses a hardcoded key;
\item Out-of-Band, which relies on an external channel (such as NFC);
\item Numeric Comparison, where the user visually compares numbers;
\item Passkey Entry, where the user inputs twice the same code on the devices.
\end{itemize}

There exists several active attacks on SSP, based on some form of Man-in-the-Middle~\citep{haataja2008man,haataja2008practical,haataja2010two} or low-entropy key generation~\citep{antonioli2019knob,antonioli2020key}.
If the pairing is done correctly and the key uses sufficient entropy, the subsequent communication should be confidential and integrity-protected (AES-CCM with a $128$-bit key).

\para{Bluetooth Eavesdropping.}
Eavesdropping on Bluetooth traffic is complex due to the frequency hopping.
Inexpensive sniffers (such as the Ubertooth~\citep{ubertooth}) attempt to follow the frequency hopping of a connection and often require observing the pairing~\citep{ryan2013bluetooth}.
However, by listening concurrently on all channels, high-end commercial scanners accurately capture all traffic without the need to observe the pairing.
Recent advances in research use coordinated Ubertooth devices to achieve a greater capture accuracy~\citep{albazrqaoe2016practical,albazrqaoe2018practical}, or use software-defined radios (SDRs) to cover the Bluetooth spectrum at a cost lower than commercial scanners~\citep{cominelli2020even,cominelli2020one}.
We discuss the practicality of Bluetooth eavesdropping in §\ref{sec:discussion}.
%Theoretically, the communication range of both flavors goes up to $100$m.

\section{System and Adversarial Model}
\label{sec:system_model}

\begin{figure}
\centering
\includegraphics[width=0.5\linewidth]{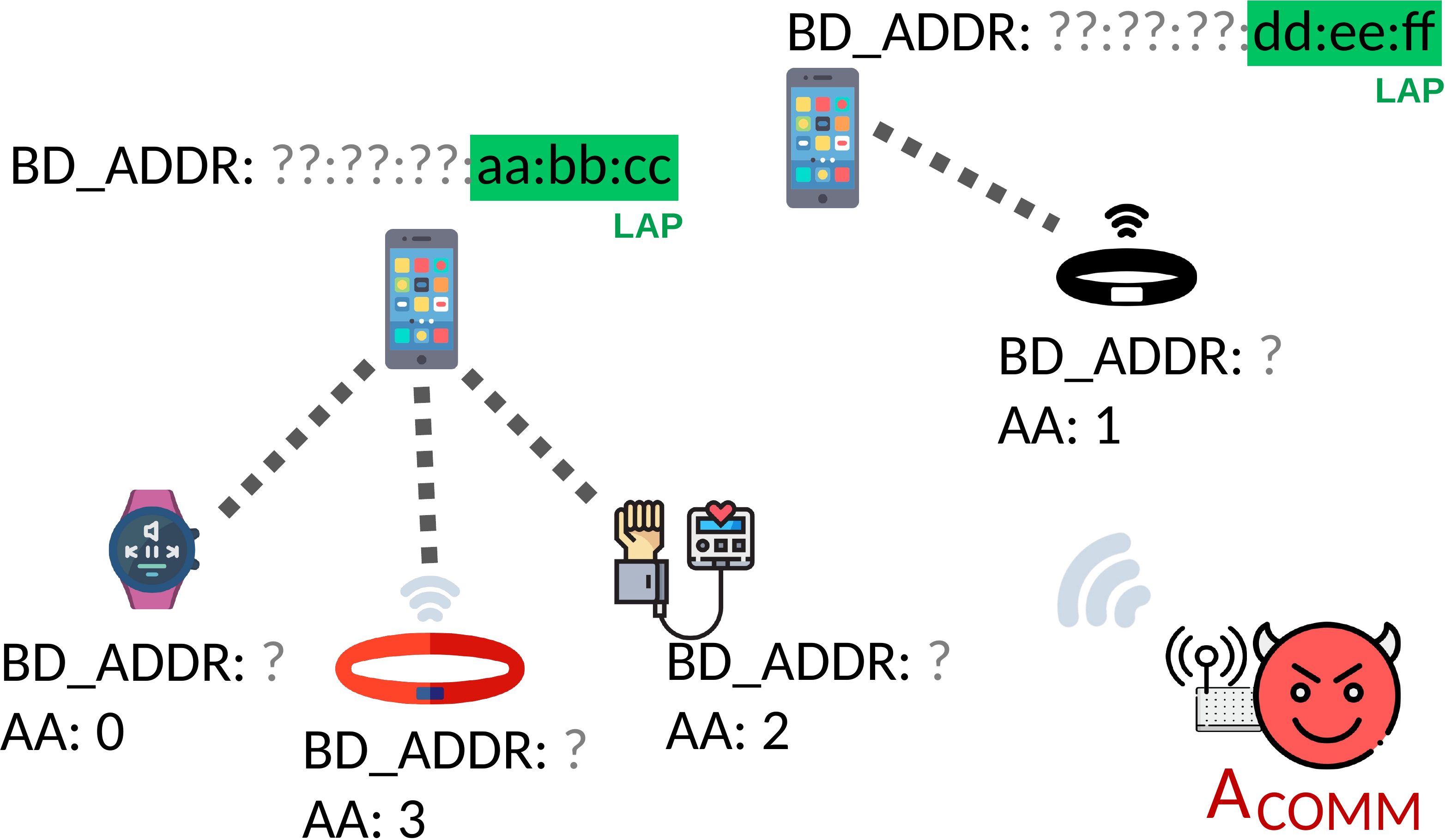}
\caption{Information visible to \adcomm{}: (1) the active connections between each pair of wearable-smartphone, identified by a random Access Address (AA); \ad{} sees all packets on these connections; (2) for each master device, the Lower Address Part (LAP) of the BD\_ADDR (sometimes called the MAC address).
\adcomm{} cannot tell what types of devices are communicating.}
\label{fig:infovisible-acomm}
\end{figure}

We consider a user $U$ that possesses a smartphone $S$ and a collection of wearable devices $\vec{W}$.
This collection $\vec{W}$ contains heterogeneous devices: some with a general-purpose OS (\eg an Android smartwatch) or a simpler firmware (\eg a step counter).
Devices in $\vec{W}$ can communicate with $S$ over Bluetooth Classic or Low Energy.

\para{Adversary.} We consider an adversary \ad{} who is a passive eavesdropper.
$\ad{}$ uses a Bluetooth sniffer to capture all Bluetooth traffic in the vicinity of $U$.
In particular, \ad{} might capture traffic from other users and devices that are also nearby.
$\ad{}$ does not compromise any devices and does not possess the keys to decrypt Bluetooth traffic.

\para{Goals and Traffic Captured.} Informally, \ad{} attempts to infer information from all communications between the wearable devices in $\vec{W}$ and $S$ (Figure~\ref{fig:infovisible-acomm}).
However, we need to precisely define the aforementioned adversary.
In practice, an eavesdropper who sees \emph{all} traffic between a wearable device and its connected smartphone is not realistic:
a sporadic or local adversary is unlikely to consistently capture one-time events such as pairings.
Therefore, we define as \adpair{} an adversary who sees all communications between a wearable-smartphone device pair, and we further define two weaker adversaries in the sense of \ad{}, but who only observe, respectively:

\squishlist
\item \adpage{}: all paging events (devices waking up from sleep and performing minimal discovery) and subsequent communications, but not the pairings;
\item \adcomm{}: all active communications, but neither pairing nor paging events.
\squishend

\noindent In terms of adversarial power, we have that $\adpair{} \ge \adpage{} \ge \adcomm{}$.
We focus on the weakest adversary $\adcomm{}$ who only observes ongoing communications between a wearable device and its connected smartphone.
This matches what a passive adversary can do consistently.

\begin{figure}
\centering
\subfigure[Structure of a BB\_PDU packet (BaseBand Packet Data Unit) used in Bluetooth Classic.
HEC (Header Error Correction) is an error-correction code.
]{\includegraphics[width=0.35\linewidth]{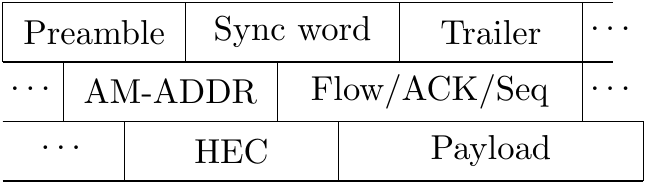}
\label{fig:info-packet-cla}
}\hspace{1cm}%
\subfigure[Structure of a Link Layer packet used in Bluetooth LE\@.]{\includegraphics[width=0.35\linewidth]{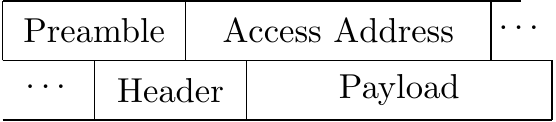}
\label{fig:info-packet-le}}
\caption{Packet Structures.
The features used in our attacks are derived from the time of reception and the size of the payload.}
\end{figure}

\para{Information Visible to \adcomm{}.}
Figure~\ref{fig:info-packet-cla} shows the information that is visible to \adcomm{} during Bluetooth Classic communications.
The sensitive/identifying fields are the Sync word and the Active-Member Address (AM-ADDR).
The Sync word reveals the 24 lower bits of the master's MAC address (LAP).
The Upper Address Part (UAP), that is not transmitted in this packet, is assigned to manufacturers and identifies a wearable device.
Finally, the Active-Member Address is a 3-bit integer (and not a MAC address) identifying a device in a given Piconet; this is similar to a connection identifier.
Figure~\ref{fig:info-packet-le} shows the information visible to \adcomm{} for Bluetooth LE.
Similar to AM-ADDR, the Access Address is a connection identifier (and not a MAC address).
On a higher-level, in both Bluetooth flavors, the information given to \adcomm{} per packet is
\[(\text{Packet type},~\text{connection ephemeral ID},~\text{time},~\text{payload})\]
where the Packet type indicates whether Bluetooth Classic or LE is used, and where the connection ephemeral ID is randomized, except for the master device in Bluetooth Classic where it is fixed.
The payload can contain upper-layer packet information, for instance the packet type, as well as the encrypted application payload.

\para{Capabilities of \adcomm{}.} In conclusion, with the information visible on the Bluetooth channel, \adcomm{} can
\squishlist
\item enumerate the devices $w \in \vec{W}$ and assign a pseudonym to each of them;
\item group them by connection between a pair of devices; and
\item collect a series $\{\left(\text{time},~\text{packet type},~\text{packet size}\right)\}_w$ for each device $w \in \vec{W}$ for the duration of the capture.
\squishend

We assume that the adversary is able to isolate a communication of interest, \eg using the Active-Member Address (AM-ADDR) of Bluetooth Classic or Access Address of Bluetooth Low Energy, possibly combined with other techniques, \eg distance estimation using the Received Signal Strength Indicator (RSSI).
The pseudonymous recipient of a packet is often known, as most packets are acknowledged.

\para{Attacker Goals.} Given the Bluetooth information available, the goal of \adcomm{} is threefold:

\begin{enumerate}
\item[G1)] \textbf{Device Identification:} for a device $w$, recognize the device brand/manufacturer;
\item[G2)] \textbf{Action Identification:} for a device $w$, recognize user actions (\ie interactions with the device);
\item[G3)] \textbf{Application Identification:} for a device $w$, recognize the running application.
\end{enumerate}

\noindent We use Goal $1$ as a stepping-stone for the following goals.
However, meeting the first goal already has privacy implications: such an adversary can recognize and track a user through the list of her wearable devices, despite MAC address randomization, and can build a profile for that particular user.

\section{Methodology \& Dataset}
\label{sec:methodology}

We describe how the adversary (defined in §\ref{sec:system_model}) eavesdropping on Bluetooth communications can perform traffic-analysis attacks to achieve its goals.
We present the methodology that applies to the traffic-analysis attacks presented in §\ref{sec:dev_id} and §\ref{sec:act_id}.
We first build a data collection framework to record Bluetooth traffic and to automate the data transmission for some Bluetooth wearable devices.
%Our framework consists of a Windows laptop piloting a Bluetooth sniffer and of a Linux laptop that issues commands to Wear-OS-based wearable devices.
For other devices, we manually trigger Bluetooth traffic by performing the appropriate human action (\eg pressing a sequence of buttons, performing a physical activity).
%The Bluetooth sniffer records all Bluetooth traffic and groups the recorded packets per connection.
Then, we process the captured traffic and use it with machine-learning classification techniques to infer information from encrypted Bluetooth traffic.

\para{Testbed.}
We set up a testbed with multiple wearable devices to account for a wide range of manufacturers, device capabilities, and functionalities.
We initially selected $18$ Bluetooth Classic and LE devices: $5$ popular smartwatches, $2$ headphones, and $11$ other wearable devices (step counters, fitness trackers, and blood-pressure monitors) from the most popular vendors.

All Bluetooth Classic devices analyzed use link-layer encryption.
To our surprise, from the Bluetooth Low Energy devices, only the Fitbit Charge 2 uses link-layer encryption.
We perform entropy tests to validate that the remaining Low Energy devices use encryption at the application layer.
We observe a high entropy ($6-8$bit of information per byte) for all devices, except for $3$ of them (the \dev{Beurer AS80}, \dev{SW170}, and \dev{PanoBike+}, $\approx 4$bit/byte).
Hence, we discard these three devices from our testbed.
We remark that though our attacks do not use plaintext contents and would work on these three devices, an attacker can achieve the same results by simply reading the payload.
We also discard two blood-pressure monitors (the \dev{Qardio} and \dev{H2-BP}) that we could not reliably use.
In total, the testbed consists of $13$ devices: $7$ Bluetooth Classic and $7$ Low Energy devices (Tables~\ref{table:devices-classic} and~\ref{table:devices-le}).

Although our testbed consists of a modest number of devices, most devices use proprietary firmware and software (\ie they are closed-source), which does not allow us to automate their Bluetooth traffic-data collection.
Except for Wear OS smartwatches that we can automate, the data collection is a sequential and manual process.

We use a Nexus 5 as the smartphone for all Android/Tizen smartwatches and wearable devices without an OS, and an iPhone 8 for the Apple Watch/AirPods.
We updated all devices with the latest firmware and OS updates.

\para{Actions Recorded.}
For each device, we manually compile a set of possible user-actions: low-end devices sometimes only \act{Sync} with the smartphone; mid-range devices support activities such as \act{Running Workout}, \act{Measure Heartbeat}, etc., whereas high-end devices such as smartwatches offer a large range of user actions through the use of additional apps (\eg \act{AppX Open}, \act{AppX DoActionY}) that users can choose to personalize their device.

\para{Capture Methodology.}
We record all Bluetooth traffic by using a wide-band scanner (an Ellisys Vanguard~\cite{ellisys_vanguard}).
For each device $w$, we pair $w$ with the phone $S$ and let some time pass to allow for an initial data synchronization.
Then, we trigger the desired action and record the corresponding Bluetooth traffic for $T=30$ seconds (we observe this to be a conservative value); this constitutes one sample.
To collect sufficient number of samples for an action, we repeat the capture process $N=25$ times (we observe this to be a conservative value, Figure~\ref{fig:number_of_samples}).
Each sample corresponds to a distinct recording; hence, samples are independent from each other.
We manually select the device of interest and discard traces from other devices.

We record real activities: for instance, for a \act{Running Workout}, we perform a short running activity in the vicinity of the sniffer; for the wearable that records the heart rate, we fit the wearable to ourselves during the experiment.
This is in contrast with some previous works (in the context of Wi-Fi traffic) that use UI fuzzing to quickly provide many traffic samples generated by smartphones~\citep{conti2015analyzing,taylor2017robust,taylor2016appscanner}.
This ensures that most values communicated by the wearable reflect what a real user would generate (\eg the number of steps, speed, and distance).
However, due to our non-portable experimental setup, the recorded GPS coordinates will show an almost-fixed position.

\para{Environment.}
We conduct the experiments in an office where both Wi-Fi devices and other Bluetooth devices communicate.
This corresponds to a noisy environment; we note that an anechoic chamber or a Faraday cage would advantage the adversary by allowing to record trace with less noise.
%We turned off our own unused wearable devices to avoid unnecessary interference.

\para{Dataset.}
We collect a dataset that consists of $10{,}700$ Bluetooth captures with a duration of $\approx 30$sec, recorded over $13$ devices, $80$ applications and $32$ user-actions (see Table~\ref{table:applications-actions}).
This amounts to $\approx 98$ hours of recording and represents $21$GB of data.
$10{,}371$ of these traces are over Bluetooth Classic, and $329$ Bluetooth Low Energy.
We note that $2{,}215$ of these captures are the result of the corresponding human action (\eg performing a short workout), and $8{,}485$ are automated actions (\eg the opening of an application) on Wear OS devices.

Additionally, we record a second dataset that contains longer captures with a duration of $\approx 20$min.
These captures contain automated actions that, when combined, represent a plausible usage of a smartwatch over a day.
We use them to model a persistent adversary (§\ref{subsubsec:long-term-capture}).
This amounts to $38$ hours of recording and $9.5$GB of data.

Overall, the datasets contain raw binary files with all captured traffic.
As the file format is proprietary, we extract the corresponding CSV files.
We make the datasets (and all code used in this paper) available for research purposes~\citep{wearable-tools}.

\para{Sample Processing.}
We process the captured raw Bluetooth traces as follows.
We follow the approach of the related work and extract only the total length of the payload to avoid relying on the presence of plaintext markers in it~\citep{taylor2016appscanner,taylor2017robust}.
From each recorded Bluetooth trace, we extract the Bluetooth packets at the Logical Link Control and Adaptation Protocol (L2CAP) layer and output a series of (arrival time, size)-tuples.
We label the trace with the \dev{device} used, the \app{application} used (in the case of a smartwatch), the \act{action} performed, and whether it used Bluetooth Classic or Low Energy.
We split the set of recorded samples $\vec{S}$ into two subsets $\vec{S}_\text{Classic}$ and $\vec{S}_\text{LE}$ depending on the Bluetooth variant used.
We remark that an adversary is able to do the same, as the frames are different.

\para{Feature Extraction.}
We design features to capture patterns based on incoming/outgoing size distributions and inter-packet timings.
We map each sample in $\vec{S}_\text{Classic}$ and $\vec{S}_\text{LE}$ to a $32$-scalar feature vector:
First, we capture global statistics about packet sizes.
We filter each sample into $3$ packet sequences: (1) from Master to Slave, (2) from Slave to Master, and (3) sequences consisting of all non-null, non-ACK packets.
From each filtered sequence, we extract the following $5$ statistics: the min/mean/max/count/standard deviation of the packet sizes in bytes.
Then, we observe that different devices/applications send packets with a distinctive size.
To this end, following the techniques used by Liberatore and Levine~\citep{liberatore2006inferring} and Herrmann \etal~\citep{herrmann2009website} in the context of website fingerprinting, we extract features corresponding to the number of packets that are in a certain range.
We select $10$ buckets that represent the size (in bytes) ranges $[0,9], [10,19], \cdots, [80, 89]$ and a last ``catch-all'' bucket for packets with $[90, \infty]$ bytes.
Finally, we examine the timings of the packets.
In more detail, we compute, in seconds, the series of inter-packet durations and extract the same $5$ statistics (min/mean/max/count/standard deviation) from it.
Furthermore, we calculate the average send/receive inter-packet times as done by Saltaformaggio \etal~\cite{saltaformaggio2016eavesdropping} in the case of TLS traffic:
$\text{AvgIPT}(P)=\frac{\sum_{i=0}^{|P|-1} \text{ts}_{i+1}-\text{ts}_{i}}{|P|-1}$,
where $P$ is the set of sent/received packets, and $\text{ts}_i$ is the timestamp of packet $i$.
Intuitively, this captures the communication bursts of each device.

\para{Classification.}
For our traffic-analysis attacks (described in §\ref{sec:dev_id} and §\ref{sec:act_id}), we use a Random Forest classifier, a popular algorithm for multi-class classification that is also widely used in the related work~\citep{taylor2016appscanner,taylor2017robust}.
We opt for a Random Forest classifier over the recently proposed deep-learning approaches~\citep{sirinam2018deep,sirinam2019triplet} due to its interpretability and the moderate size of our dataset.
We split the samples of each device/application/action (depending on the adversarial goal) into $80\%$ training and $20\%$ testing sets and perform 10-fold random-stratified cross-validation.
We also retain the most important features according to the classifier's importance using Recursive Feature Elimination (RFE).

%\para{Reproducibility.}
%Our results are easily reproducible.
%The code is available upon request for research use~\citep{wearable-tools}. Each result is reproducible with a single command.
%The datasets, code, plots, and tables are under version control.

\begin{table}
\caption*{Bluetooth wearable devices used in our experiments.
BT indicates the Bluetooth version.
The \dev{AppleWatch} uses both flavors of Bluetooth.}
\small
\begin{minipage}[t]{0.49\textwidth}
\caption{Bluetooth Classic devices.}
\begin{tabular}{lllll|}
\textbf{Vendor} & \textbf{Model} & \textbf{OS} & \textbf{BT} & \textbf{Chipset} \\
Samsung          & Galaxy Watch          & Tizen OS    & 5.0                & Broadcom        \\
Fossil          & Explorist HR   & Wear OS     & 4.2                & Qualcomm         \\
Apple           & Watch 4        & Watch OS 5  & 5.0                & Apple            \\
Huawei          & Watch 2        & Wear OS 2   & 4.1                & Broadcom        \\
Fitbit          & Versa 2        & Fitbit OS 4 & 5.0                & Cypress S.   \\
Sony            & MDR-XB9        & ---         & 4.1                & Qualcomm         \\
Apple           & AirPods 2      & ---         & 5.0                & Apple            \\
\end{tabular}
\label{table:devices-classic}
\end{minipage}\hfill
\begin{minipage}[t]{0.49\textwidth}
\caption{Bluetooth Low Energy devices.}
\begin{tabular}{llll}
\textbf{Vendor} & \textbf{Model} & \textbf{BT} & \textbf{Chipset}       \\
Apple           & Watch 4        & 5.0                & Apple            \\
%Beurer                & AS80           & 4.0                & Texas Instruments      \\
Fitbit                & Charge 2       & 4.1                & Microelectronics       \\
Fitbit                & Charge 3       & 5.0                & Cypress Semiconductor  \\
%			Q-Charm               & H2-BP          & 4.2                & Nordic Semiconductor   \\
Huawei                & Band 3e        & 4.2                & RivieraWaves           \\
Mi           & Band 2         & 4.1                & Dialog Semiconductor   \\
Mi           & Band 3         & 4.1                & Dialog Semiconductor   \\
Mi           & Band 4         & 5.0                & Dialog Semiconductor   \\
%			Topeak                & PanoBike+      & 4.0                & Texas Instruments       \\
%			Qardio, Inc           & Qardio         & 4.1                & Qualcomm               \\
%Denver                & SW170          & 5.0                & Nordic Semiconductor   \\
\end{tabular}
\label{table:devices-le}
\end{minipage}
\end{table}

\begin{table}
\caption{Applications and Actions captured in the dataset. Some actions are application-specific, \eg \act{DiabetesM_AddCalorie}.
Other actions, \eg \act{Workout}, exist in different apps and in the firmware of wearable devices.}
\small
\begin{tabularx}{\linewidth}{ r X }
\textbf{Applications} & 20Min, ASB, Alarm, AppInTheAir, AthanPro, AthkarOfPrayer, Battery, BeurerApp, Bild, Bring, Calm, Camera, ChinaDaily, Citymapper, DCLMRadio, DailyTracking, DenverApp, DiabetesM, DuaKhatqmAlQuran, Endomondo, FITIVApp, FITIVPlus, FindMyPhone, Fit, FitBreathe, FitWorkout, Fitbit, Flashlight, FoursquareCityGuide, Glide, GooglePay, GooglePlayMusic, HealthyRecipes, HeartRate, HuaweiApp, Kaia, KeepNotes, Krone, Lifesum, MapMyFitness, MapMyRun, Maps, Medisafe, Meduza, MiApp, Mobills, Music, MyFitnessPalApp, NYT, NoApp, Outlook, PearApp, Phone, PhotoApp, PillReminder, PlayMusic, PlayStore, Qardio, RamadanTime, Reminders, Running, SalatTime, SamsungHealthApp, Shazam, Sleep, SleepTracking, SmartZmanim, SmokingLog, Spotify, Strava, Telegram, Timer, Translate, Walgreens, WashPost, WearCasts, Weather, Workout. \\

\textbf{Actions} & AddCalorie, AddCarbs, AddFat, AddFood, AddGlucose, AddInsulin, AddProteins, AddWater, Browse, BrowseMap, CaloriesAdd, Coffees, EmailReceived, HeartRate, Leisure, LiveStream, NightLife, Open, PhoneCallMissed, PhotoTransfer, Play, Restaurants, Running, SearchRecipe, Shopping, Skip, SMSReceived, Sync, Walking, Workout. \\
\end{tabularx}
\label{table:applications-actions}
\end{table}

\section{Device identification}
\label{sec:dev_id}

We first focus on the adversarial goal G1: recognizing a device name/brand from the metadata of its encrypted Bluetooth traffic.
This attack is a stepping-stone for other attacks that aim to infer more fine-grained information such as actions performed by the wearer or applications installed on her device (§\ref{sec:act_id}).
We recall our assumption that the adversary \adcomm{} does not observe the pairing event between a wearable $w$ and the smartphone $S$, which would give him the device information in plaintext.
Instead, we demonstrate how \adcomm{} can infer this information from encrypted communication patterns, for instance from devices that are already paired and communicating.
Some recent related works already highlight that current BLE devices do not rotate their MAC addresses sufficiently or at appropriate times, enabling tracking~\citep{celosia2019saving, becker2019tracking, martin2019handoff}.
We highlight a deeper problem: the communication patterns (\eg inter-packet timings) are sufficient to accurately identify and track devices.

\para{Attack.}
We use the methodology described in §\ref{sec:methodology} and train two classifiers: one for identifying Bluetooth Classic devices and one for distinguishing BLE devices from their encrypted traffic.
We use our captured dataset of encrypted traces with the \dev{device} label as the classifier's target.
Initially, we have $10{,}371$ feature vectors across diverse applications/actions of $7$ Bluetooth Classic devices, and $329$ feature vectors for $7$ Bluetooth Low Energy devices.
We note that the \dev{AppleWatch} is in both categories as it uses both Bluetooth flavors.

The large sample difference in Bluetooth Classic is due to the automation of two Wear OS smartwatches (§\ref{subsec:app-id-wearos}).
As the Bluetooth Classic dataset is imbalanced with respect to the number of samples per device, we balance the samples per device label and maintain an equal representation of each device's actions.
This gives us between $20$ and $60$ samples per device, except for the \dev{HuaweiWatch} that has $125$ samples (1 per class).
In total, the equalized Bluetooth Classic dataset consists of $326$ feature vectors.
We use a Random Forest classifier with $10$ trees (our experiments showed that additional trees were not necessary - Figure~\ref{fig:number_of_features}) that we train using the features described in §\ref{sec:methodology}, and we do not limit their depth.
We apply Recursive Feature Elimination and keep the most significant $10$ features (Figure~\ref{fig:action-id-wearable-nfeatures}).

\begin{figure}
\centering
\subfigure[Classic devices.]{\includegraphics[width=0.49\linewidth]{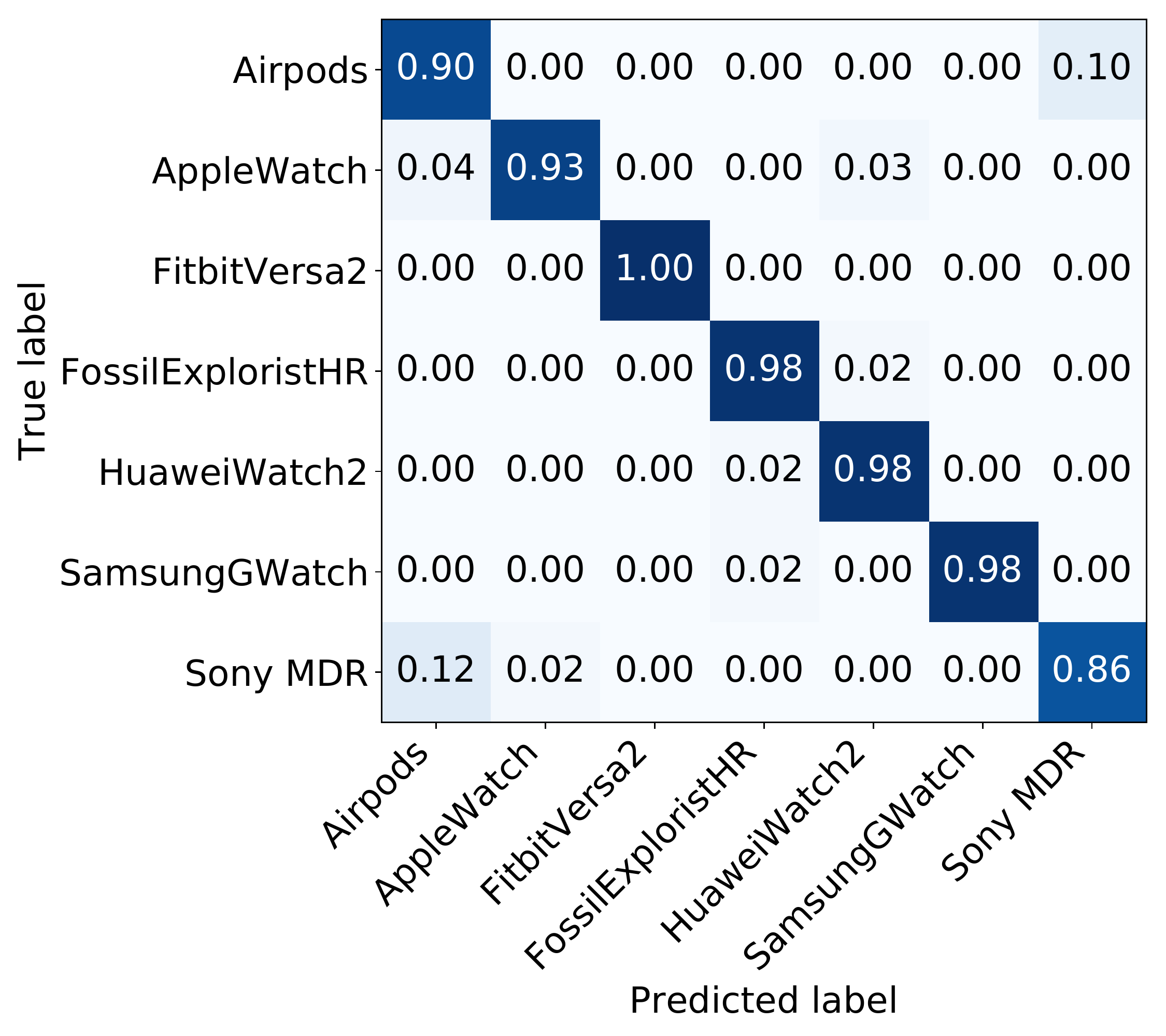}
\label{fig:confusion-matrix-classic}
} %
\subfigure[LE devices.]{\includegraphics[width=0.49\linewidth]{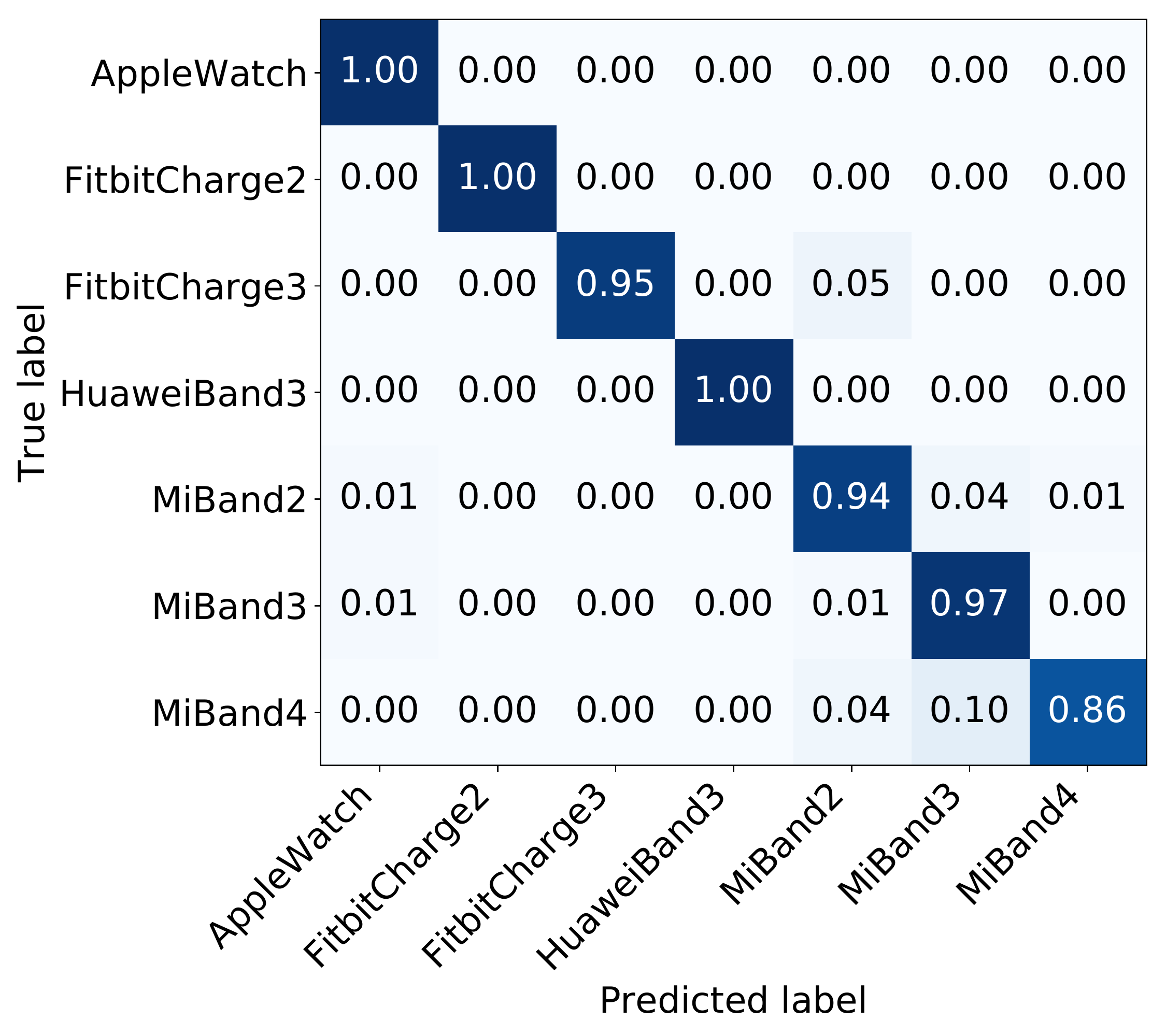}
\label{fig:confusion-matrix-le}}
\caption{Normalized confusion matrix per true label for device identification. Values in the diagonal are the recall per class.}
\label{fig:confusion-matrix}
\end{figure}

\para{Results.}
For the multi-class classification problems with $7$ Classic and $7$ LE devices, the classifier's precision/recall/F1 score is $0.96$ for Bluetooth Classic and $0.97$ for Low Energy (Tables~\ref{table:device-id-cm-cla} and~\ref{table:device-id-cm-ble}), thus showing that identifying a wearable device from its encrypted traffic is successful.
We also find that our classifier accurately distinguishes between different models from the same vendor (\ie Mi Band 2, 3 or 4, and Fitbit Charge 2 or 3) indicating that each model has distinctive traffic patterns.
Figures~\ref{fig:confusion-matrix-classic} and~\ref{fig:confusion-matrix-le} show the confusion matrices.
The values in the diagonals are the recall per class.

\begin{figure}
\centering
\subfigure[Classic devices.]{\includegraphics[width=0.49\linewidth]{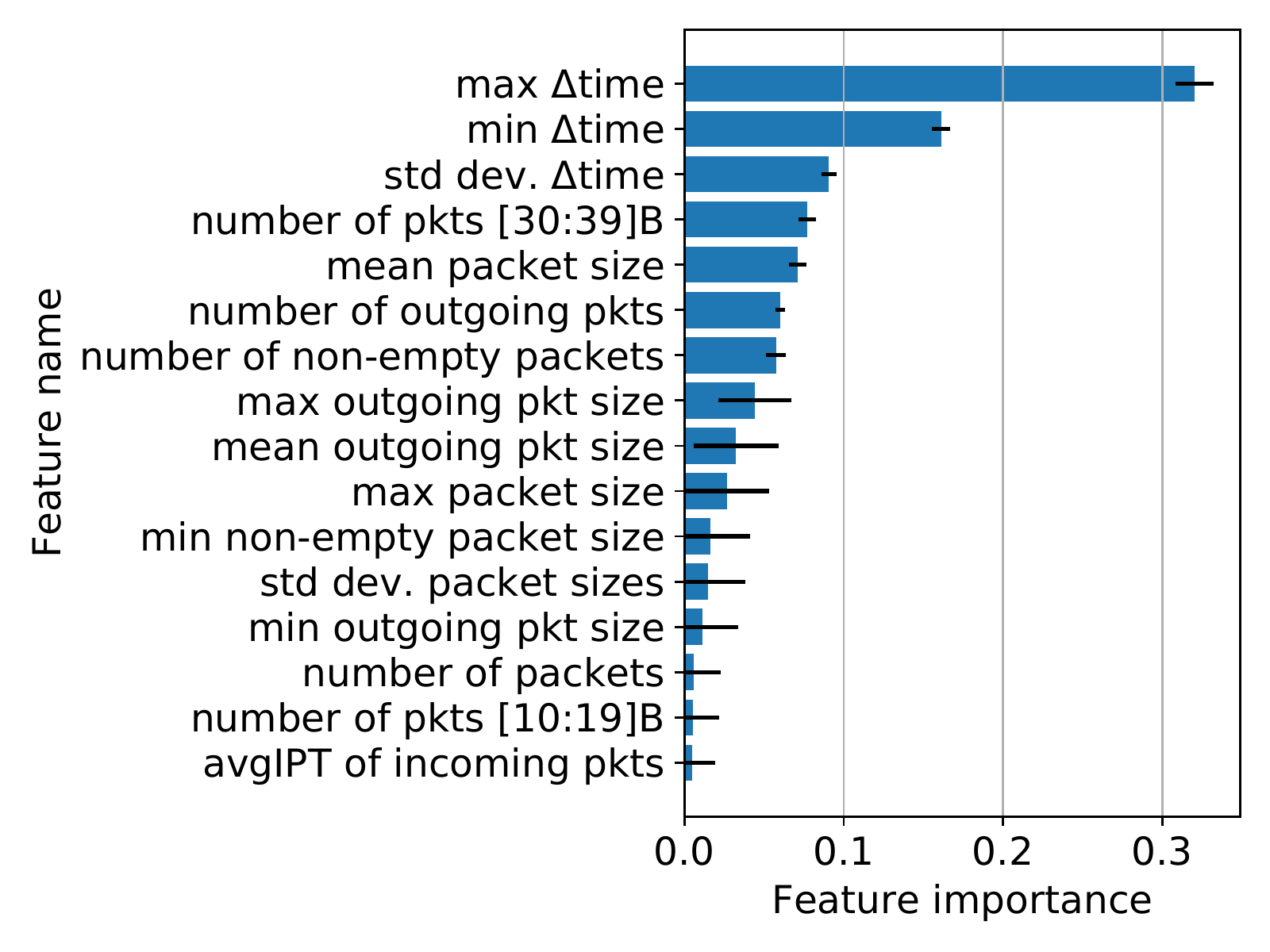}
\label{fig:dev-id-fi-cla}
} %
\subfigure[LE devices.]{\includegraphics[width=0.49\linewidth]{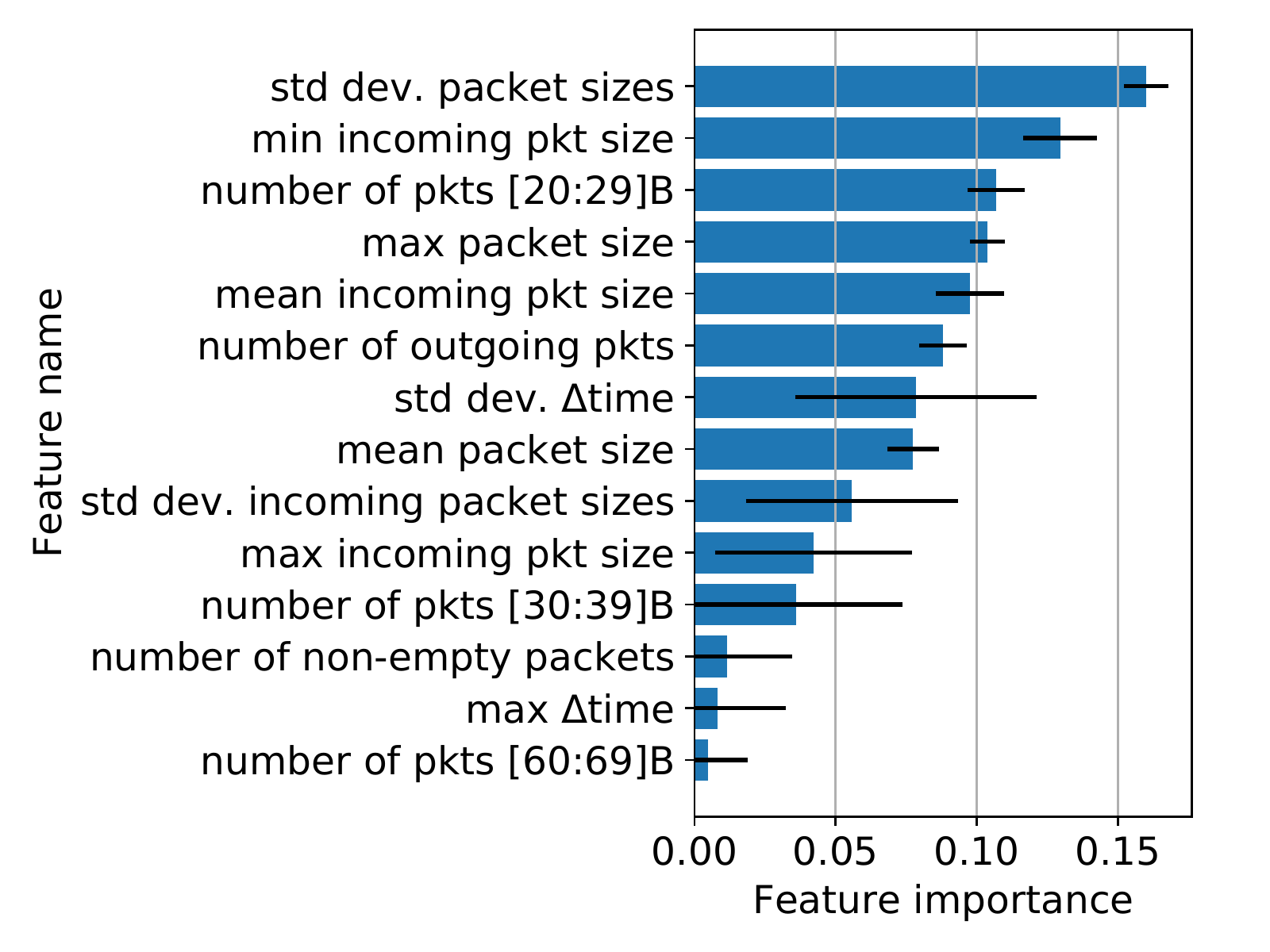}
\label{fig:dev-id-fi-le}}
\caption{Feature Importance for device identification. $\Delta$\feat{time} is the sequence of inter-arrival times.
\feat{avgIPT_X} are the features from Saltaformaggio \etal~\citep{saltaformaggio2016eavesdropping} also describing inter-arrival times.
The x-axis  indicates the relative feature importance.}
\label{fig:dev-id-fi}
\end{figure}

We perform a feature importance analysis and find that timings are crucial for discriminating among the Bluetooth Classic devices: all three most important features are related to inter-packet timings (Figure~\ref{fig:dev-id-fi-cla}).
This corroborates the findings of Aksu \etal~\citep{aksu2018identification} who show that the traffic from $6$ smartwatches has a distinct inter-packet timing distribution, and is in agreement with fingerprinting results in other domains (\eg website fingerprinting based on Tor traffic~\citep{rahman2020tik}).
In the case of Bluetooth Low Energy, the classifier selects primarily size-based features (Figure~\ref{fig:dev-id-fi-le}).
We postulate that this difference is due to the increased capabilities of Bluetooth Classic devices (\ie high-end smartwatches) compared to LE devices that consist of simpler devices.
Smartwatches support a wide range of possible actions that can generate small or large amounts of data, which makes global-volume-based features less stable than timings that are inherently tied to the OS and the hardware.
On the contrary, LE devices support only limited functionalities (\eg activity tracking or heart rate monitoring), and their communication pattern is inherent to the nature of the application, making size-based features discriminative across devices.
Another possible explanation is that due to the absence of a block cipher on the link-layer, BLE packet sizes reveal more information to an eavesdropper.
Finally, we observe that the relative feature importance is less pronounced than in the case of Bluetooth classic, thus indicating that the classifier needs more features to successfully distinguish the samples.

\para{Chipset Fingerprinting.}
We note that our classifier described above fingerprints a combination of the hardware, the firmware, the OS, and the applications installed.
To specifically target the hardware (\ie chipset manufacturer), we reproduce the same experiment, this time using the Bluetooth chipset manufacturer as the classifier's label.
Our goal is to investigate if our classifier learns common patterns from the encrypted traffic recorded by devices that have the same chipset manufacturer, \eg \dev{Samsung Galaxy Watch} and \dev{Huawei Watch 2}, which both have a \dev{Broadcom} Bluetooth chip or \dev{Mi Band 2, 3} and \dev{4} all equipped with a Dialog Semiconductor chip.
Once again, we find that the classifier's performance is high ($96\%$ precision/recall/F1 score, Figure~\ref{fig:chipset-id-cm}, Table~\ref{table:chipset-identification-accuracy}), which indicates that there exist stable communication patterns across chipsets.
We remark that the limited sample size (\ie $1-3$ devices per chipset manufacturer) limits the generalization of our conclusions; further analysis is needed in this specific direction.
Nonetheless, this result demonstrates the robustness of our methodology, as our earlier results show that our device-identification classifier can distinguish between devices from the same vendor.

\para{Take-Aways.}
Our experiments confirm that there exist discriminating features across the encrypted traffic of Bluetooth wearable devices; an eavesdropper can use these features to differentiate devices.
This raises a question about the level of protection offered by tracking countermeasures, \eg MAC address randomization, employed by the Bluetooth LE protocol.
Moreover, our results show that Bluetooth Classic devices are distinguishable due to their communication time patterns, whereas LE ones have distinct size patterns.
We also find that the traffic of devices with the same chipset manufacturer have common patterns.
Despite these common patterns, devices from the same vendor can still be distinguished from each other using the device-identification methodology, demonstrating the robustness of our approach.
Finally, we remark that device identification is an entry-level attack to other attacks (\eg action or application identification described next) that aim at inferring more sensitive information about the owners of wearable devices.

%Aksu \etal~\citep{aksu2018identification} reach similar conclusions using $6$ smartwaches and a more heavy-weight machine-learning pipeline that trains 20 algorithms for classification.

\section{Action \& Application identification}
\label{sec:act_id}

We now consider the adversarial goals G2/G3 and focus on recognizing user-actions from their corresponding encrypted Bluetooth communications.
Actions are related to the capabilities of wearable devices: measuring a heartbeat, beginning a workout, tracking a meal or medicine intake, playing music, etc.
Our dataset analysis shows that most user-actions — even the mere opening of an application on a smartwatch —  result in Bluetooth communication with the paired phone.
In this section, we train a classifier to recognize user-actions from the metadata of these encrypted Bluetooth communications.

\noindent Due to technical constraints, our experimental evaluation is two-fold and ``T-shaped'':
\begin{enumerate}
\item \textbf{Wide-part} (§\ref{subsec:app-id-wearables}): We run the attack on all the wearable devices of our testbed (Tables~\ref{table:devices-classic} and~\ref{table:devices-le}). Since most of them are not automatable due to their proprietary OS/firmware, we use the samples that we manually trigger by performing the appropriate action (\eg clicking a button on the wearable or performing a short running workout).
\item \textbf{Deep-part} (§\ref{subsec:app-id-wearos}): We build and use an automation pipeline for Wear OS devices to generate more data.
We use this enhanced dataset (1) to identify applications running on a smartwatch, for instance religious or medical applications, (2) to identify fine-grained actions within specific applications, for instance the action ``record an insulin injection'' in an application that is used to manage diabetes, and (3) to model an attacker capturing traffic over a longer period of time (hours) and attempting to recognize actions and applications in the trace.
\end{enumerate}

\para{Methodology.} For our two-fold experimental evaluation, we follow the methodology described in §\ref{sec:methodology} except that we enhance the set of extracted features (see next point).
We use the \act{Action} (or \act{Application}) label as our classifier's target.
In subsections §\ref{subsec:app-id-wearables} and~§\ref{subsec:app-id-wearos}, we include further details that depend on the specific experimental settings.

\para{Feature Extraction \& Classification.}
We slightly modify the features described in §\ref{sec:methodology}.
First, we replicate the $3$ packet sequences: (1) from Master to Slave, (2) from Slave to Master, and (3) sequences consisting of all non-null, non-ACK packets, and we remove small packets from the copies.
We empirically observe that small packets are not useful to the classifier because they are common across applications and actions.
After experimenting with different thresholds, we filter out packets smaller than $46$B from the $3$ new time-series.
As before, we extract the min/mean/max/count/standard deviation from these time-series, which leads to $15$ additional scalar features.
Furthermore, we tweak the features proposed by Liberatore and Levine~\citep{liberatore2006inferring} regarding the counts of packets in certain size ranges.
Recall that, for device identification (§\ref{sec:dev_id}), we used coarse, 10-byte wide buckets.
However, our analysis of the Bluetooth traffic concerning user actions shows that some of them produce consistent and unique packet sizes.
As a result, we define more fine-grained buckets and record the number of packets with size $x$, for $x\in[46; ~1,005]$ bytes.
We ignore packets above $1{,}005$~B, the maximum payload size in our dataset.
This leads to $960$ scalar features replacing the $10$ described in §\ref{sec:methodology}.
Finally, we retain the same timing features as those described in §\ref{sec:methodology}.
Overall, with our modified approach, we extract $997$ features that we feed to our random forest classifier.
To cope with the increased number of labels (\ie $49$ actions in the ``wide' experiment of §\ref{subsec:app-id-wearables} and $56$ applications for the ``deep'' experiment of §\ref{subsec:app-id-wearos}), we set the number of trees in the Random Forest algorithm to $30$ and we use Recursive Feature Elimination to retain the $50$ most important features.

\subsection{``Wide'' Experiment on All Wearable Devices of our Testbed}
\label{subsec:app-id-wearables}

We first focus on devices whose Bluetooth traffic capture can not be automated: step counters and fitness trackers (\dev{Fitbit Charge 2-3, Huawei Band 3e, Mi Band 2-3-4}), and smartwatches (\dev{Fitbit Versa 2, Apple Watch, Samsung Galaxy Watch}), all with proprietary OS/firmware.
We also include manually-triggered actions from Wear OS watches (\dev{Huawei Watch 2, Fossil Explorist}), but not machine-automated ones that we use in §\ref{subsec:app-id-wearos}.
The subset of the dataset that we use in this section consists solely of Bluetooth traces triggered by the corresponding human action.
We demonstrate that most actions performed on the wearable devices under examination result in a distinctive encrypted Bluetooth communication, and that an eavesdropper can automatically and passively recognize user actions (\eg starting a workout) and various events (\eg the reception of an e-mail/SMS/phone call).

\begin{figure}
\centering
\subfigure[Normalized confusion matrix per true label.
Red squares regroup actions within one application.
Full figure: Figure~\ref{fig:application-identification-wearables-full}.]{\includegraphics[width=0.49\linewidth]{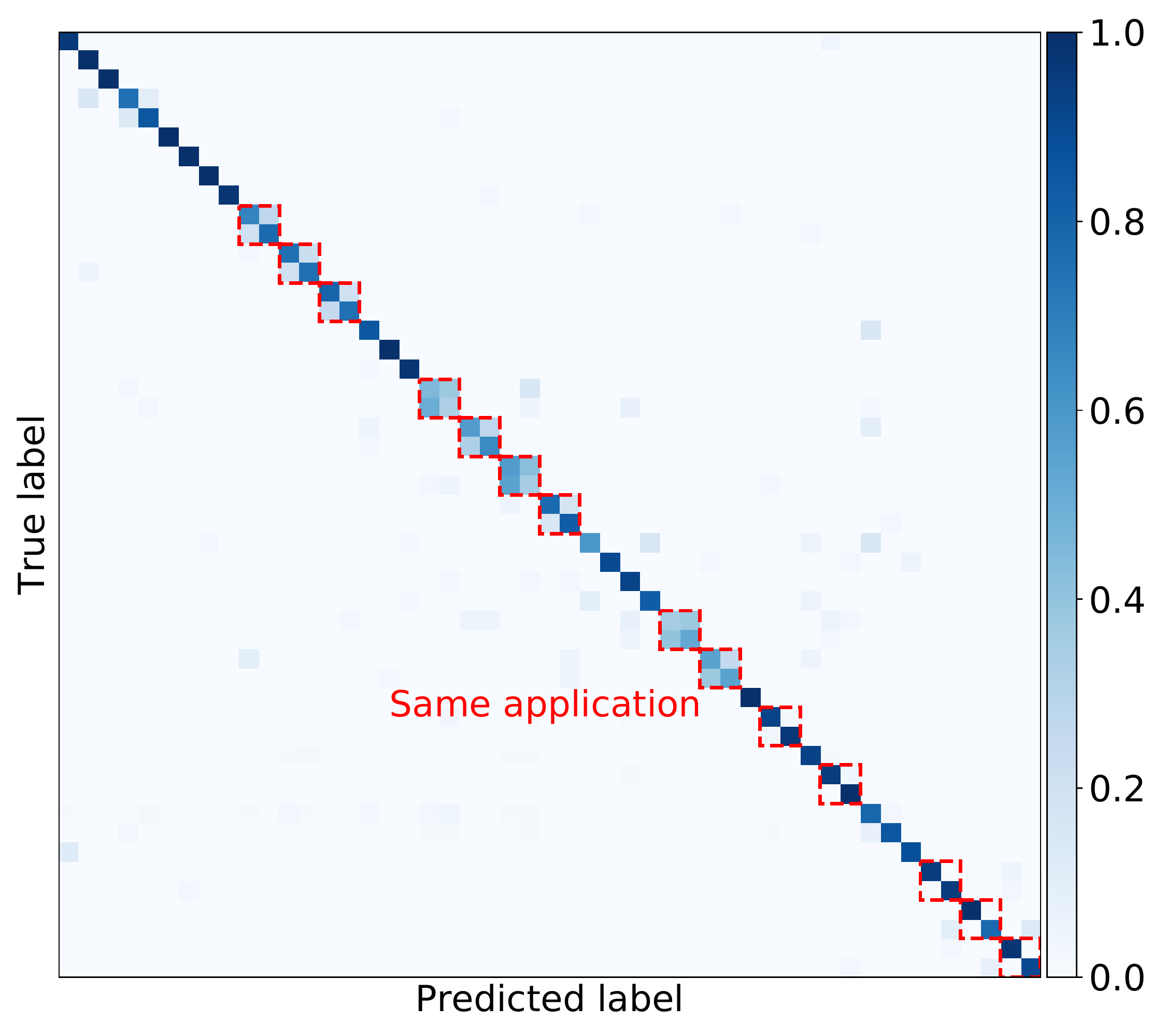}
\label{fig:application-identification-wearables-cm}
} %
\subfigure[Feature importance.]{\includegraphics[width=0.49\linewidth]{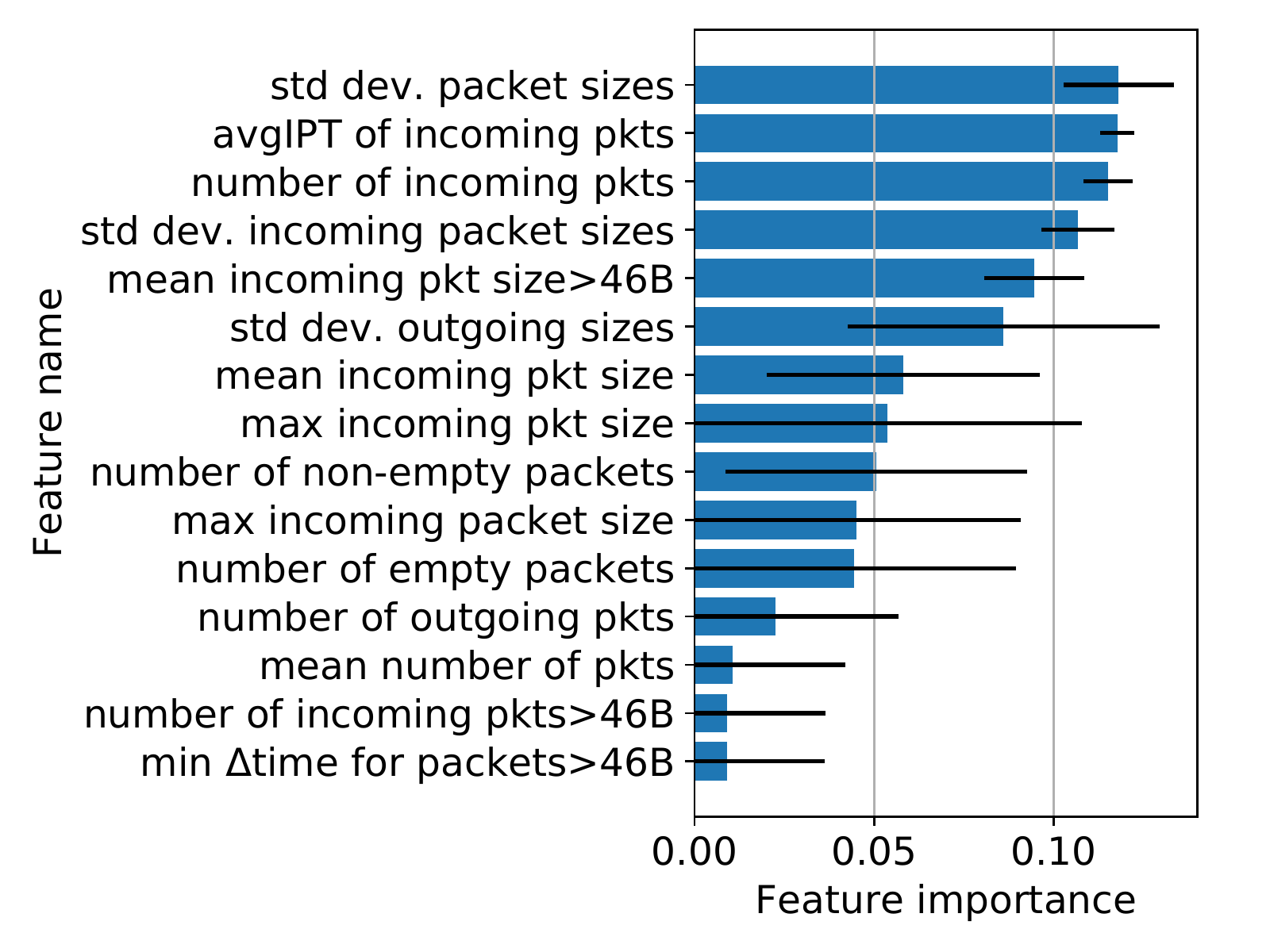}
\label{fig:application-identification-wearables-fi}}
\caption{Application and Action Identification for human-triggered actions (``wide'' experiment).}
\label{fig:application-identification-wearables}
\end{figure}

\para{Attack.} We do not assume that the adversary knows the device; we train a classifier to infer both the device and the action in one step.
A true positive occurs when both the device and the action are guessed correctly.
We enumerate and collect $49$ actions from the set of wearable devices considered and the companion apps installed on the phone (\eg \act{Measure Heartbeat}, \act{Record Food/Water} intake).
The dataset is balanced: for each action, our dataset contains between $20$ and $25$ samples.

\para{Results.}
The classifier performance over $49$ actions is $82\%$ precision/recall/F1 score (Figure~\ref{fig:application-identification-wearables-cm}).
We observe that most actions are recognized with a recall close to one which demonstrates that their corresponding Bluetooth communications have distinct patterns.
This includes potentially sensitive user actions such as measuring heart rate, beginning a workout, receiving an e-mail or phone call.
Moreover, we note that there exist few classes with lower accuracy compared to others (with precision/recall between $40\%$ and $60\%$).
Our analysis shows that these classes concern actions within the same application, \eg \act{EndomondoApp_Running} or \act{EndomondoApp_Walking}, on the \dev{SamsungGalaxyWatch}.
However, this can be a limitation due to the number of samples ($20$--$25$ per class); we show in further experiments that fine-grained actions within one application can be inferred with more data (§\ref{subsubsec:diabetesm}).
%Surprisingly, the same applications whose actions are confused among themselves (\app{Endomondo}, \app{FITIVApp}, \app{MapMyRun}) on the \dev{SamsungGalaxyWatch} are more accurately recognized on the \dev{FossilExploristHQ} smartwatch. \todo{find out why - yeah, not clear what we want to say here maybe talk about workout apps that we can not exactly tell if the user is running or not?even though the accuracy might not be that bad - we could sell it as a great result, look we can even find details.}
We also observe that the classifier accurately labels the same action \act{Running} on devices from the same line of products: \dev{MiBand2}, \dev{MiBand3}, \dev{MiBand4} demonstrating how it performs device and action identification in one step.
Finally, we observe that our classifier yields the same accuracy for the devices \dev{Fitbit2} and \dev{Fitbit3} -- which only \act{Sync} with their connected smartphone -- thus confirming our Device Identification results (§\ref{sec:dev_id}): an eavesdropper can recognize the communicating devices based on their metadata.

We perform a feature analysis and find that the most important features are based on communication volumes (\feat{number of incoming}, Figure~\ref{fig:application-identification-wearables-fi}) and packet sizes (features 3-8 of Figure~\ref{fig:application-identification-wearables-fi}), with a single highly-ranked feature about timings (\feat{avgIPT of incoming packets}).
This is consistent with our initial dataset analysis that demonstrated that various user actions trigger Bluetooth traffic with distinct packet sizes.
Finally, we remark that in this experiment, the classifier recognizes the device and the application in one step.
In §\ref{subsubsec:diabetesm}, we improve the classifier's accuracy for identifying actions within the same-application by assuming that device and application identification has already taken place and by tailoring the training of our classifier to one specific application.

\subsection{``Deep'' Experiment on Wear OS devices}
\label{subsec:app-id-wearos}

We now perform an in-depth analysis of Bluetooth communications' metadata on Wear OS smartwatches.
We use automation to increase the size of our dataset, and we demonstrate the performance of our methodology on various adversarial tasks.
To ensure that our synthetic (\ie computer triggered) actions generate plausible Bluetooth traffic, we restrict ourselves to a one-click action: the opening of an application on the smartwatch.
We argue that this should be independent of the wearer's data and inputs and should generalize across users.
The first purpose of this section is to demonstrate that the simple opening of a smartwatch application generates (encrypted) Bluetooth traffic patterns that can be accurately recognized by an eavesdropper (§\ref{subsubsec:app-id-wearos}).
Then, we investigate if the classifier trained by an adversary generalizes and transfers across different smartwatch-smartphone pairs, \ie to a different setup than that used offline by the adversary (§\ref{subsubsec:transferability}).
Furthermore, we build upon our device and application identification attacks, and we target a specific application used to manage diabetes to infer actions within that application.
In~§\ref{subsubsec:diabetesm}, we demonstrate how an eavesdropper can passively recognize sensitive, health-related actions (\eg record an insulin injection) within this specific application.
Subsequently, we show how a persistent adversary that continuously monitors the Bluetooth traffic of a user can extract her profile by inferring her application openings and actions over the course of a day (§\ref{subsubsec:long-term-capture}).
Finally, we briefly investigate the effect of dataset \textit{aging} on the attacker's classification performance (§\ref{subsubsec:aging}).

\subsubsection{Application Identification}
\label{subsubsec:app-id-wearos}

\begin{figure}
\centering
\subfigure[Normalized confusion matrix for $56$ apps, Wear OS\@. Apps are sorted by increasing median transmitted volume.]{\includegraphics[width=0.49\linewidth]{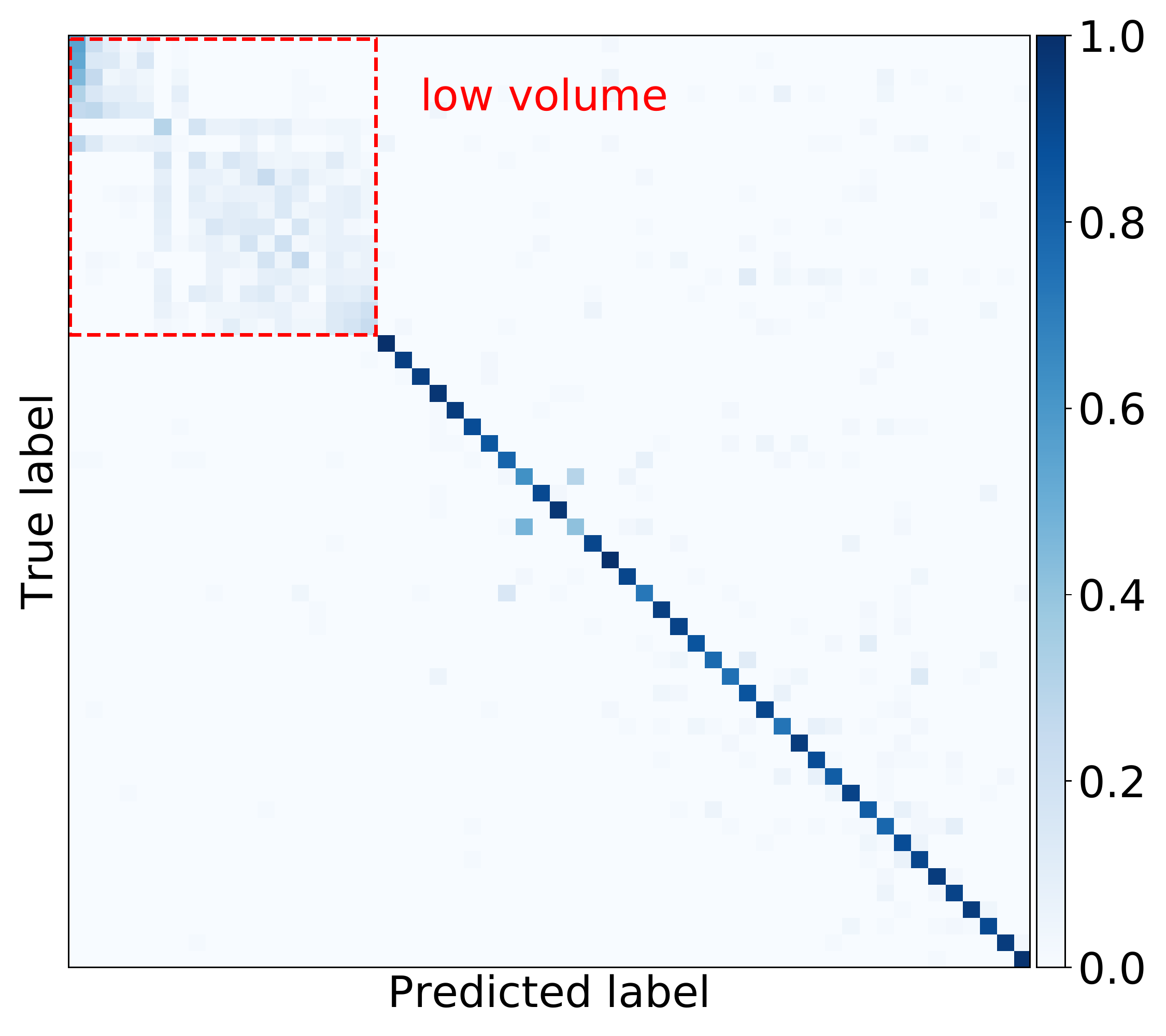}
\label{fig:wearos-app-id-confusion-matrix}
} %
\subfigure[Feature importance for application identification, ``deep'' experiment.]{\includegraphics[width=0.49\linewidth]{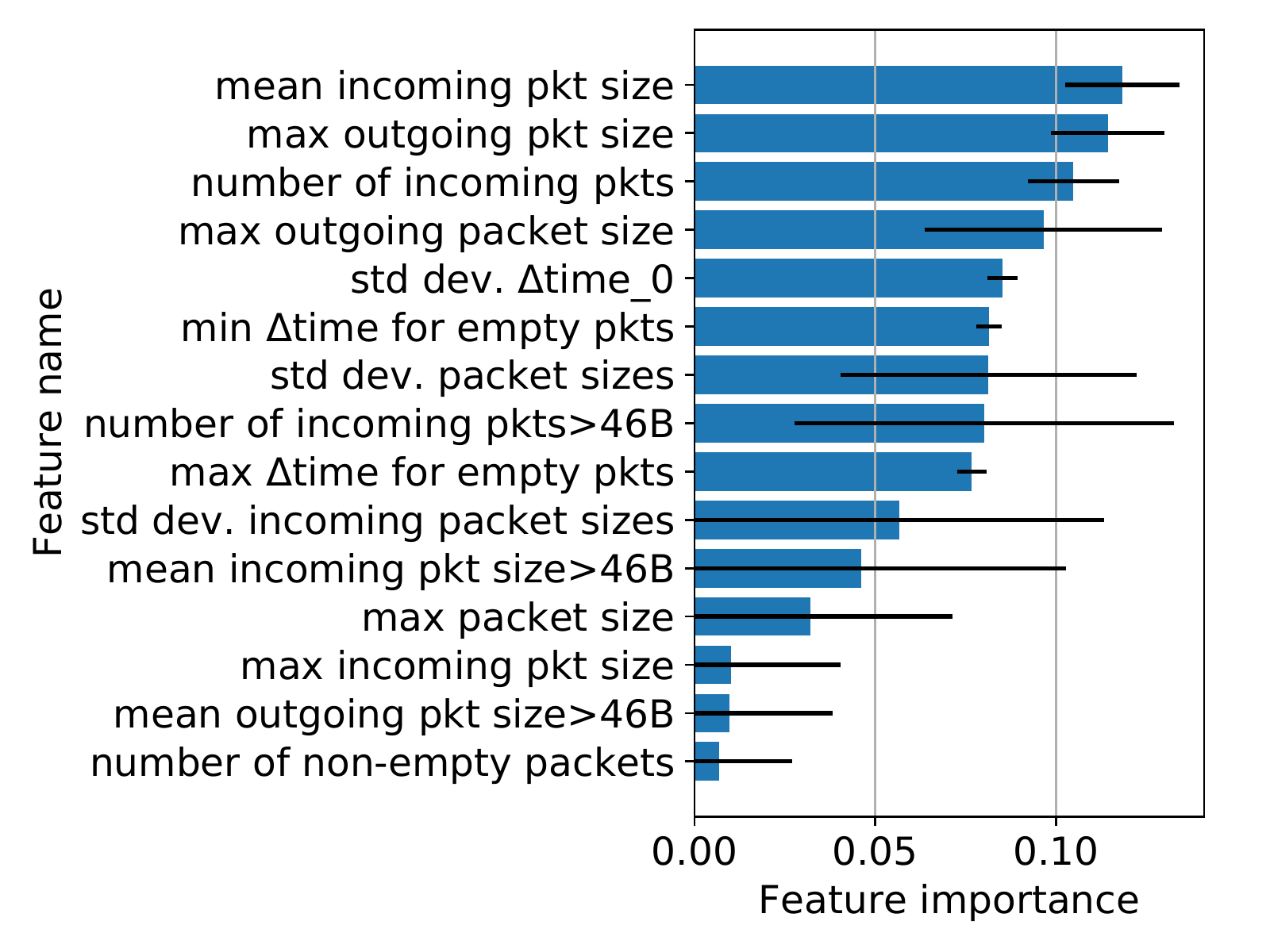}
\label{fig:app-id-fi}}
\label{fig:classifier_accuracy_apps_openings}
\caption{Normalized confusion matrices per true label for recognizing smartwatch application openings.}
\end{figure}

We train our classifier to infer the opening of specific applications on a smartwatch.
Identifying app openings has the benefit of being (1) independent of user actions and data more than of the recognition of actions (§\ref{subsec:app-id-wearables}), and (2) easier to automate (and hence to train a classifier upon) for an adversary.

We remark that the mere presence of an app can leak sensitive information about the wearer.
For instance, medical and well-being applications (\eg medication reminders, applications to stop smoking) hint that the wearer is concerned with a medical condition, and religious (\eg prayer time reminders) or news applications with a political orientation can reveal information about the users' beliefs.
Even less revealing apps can be useful to a long-term adversary; users naturally install applications based on their interests and behaviors, and the list of apps on their wearable device can be exploited to build personal profiles. We argue that it is difficult to foresee whether the presence of an application is sensitive or not, especially when considering long-term profile building based on data from multiple sources, \eg for machine-learning-based advertising~\citep{perlich2014machine,dave2014computational}.
We envision that as Bluetooth sniffing technologies are becoming less expensive (see our discussion in §\ref{sec:discussion}), Bluetooth traffic could become a valuable source of information.
We remind the reader that companies are currently experimenting with Bluetooth-based ``proximity advertising'', a technology used to track users and display local targeted advertisements in transportation systems, airports, and supermarkets~\citep{the_drum_london_cabs, proxbook, stores_secret_surveillance}.

\para{Automation Pipeline \& Applications.}
We use the automation part of our Bluetooth traffic capture pipeline (§\ref{sec:methodology}) that consists of a Linux laptop that coordinates with a Windows machine that records traces via a Bluetooth sniffer.
Using adb and monkeyrunner~\citep{monkeyrunner}, the Linux laptop issues synthetic clicks and swipes to a Wear OS smartwatch connected over Wi-Fi.
We force the watch to send data over Bluetooth (rather than Wi-Fi) by making sure that the Wi-Fi network does not have Internet access.
We do not perform UI fuzzing; we manually specify the clicks and swipes needed to perform the desired actions on the watch.
The exact same clicks and swipes are reused to repeat an action.
At the time of writing, due to the lack of debugging tools, only Wear OS smartwatches could be automated in the desired way.
Devices running Tizen OS should support automation using a variant of adb called sdb~\citep{sdb}, however, the current API does not enable it. Our first experiment uses a Huawei Watch 2 (LEO-BX9)\footnote{We emphasize that the issue we present appears to be generic to wearable devices and not specific to the device selected in this experiment.} running Wear OS 2.16, paired with a Pixel 2 running Android 9.

We select $56$ applications from the Google (Wear OS) Play store, favoring top-rated applications per category.
Our choice of applications was constrained by the availabilities of apps on the Swiss Play store.
Our list includes:
\begin{itemize}
\item Religious apps (\app{DuaKhatqmAlQuran}, \app{AthkarOfPrayer}, \app{SalatTime}, \app{DCLMRadio})
\item Health-related apps (\app{DiabetesM}, \app{Medisafe}, \app{SmokingLog}, \app{Qardio}, \app{HeartRate})
\item Lifestyle-related apps (\app{Lifesum}, \app{Calm}, \app{DailyTracking}, \app{HealthyRecipes}, \app{SleepTracking}, etc)
\item Sport/Activity-related apps (\app{Endomondo}, \app{FitWorkout}, \app{FITIVPlus}, etc)
\item Local newspapers (\app{ChinaDaily}, \app{WashingtonPost}, \app{Meduza}, \app{Krone})
\item Mobile banking and finances (\app{ABS}, \app{Mobills})
\item ``Local guides''/maps (\app{Citymapper}, \app{Foursquare})
\item Communication apps (\app{Telegram}, \app{Glide}, \app{Outlook})
\end{itemize}

\noindent We also include stock applications (\eg \app{Reminders}, \app{Weather}, \app{Phone}), as well as common applications (\eg \app{Telegram}, \app{Translate}) as a control for the sensitive groups and to increase the number of applications.
Finally, we include a \app{NoApp} label that corresponds to background communications between the smartwatch and the phone.

\para{Results.}
The classifier's performance (precision/recall/F1 score) over the $56$ apps is $64\%$.
However, we find that the precision and recall per class varies greatly (Tables~\ref{table:app-id-huaweiwatch-cm-low} and~\ref{table:app-id-huaweiwatch-cm-high}).
In particular, the majority of apps ($38$ out of $56$ apps) is classified with a mean accuracy close to $1$, whereas a smaller subset of the apps is confused among each other by the classifier.
Our analysis shows that this concerns apps that trigger minimal Bluetooth communications upon their opening (\eg \app{Battery}, \app{Flashlight}); this fact is visible when we order the confusion matrix by the median transmitted volume (Figure~\ref{fig:wearos-app-id-confusion-matrix}).
We call these apps ``low-volume'', as they communicate less than $200$ bytes for at least $75\%$ of their samples (red dashed square of Figure~\ref{fig:wearos-app-id-confusion-matrix}).
However, except for \app{Battery} and \app{Reminders}, which are the unique apps that trigger absolutely no communication, we find that low-volume apps communicate a small amount of information on the Bluetooth layer (Table~\ref{table:low-volume-apps-details}).
To further investigate the difference across the two subsets of apps, we train two separate classifiers: one specifically targeting the $18$ low-volume apps and another one to distinguish among the remaining $38$ non-low-volume apps (Figures~\ref{fig:wearos-app-id-confusion-matrix-filtered2} and~\ref{fig:wearos-app-id-confusion-matrix-filtered}).
The mean accuracy of the first classifier reaches $17\%$ with high variance across the labels, which shows that the data exchanged by these apps is simply too small and/or too variable to be learned by the model.
We find that the \app{NoApp} label that corresponds to background communications between the smartwatch and the smartphone is part of the low-volume group.
This indicates that low-volume apps are not only hard to distinguish among their peers but are also difficult to differentiate from the absence of activity.
On the contrary, the non-low-volume apps are recognized by the second classifier with a high accuracy of $90\%$, demonstrating the practicality of the attack in this case.
We observe that — as before — the important features are based on the sizes of packets and the timings, with an unsurprising emphasis on the direction smartwatch$\to$smartphone for the timings (Figure~\ref{fig:app-id-fi}).

\begin{figure}
\centering
\subfigure[Normalized confusion matrix per true label.]{\includegraphics[width=0.49\linewidth]{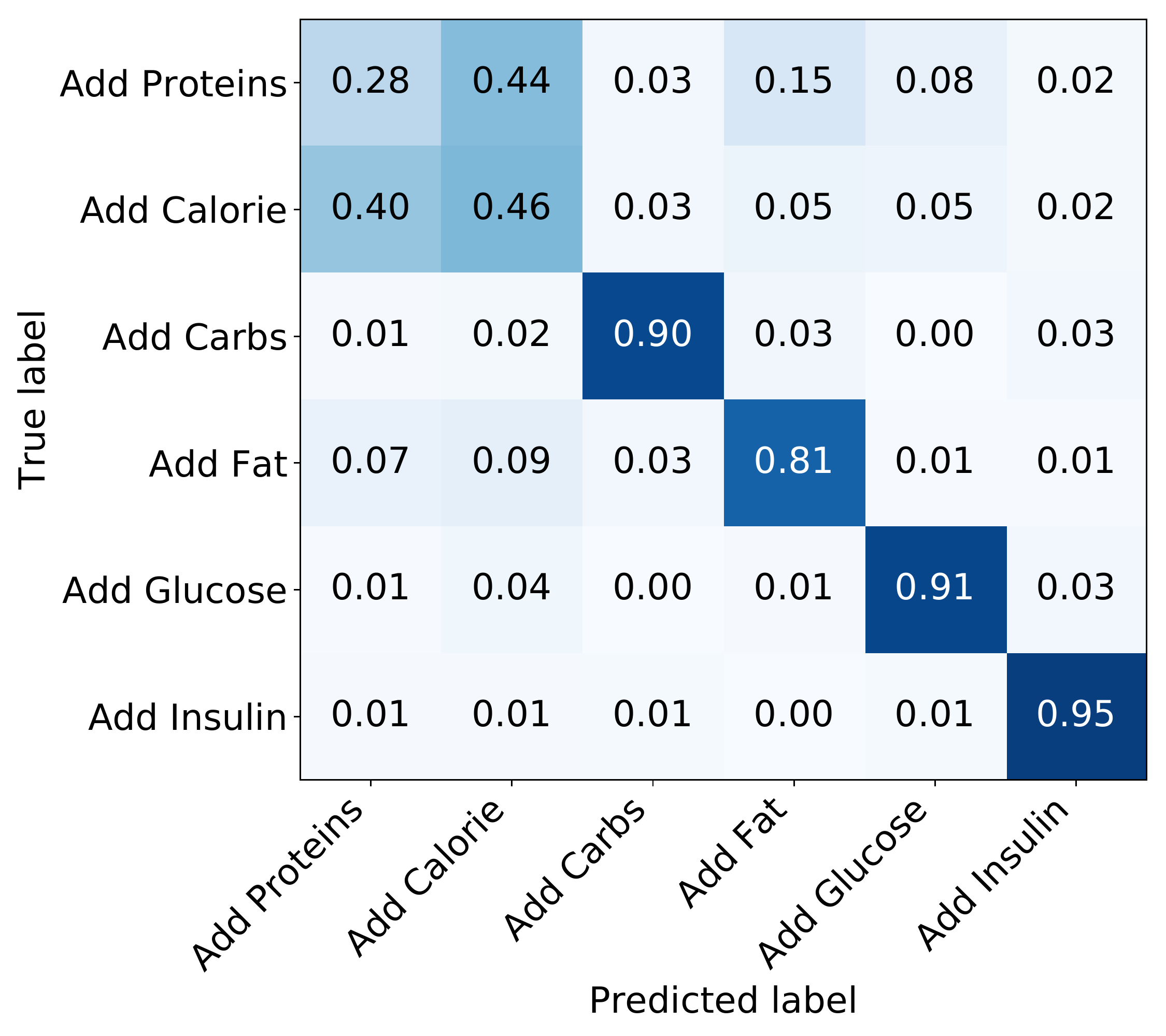}
\label{fig:action-id-diabetesm-cm}
} %
\subfigure[Feature importance.]{\includegraphics[width=0.49\linewidth]{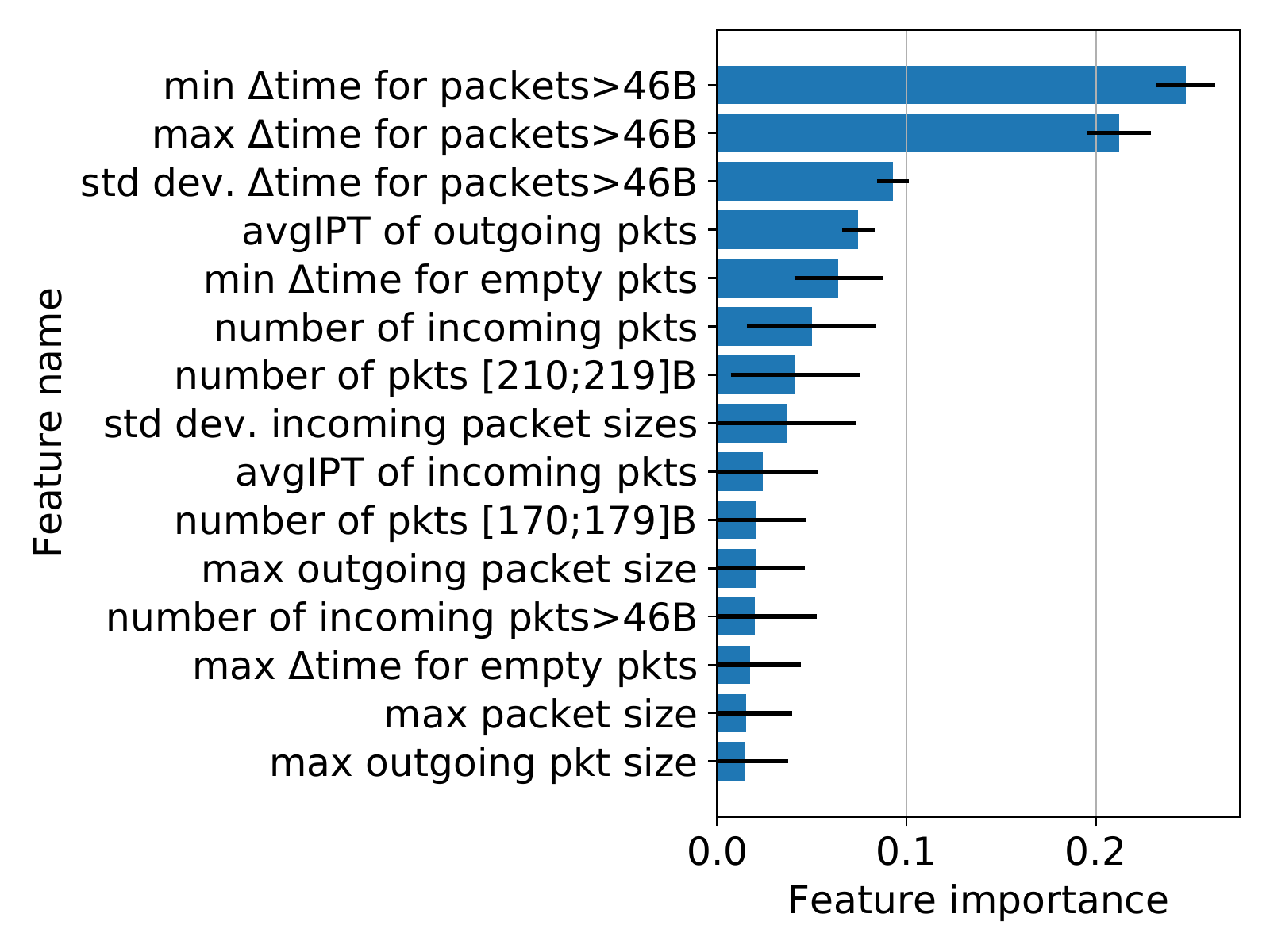}
\label{fig:action-id-diabetesm-fi}}
\caption{Fingerprinting fine-grained action within on application: \app{DiabetesM}.}
\end{figure}

\subsubsection{Model Transferability}
\label{subsubsec:transferability}

Our application-identification attack (§\ref{subsubsec:app-id-wearos}) was successful on a single smartwatch-smartphone pair (\ie Huawei Watch 2 - Pixel 2).
However, our attack would require the adversary to train a classifier on the particular pair of devices that the target possesses.
In this subsection, we investigate if the trained model generalizes to other devices, upon which the adversary has never used for training.

In more detail, in the previous experiment we use a Huawei Watch 2 (LEO-BX9) running Wear OS 2.16, paired with a Pixel 2 phone running Android 9.
In this experiment, we include a new pair of devices: a Fossil Q Explorist HR smartwatch running Wear OS 2.16, paired with a Nexus 5 phone running Android 6.
We could not downgrade one Wear OS device to a different version to introduce more variability to our experiment.
Also, we could not include the Apple Watch / iPhone pair in the transfer experiment due to a lack of overlap in the apps that we selected and the apps available in the Apple Store.
Hence, we leave the question of cross-OS transferability as an interesting future work.

We select $34$ apps from the Huawei-Pixel pair that could also be installed on the Fossil-Nexus pair (Table~\ref{table:transfer-apps}).
We follow the same methodology, except that we use all the samples collected from one pair of devices as the classifier's training set and the data collected from the other pair as the testing set.
We perform our experiments in both directions.

Our results show that the trained model generalizes well: it has a precision/recall/F1 score of $81\%$ when the data collected from the Huawei-Pixel pair is part of the training set and the data collected from the Fossil-Nexus pair is the testing set (Figure~\ref{fig:transfer1}, Table~\ref{table:app-id-transfer1-cm}).
The classifier's precision/recall/F1 score reaches $86\%$ when we perform the experiment in the opposite direction (Figure~\ref{fig:transfer2}, Table~\ref{table:app-id-transfer2-cm}).
Some apps are misclassified: our analysis shows that these are applications that are native to the OS (\app{Fit Packages}).
This is not surprising, given the differences in major Android versions.
However, and more importantly, our experiment shows that fingerprinting non-native applications by their (encrypted) Bluetooth traffic is possible independently of the smartphone's OS version, which demonstrates the robustness of our attack methodology.

\subsubsection{Fingerprinting Fine-grained Actions Within an Application}
\label{subsubsec:diabetesm}

We now use the application-opening identification (§\ref{subsubsec:app-id-wearos}) as a stepping stone to another attack that aims at inferring potentially sensitive actions within one particular application.
For a use-case, we choose the app \app{DiabetesM} that helps people diagnosed with diabetes to keep track of their meals and medicine intakes.
Although the Huawei smartwatch that we use for this experiment is not marketed as a medical device, this information is clearly of medical nature.

We follow the previous methodology and use the automation tool to generate traffic that corresponds to the usage of the \app{DiabetesM} app.
We manually program user interactions (\ie pressing buttons) within the application and capture the traffic of $6$ actions related to the management of meals and medicine intakes: \act{Add Calorie}, \act{Add Carbs}, \act{Add Fat}, \act{Add Glucose}, \act{Add Insulin}, and \act{Add Proteins}.
We collect $150$ samples per action, map them to feature vectors using the features described in §\ref{sec:act_id} and split them into $80\%$ for training and $20\%$ testing.
Then, we train a classifier tailored to this particular app that aims to distinguish among the $6$ possible user actions, \ie we assume that the application's traffic has been already classified as \app{DiabetesM}.

The overall accuracy of the classifier over the $6$ actions is $70\%$ (Figure~\ref{fig:action-id-diabetesm-cm}, Table~\ref{table:action-id-diabetesm-cm}), which is significantly better than guessing at random.
This indicates that actions within a specific application generate distinct Bluetooth traffic that can be fingerprinted by an eavesdropper.
More importantly, our results show that the sensitive action \act{Add Insulin} is recognized with a precision/recall/F1 score of $90/95/92\%$, respectively, \ie the encrypted communication patterns generated by this sensitive action ``stand out'' from other actions of the \app{DiabetesM} app.

We here remark that all of these actions are semantically similar: they all update a variable in a database stored on the paired smartphone.
We expect that developers could prevent the traffic-analysis attacks by simply padding the traffic corresponding to all of these actions to a constant size.
However, the feature analysis reveals that \emph{timings} matter most, not sizes (Figure~\ref{fig:action-id-diabetesm-fi}).
A closer inspection reveals that a human has to interact with \app{DiabetesM} in two consecutive steps: (a) by increasing a value (\eg pressing ``record insulin injection''), and, (b) by clicking on a ``save'' button on a different screen.
To our surprise, both actions generate traffic, and the classifier detects the timing differences between the two actions.

First, this highlights a limitation of our experiment.
In our methodology, the duration between the press of the first button, the swipes, and then the press of the ``save'' button are constant and precisely reproduced by the automation framework.
A human would produce more variable durations that would be harder to fingerprint by their encrypted traffic.
However, we believe that this finding is still an interesting showcase for the capabilities of traffic-analysis attacks: padding each action into a similar message is not sufficient, because the classifier can ``count'' the number of swipes from the screen containing the save button back to the screen containing the target action.
Hence, it is necessary to obfuscate this duration, or rethink the strategy that the application employs to synchronize data with the smartphone.

\subsubsection{A Persistent Adversary}
\label{subsubsec:long-term-capture}

We now consider a longer-term adversary that aims at identifying the actions performed on a smartwatch by its wearer over the course of a day.
This adversary could be a nosy neighbor or an office eavesdropper, capturing Bluetooth traffic continuously over a long period, and attempting to monitor the habits of the target.
This experiment drops a simplifying assumption made in the previous experiments, \ie that the adversary knew when the traffic related to an action starts and stops.
Indeed, all previous experiments used $30$-second traffic samples, each corresponding to exactly one action.
In this new experiment, the adversary records a continuous traffic capture over $24$ hours, and its goal is to output a series of predictions over this period.
Furthermore, the adversarial task is slightly tweaked: the prediction can be either a user-action, or the \act{NoAction} label corresponding to the absence of user activity.
Finally, we note that due to their long duration, these traffic captures contain background traffic (updates, synchronization) more than the short $30$-sec samples used previously.
The goal of the adversary is therefore to distinguish specific app openings from background noise and OS communications.
%We remark that due to battery constraints, the wearable devices used in our experiments do not run multiple apps at the same time.

\para{Methodology.}
We generate application openings and user actions with the \dev{HuaweiWatch2} smartwatch, using the automation framework presented in the ``deep'' experiment (§\ref{subsubsec:app-id-wearos}).
We simulate one user who wears the watch for $24$ hours.
Over the course of a day, we model the user's interactions with her smartwatch following a recent user study that quantifies smartwatch usage in the wild~\citep{liu2017characterizing}.
In particular, for each $1$-hour slot, the number of interactions is drawn from a probability distribution favoring daytime hours over nighttime ones (Figure $4$ of~\citep{liu2017characterizing}, page $389$).
%In short, the priors of the adversary will be different during the day and the night.
User actions are not triggered with an equal probability: popular applications such as messaging/e-mails, maps, alarms/clocks, and fitness trackers, are more likely to be triggered than others (Table $6$ of~\citep{liu2017characterizing}, page $390$).
We model this by updating their prior probability to $2\times$ compared to that of non-popular applications.
We select $33$ high-volume applications from popular categories and enumerate $17$ user actions within these applications, \eg \act{DiabetesM_AddInsulin}, \act{HealthyRecipes_SearchRecipe}, \act{FitWorkout_Open} (Table~\ref{table:applications-actions-longrun}).
Individually, each of these actions has a short duration ($\le 20$sec).
However, these fine-grained actions follow each others in semantic sequences: \eg we automate the sequence \act{Endomondo_Open}, \act{Endomondo_Running}, waiting $2$min, \act{Endomondo_Close}, which is the equivalent of a $2$-min workout.
The classifier attempts to recognize each individual action (except for \act{_Close} actions).

Then, we automate the recording and triggering of actions.
Due to technical constraints, we record $20$min-captures that we concatenate to form a $24$h capture.
The parameters of each $20$min-capture are drawn from the distribution of the modeled time of the day.
Within one capture, the simulation is a simple state machine that loops over the following actions: (1) it flips a biased coin deciding whether to trigger an action or not, and (2) if the outcome is positive, it draws one action at random following the biased probability distribution, runs the action, and waits for a random time defined by the expected number of events in the modeled time of the day.
In parallel, the smartwatch and smartphone normally exchange background data.

\para{Attack.}
To train its classifier, the attacker uses the short $30$-sec captures corresponding to the $50$ classes (\ie applications and actions) selected.
Moreover, it employs a $51$st class that it uses to model noise: this class contains the background communications of the wearable, recorded as \act{NoApp_NoLabel}.
In addition, we notice that closing an application also generates network communications.
The volume exchanged is low, hence they are difficult to classify.
However, we observe that when treated as noise, they are useful in helping the classifier distinguish between actions of interest and background traffic.
Therefore, we add the classes \act{AppX_Close} for all $33$ applications of interest.
We note that the adversary's dataset is balanced: it contains between $30$ and $40$ samples per class.
Finally, the attacker extracts features, as in the previous experiments (§\ref{subsec:app-id-wearos}), and trains a Random Forest classifier.

During testing, the attacker is provided with the uncut $24$-hour traffic capture.
First, it runs a splitting algorithm that identifies sequences in the capture that possibly correspond to user-actions.
This splitting algorithm uses a sliding window and records the times at which the sum of bytes exchanged in the window is greater than $200$ bytes, following the criteria for ``high-volume'' apps presented in §\ref{subsec:app-id-wearos}.
Then, it classifies the contents of each window; however, it only outputs the most likely class if its probability is greater than a confidence threshold $T$.
If no class meets this criteria, it outputs \act{NoAction}.

\para{Evaluation Metrics.}
We adapt the attack's evaluation metrics to the new task.
A true positive corresponds to a correct prediction in the ``correct'' time-interval, \ie if the time intervals of the real action and the predicted one overlap.
A false positive occurs when the attacker's classifier outputs a label other than \act{NoAction} that does not overlap with a real action of the same label.
Finally, a false negative is when the classifier misses a real action.
Following these definitions, we calculate the classifier's precision/recall/F1 score as usual.

\begin{figure}
\centering
\includegraphics[width=0.5\linewidth]{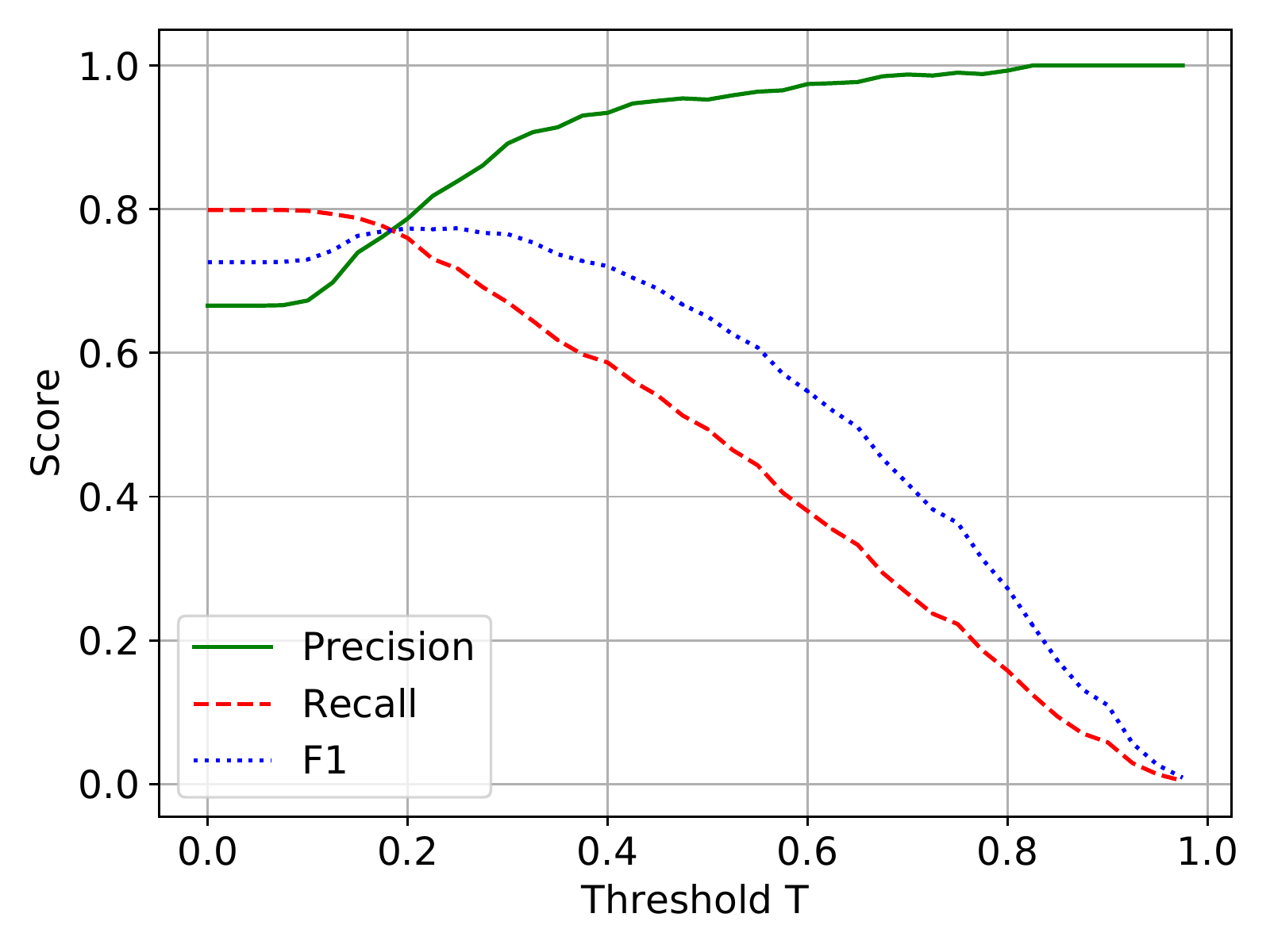}
\caption{Attacker's score when classifying events over $24$-hour captures. The threshold is the minimum confidence needed to output a prediction.}
\label{fig:longrun-threshold}
\end{figure}

\para{Results.}
The results are computed over the $72\times20$min = $24$ hours of the experiment.
We parameterize them with the confidence threshold $T$, which impacts the overall sensitivity: lower $T$ values result in a higher recall and lower precision, and vice-versa.
For the classification task with $51$ classes, the classifier's average precision ranges from $0.65$ to $1$, and its mean recall per class from $0.7$ down to $0$ (Figure~\ref{fig:longrun-threshold}).
The maximum recall is $83.5\%$, for a threshold $T$ of $0.1$, and the corresponding precision is $74.9\%$.
The maximum precision is $1.0\%$, for $T=0.6$, and the corresponding recall is $23.5\%$.
The best F1 score is $0.83$ and is achieved at $T=0.25$.
This experiment demonstrates that a persistent adversary successfully recognizes high-volume applications from the absence of activity and accurately classifies them over the course of the day.
The different values for the confidence threshold $T$ indicate a precision/recall trade-off.
An adversary can choose to optimize its strategy towards one metric or the other (or both) depending on its goals.
For instance, an adversary who aims at recognizing, with high precision, an application of interest or a particularly sensitive action (\eg, \act{AddInsulin}) could do so at the cost of more false negatives (and lower recall).
Whereas, another adversary, \eg a smart billboard displaying an advertisement to a passer-by, aiming to identify the set of applications and actions of its target (that can help to build a profile) could choose a lower decision threshold and achieve better overall performance.

\subsubsection{Dataset Aging}
\label{subsubsec:aging}

\begin{figure}
\centering
\subfigure[Training on day $0$, testing on day $i$.]{\includegraphics[width=0.5\linewidth]{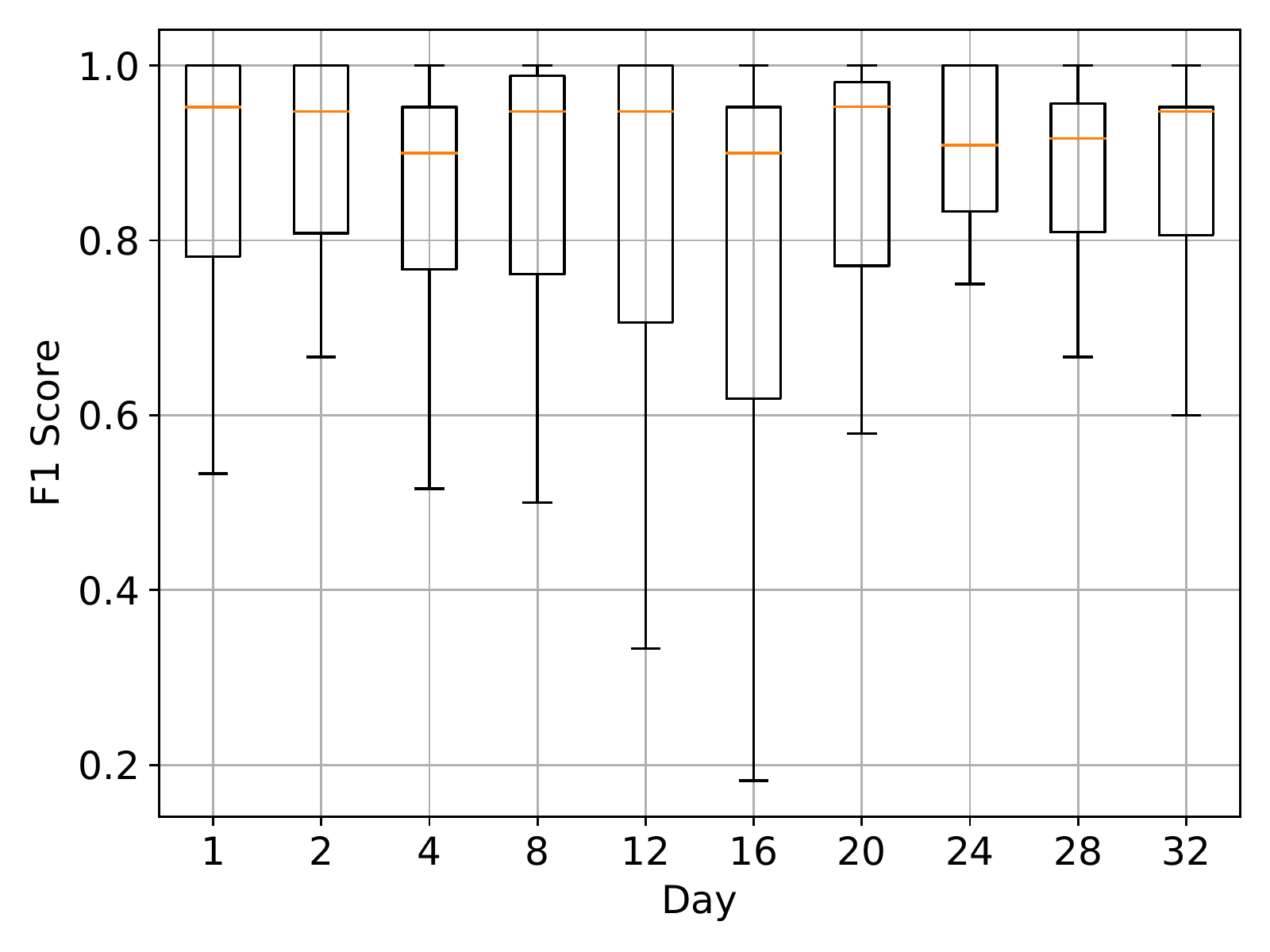}
\label{fig:aging}
}%
\subfigure[Evolution of the F1 score per class, averaged over $32$ days.]{\includegraphics[width=0.5\linewidth]{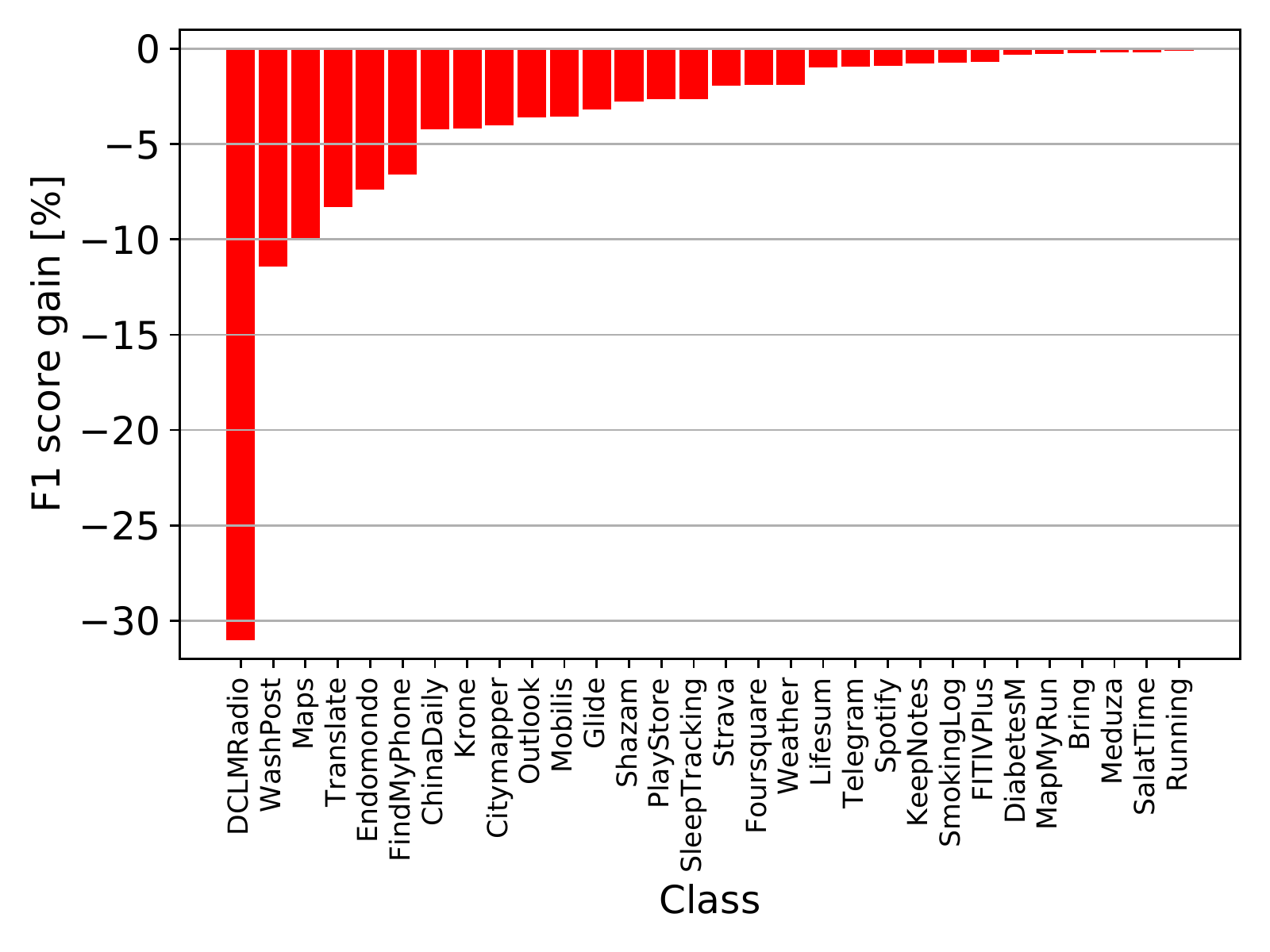}
\label{fig:aging2}}
\caption{Effect of dataset aging on accuracy.}
\end{figure}

Finally, we briefly explore the effect of dataset \emph{aging}, that is, the loss of accuracy incurred as the adversary uses older datasets.
We perform an experiment in which we collect data from applications over $32$ days; then, we measure the classifier accuracy when classifying each day's samples with the model trained on the data collected at day 0.
We use the $38$ high-volume applications presented in Section~\ref{subsec:app-id-wearos}, from which we exclude $5$ applications that were not successfully automated and did not produce traces over the multiple days of this experiment.
We also exclude $3$ applications that stopped communicating data after an update over the month.
Figure~\ref{fig:aging} shows a boxplot of the F1 score per application over the duration of the experiment.
First, we observe that the median F1 score is stable through the month, with small variations between $90\%$ and $95\%$.
The overall performance of the classifier did not vary significantly, even when using traces that are a month old.
We do observe that the F1 score of some classes varies through the month (\eg on days $12$ and $16$, the lower whiskers indicate that some apps were misclassified more than usual).
We investigate the F1 score per class, averaged over the month (Figure~\ref{fig:aging2}).
We see that the different classes have variable F1 scores, in particular with one application being harder to classify: \app{DCLMRadio}.
This a web radio that loads a large amount of content updated daily.
It is unsurprising that training on a single day is not representative of the whole month.
Nonetheless, other applications are still classified with high accuracy after $32$ days.

Finally, to improve the classification, we also experiment with training the classifier using the data of several days.
We use the data collected during the initial three days (instead of just the first), while keeping the same amount of training data (corresponding to a single day).
We observe that training on just a few days leads to a performance increase of $3\%$ mean accuracy (from $92\%$ to $95\%$) over the month.
The F1 score of \app{DCLMRadio}, the worst class, also increases by $14$ percentage point (from $68\%$ to $82\%$; Appendix, Figure~\ref{fig:aging-3-days}) .
This suggests that an adversary does not need a fresh dataset to perform the attack.
Thus, this lowers the cost of the attack by reducing the amount of training needed and by facilitating dataset reuse.

\color{black}

\subsection{Summary of the Attacks}

% A long-term eavesdropper  can recognize nearby activities on wearable devices.
Overall, the experimental results of this section demonstrate that different actions (\eg declaring an insulin shot on a diabetes monitoring application) performed on a wearable device trigger unique Bluetooth traffic.
These communication patterns can be recognized by an eavesdropper (\eg a nosy neighbor or a proximity-based advertiser) to infer the action performed despite the encryption.
This holds across the multiple wearable devices such as smartwatches and fitness trackers from diverse vendors.
We also find that the mere opening of an app on a smartwatch generates distinguishable traffic, leaking potentially sensitive information that is associated with the presence of the app (\eg a religious or a political app).
For application-openings, we identify a subset of applications that is inherently well-protected against traffic analysis: ``low-volume'' applications.
This hints that minimizing data exchanges is an obvious natural defense against our attacks.
Furthermore, our results demonstrate that our attacks generalize well across different devices:
we show that a model trained for recognizing application openings on one smartwatch-smartphone pair can be applied with high accuracy on another smartwatch-smartphone pair.
This suggests that our methodology can be cost-effective for an adversary.
Finally, we demonstrated how a persistent long-term tracking adversary can benefit from our traffic-analysis attacks by employing them to profile users, \ie infer their habits and actions on their daily lives, and that dataset aging does not significantly affect the attack's performance.

\section{Protections}
\label{sec:protections}

In the previous sections (§\ref{sec:dev_id}, §\ref{sec:act_id}), we demonstrated how an eavesdropper capturing Bluetooth communications between a wearable device and its connected smartphone can perform traffic-analysis attacks and infer information such as the device model, the applications installed on it, and the actions performed by the wearer.
In this section, we investigate defense strategies against Bluetooth traffic-analysis attacks.
We first review the existing types of defenses against traffic-analysis attacks and identify the most popular approaches.
Then, we implement these strategies and evaluate them against our traffic-analysis attacks.

\para{High-level Defense Strategies.}
Before diving into the design of a defense, we remark that \textit{data minimization} is a simple, inexpensive, and valid approach: data that is not exchanged cannot be fingerprinted.
Indeed, our ``deep'' experiment shows that applications with low traffic volumes are naturally better protected against traffic analysis (§\ref{subsec:app-id-wearos}).
In a similar vein, infrequent ``bulky'' communications (\eg syncing a step counter only once a day at midnight) make the adversary's task harder in two ways: by leaking less metadata about timings, and by requiring the adversary to observe the communication at the right time.
However, such high-level strategies do not apply to all applications (\eg interactive applications such as newspapers, radios or music players).

\para{Defense Design.}
The purpose of this section is not to discover a \textit{perfect} defense: it might not exist or its cost might be unbounded.
Rather, we evaluate the effectiveness of common protections against traffic analysis with respect to the attacks that we presented earlier and study the communication overheads that they introduce.
Fundamentally, a classifier properly trained can detect even small differences between two network traces.
Ensuring that all network traces are perfectly indistinguishable from each other is infeasible.
Therefore, we analyze the effectiveness/cost trade-offs introduced by standard defenses and discuss their feasibility for the protection of Bluetooth communications.

\para{Defense Categories.} We first perform a brief taxonomy of defenses against traffic analysis.
The most active research fields are focused on the Tor network~\citep{panchenko2011website,juarez2016toward,gong2020zero,dyer2012peek,wang2014effective,cai2014systematic,wang2017walkie,wright2009traffic,luo2011httpos} and on IoT traffic/smart-homes~\citep{trimananda2020packet,apthorpe2017spying,apthorpe2019keeping,alshehri2020attacking,dyer2012peek}.
In a different category, recent anonymous messaging protocols often resist traffic analysis against a much stronger global adversary, at the cost of having a high bandwidth~\citep{barman2020prifi,chen2018taranet} or a high latency~\citep{corrigan2015riposte,angel2016unobservable,kwon2016riffle}; indeed, spending time or bandwidth is a fundamental trade-off for achieving traffic-analysis resistant communications~\citep{das2018anonymity}.
To the best of our knowledge, traffic-analysis defenses have not yet been explored on wearable devices.

\noindent We distinguish three defense categories~\citep{gong2020zero}: regularization, obfuscation, and randomization:
\begin{itemize}
\item \textbf{Regularization} defenses make packet traces harder to distinguish by removing their differences, \eg by enforcing constant bit-rates and packet sizes~\citep{dyer2012peek,cai2014cs,cai2014systematic,apthorpe2017spying}, by altering the traces into the common closest ``super-sequence'' of packets~\citep{wang2014effective}, or by forbidding duplex communications~\citep{wang2017walkie}.
\item \textbf{Obfuscation} approaches aim at confusing the adversary by tweaking the setting, \eg by hiding traffic into another protocol~\citep{wright2009traffic,luo2011httpos}, or by loading two web pages at the same time, as in the case of website fingerprinting~\citep{panchenko2011website}.
\item \textbf{Randomization} defenses confuse the adversary by adding randomized dummy traffic~\citep{juarez2016toward,gong2020zero,apthorpe2019keeping,alshehri2020attacking, dyer2012peek}.
\end{itemize}

\noindent Defenses based on regularization are often easier to reason about and to analyze their formal guarantees, but they have the downside of being more costly than the others.
On the contrary, obfuscation approaches consist of more practical defenses that often assume a certain type of adversary, \eg that cannot de-multiplex encrypted web pages, or recognize Tor traffic hidden as Skype traffic.
We do not explore obfuscation strategies as this category does not apply well to wearable devices that only support a few classes of traffic.
One possible obfuscation strategy (not explored in this work) could be to split the traffic between Wi-Fi and Bluetooth, for smartwatches that are capable of both.
Finally, randomization defenses tend to be the most lightweight.
However, their efficacy evaluation is harder and is typically done using the success rate of the state-of-the-art attacks~\citep{juarez2016toward,gong2020zero}.
%For instance, on Tor, WTF-PAD~\citep{juarez2016toward} and Front/Glue~\citep{gongzero}, are two zero-delay lightweight randomized defenses: user traffic never experiences delay due to the defense.

\para{Defense Evaluation.}
Our goal is to investigate and understand what degree of protection against Bluetooth traffic-analysis attacks would be provided by a practical and lightweight defense.
We remind the reader that our classifiers use features based on timings (Figures~\ref{fig:dev-id-fi-cla},\ref{fig:app-id-fi},\ref{fig:action-id-diabetesm-fi}) and packet size distributions (Figures~\ref{fig:dev-id-fi-le},\ref{fig:application-identification-wearables-fi}).
Thus, we evaluate three orthogonal defenses that mask real sizes and timings, and that inject dummy packets (which achieves both):

\begin{enumerate}
\item \defense{pad}: A lightweight regularization defense.
Each Bluetooth packet is individually padded to a maximum size ($255$B for BLE packets and $1{,}021$B\footnote{This corresponds to the max payload of a 3-DH5 ACL packet in Bluetooth Classic.} for Bluetooth Classic packets).
Per-packet padding hides specific sizes and unlike per-flow padding, it incurs no delay (ignoring the small delay due to transmitted larger packets).

\item \defense{delay_group}: A regularization defense that delays and groups packets to the next second.
This obfuscates fine-grained timing information by imposing a pace.
We note that this approach is clearly incompatible with latency-sensitive Bluetooth communications such as audio streaming, real-time and interactive applications.
It does not incur bandwidth overhead.

\item \defense{add_dummies}: A randomization defense that injects packets at times drawn from a Rayleigh probability distribution.
The use of the Rayleigh distribution is inspired by the ``Front'' part of Front/Glue~\citep{gong2020zero}, a state-of-the-art lightweight randomization defense designed for website fingerprinting.
%Unlike Front, we do not randomize the parameters of the Rayleigh distribution.%: we experimentally observe that this does not make the defense more efficient in our scenario.
We experimentally select $6$s for the mean of the Rayleigh distribution, and $300$ for the number of dummies we generate (Figure~\ref{fig:add-dummies-params}).
Finally, we sample the size of each dummy from a distribution created with the collected samples.
Therefore, this defense assumes that the defender knows a priori the distribution of packet sizes.

\end{enumerate}

To assess the protection level provided by the defenses, we measure the accuracy achieved by the classifier trained by the adversary.
We assume that the adversary knows the defense in use and is able to adapt the training of the classifier.
To quantify the cost of each defense, we use $5$ metrics: the mean delay introduced per packet, the total added duration to the sample, the number of bytes added (both in terms of padding and dummy messages), and the total size overhead in percentage.

\subsection{Experimental Results}
\label{subsec:defenses-analysis}

We apply the various defenses on the Bluetooth traffic traces used for the device identification attack (§\ref{sec:dev_id}), the ``wide'' experiment consisting of human-triggered actions on all wearable devices (§\ref{subsec:app-id-wearables}), the ``deep'' experiment consisting of automated apps openings on Huawei Watch 2 (§\ref{subsec:app-id-wearos}), and the fine-grained action recognition within the \app{DiabetesM} application (§\ref{subsubsec:diabetesm}).
This enables us to evaluate the performance of the defenses against multiple adversarial goals and various traffic settings.

Tables~\ref{table:device-id-def-cla}, \ref{table:device-id-def-ble}, \ref{table:action-id-wearables-def}, \ref{table:app-id-huaweiwatch-def}, and~\ref{table:action-id-diabetesm} display the performance and cost of each defense against the attacks considered.
The accuracy of the attack is averaged over the possible classes, and the defense costs are averaged per traffic sample.
Our first immediate observation is that regardless of the attack and the defense, the mean accuracy achieved by the adversary's classifier is still significantly better than random guessing, which indicates that the defenses are far from being ``strong'' ones that provide cryptographic guarantees.
Overall, the cost of each defense lies between $1$--$23\times$ in terms of data transmission (at most $\approx 400$ KB of extra data).

\begin{table}[t]
\small
\caption{Analysis of the defenses against device identification, Bluetooth Classic devices.}
\vspace{-0.3cm}
\begin{tabular}{lrrrrrr}
\textbf{Defense} & \textbf{Accuracy [\%]} & \textbf{Delay/pkt [s]} & \textbf{Extra dur. [s]} & \textbf{Padding [KB]} & \textbf{Dummy [KB]} & \textbf{Overhead [\%]} \\
\defense{No defense} & 96.3 & - & - & - & - & - \\
\defense{pad} & 93.8 & - & - & 401.6 & - & 203.2 \\
\defense{delay_group} & 67.7 & 0.5 & 0.2 & - & - & - \\
\defense{add_dummies} & 78.0 & - & - & - & 92.9 & 47.0 \\
\end{tabular}
\label{table:device-id-def-cla}
\vspace{0.1cm}

\caption{Analysis of the defenses against device identification, Bluetooth LE devices.}
\vspace{-0.3cm}
\begin{tabular}{lrrrrrr}
\textbf{Defense} & \textbf{Accuracy [\%]} & \textbf{Delay/pkt [s]} & \textbf{Extra dur. [s]} & \textbf{Padding [KB]} & \textbf{Dummy [KB]} & \textbf{Overhead [\%]} \\
\defense{No defense} & 97.7 & - & - & - & - & - \\
\defense{pad} & 94.5 & - & - & 139.0 & - & 277.1 \\
\defense{delay_group} & 80.6 & 0.5 & 0.1 & - & - & - \\
\defense{add_dummies} & 85.8 & - & - & - & 20.4 & 40.6 \\
\end{tabular}
\label{table:device-id-def-ble}
\vspace{0.1cm}

\caption{Analysis of the defenses against action identification, ``wide'' experiment.}
\vspace{-0.3cm}
\begin{tabular}{lrrrrrr}
\textbf{Defense} & \textbf{Accuracy [\%]} & \textbf{Delay/pkt [s]} & \textbf{Extra dur. [s]} & \textbf{Padding [KB]} & \textbf{Dummy [KB]} & \textbf{Overhead [\%]} \\
\defense{No defense} & 82.3 & - & - & - & - & - \\
\defense{pad} & 64.1 & - & - & 272.0 & - & 270.5 \\
\defense{delay_group} & 52.0 & 0.5 & 0.2 & - & - & - \\
\defense{add_dummies} & 64.0 & - & - & - & 62.5 & 62.2 \\
\end{tabular}
\label{table:action-id-wearables-def}
\vspace{0.1cm}

\caption{Analysis of the defenses against application identification, ``deep'' experiment.}
\vspace{-0.3cm}
\begin{tabular}{lrrrrrr}
\textbf{Defense} & \textbf{Accuracy [\%]} & \textbf{Delay/pkt [s]} & \textbf{Extra dur. [s]} & \textbf{Padding [KB]} & \textbf{Dummy [KB]} & \textbf{Overhead [\%]} \\
\defense{No defense} & 64.4 & - & - & - & - & - \\
\defense{pad} & 27.9 & - & - & 150.5 & - & 585.0 \\
\defense{delay_group} & 37.3 & 0.5 & 0.4 & - & - & - \\
\defense{add_dummies} & 33.8 & - & - & - & 46.8 & 182.0 \\
\end{tabular}
\label{table:app-id-huaweiwatch-def}
\vspace{0.1cm}

\caption{Analysis of the defenses against action-identification in \app{DiabetesM} application.}
\vspace{-0.3cm}
\begin{tabular}{lrrrrrr}
\textbf{Defense} & \textbf{Accuracy [\%]} & \textbf{Delay/pkt [s]} & \textbf{Extra dur. [s]} & \textbf{Padding [KB]} & \textbf{Dummy [KB]} & \textbf{Overhead [\%]} \\
\defense{No defense} & 70.4 & - & - & - & - & - \\
\defense{pad} & 61.1 & - & - & 77.8 & - & 2374.8 \\
\defense{delay_group} & 60.1 & 0.5 & 0.1 & - & - & - \\
\defense{add_dummies} & 61.7 & - & - & - & 11.8 & 360.2 \\
\end{tabular}
\label{table:action-id-diabetesm}
\end{table}

\para{Device Identification.}
Tables~\ref{table:device-id-def-cla} and~\ref{table:device-id-def-ble} show that all defenses yield, at best, a moderate drop in this attack's accuracy.
However, both flavors of the device identification attack are performed on a small number of devices ($7$ both for Bluetooth Classic and Low Energy).
It is therefore not surprising that hiding the traffic of a device into that of another is a difficult task for the defenses.
Among the evaluated defenses, we observe that the \defense{delay_group} is the most effective one: it diminishes the attack's accuracy by $29$ percentage points in the case of Bluetooth Classic, and $17$ for Bluetooth Low Energy.
However, the cost of \defense{delay_group} is prohibitively high ($\approx0.5$s delay added per packet) for a defense that is meant to be applied to all the communications performed by a device.
Moreover, we find that the defense \defense{pad} is ineffective with both Bluetooth flavors.
This is not surprising for the case of Bluetooth Classic where the attacker's classifier relies mostly on timing features (Figure~\ref{fig:dev-id-fi-cla}).
For Low Energy, features that relied on packet sizes (Figure~\ref{fig:dev-id-fi-le}) have been replaced by the same top-rated timing features as in Bluetooth classic (\feat{max/min/std of }$\Delta$\feat{time}, Figure~\ref{fig:device-id-def-ble-pad-fi}).
This suggests that these three features are important for device identification, regardless of the Bluetooth flavor and corroborates the findings of Aksu \etal~\citep{aksu2018identification} on smartwatches.
We observe that \defense{add_dummies} is lightweight and reduces the attacker's accuracy by $18$ and $12$ percentage points for Bluetooth Classic and LE, respectively.
However, its performance is not uniform across the classes (Figure~\ref{fig:device-id-front-cm}).
The confusion matrix shows that \defense{add_dummies} only moderately protects the $3$ \dev{Mi Band 2-3-4} devices and does not protect the others.

\para{``Wide''-Experiment.}
We evaluate the performance of the defenses against the action identification attack demonstrated in §\ref{subsec:app-id-wearables}.
Compared to the previous attack, the task is different, and the classifier has to account for more labels (\ie actions).
We find that all defenses perform better in general, reducing the attacker's accuracy between $18$ and $30$ percentage points (Table~\ref{table:action-id-wearables-def}).
In particular, the difference in efficiency between \defense{pad} and \defense{delay_group} is now of only $12$ percentage points, for a cost of $272$KB per sample for \defense{pad}, and $0.5$s delay for \defense{delay_group}.
This result suggests that both masking individual sizes or masking fine-grained timing information can help; developers could select the appropriate defense based on the cost (either in bandwidth or latency) that best matches their requirements.
Finally, \defense{add_dummies} performs similarly to \defense{pad} but with a lower cost ($62.5$KB per sample versus $272$KB for \defense{pad}).
However, \defense{add_dummies} requires that the distribution of packet sizes is a priori known to generate dummies of plausible size.
It is unclear how to compute this distribution for a defense meant to be applied to multiple devices from different vendors.
One option would be to combine \defense{add_dummies} and \defense{pad} to avoid this requirement (we experimented with it and observed better effectiveness at a higher cost).
Finally, we note that the protection provided by the defenses is not uniform (we provide an example with \defense{add-dummies} in Figure~\ref{fig:action-id-wearables-def-front-cm}, other defenses yield similar results), but unlike the ``deep'' experiment (§\ref{subsec:app-id-wearos}), the precision/recall per class does not seem correlated with the transmitted size.
Similarly, the cost of \defense{padding} varies greatly with the classes: it has a mean of $272$KB added, but a median of only $96$KB and a standard deviation of $650$KB.
The costs soar up to $4.3$MB with streaming applications such as \act{AppleWatch_PhotoApp_LiveStream} or \act{PhoneCallMissed}.
We note that the defenses \defense{delay_group} and \defense{add_dummies} are more consistent, with a standard deviation of, respectively, $0.04$s of delay per packet and $7$KB of dummy traffic.

\para{``Deep''-Experiment.}
The traffic in this experiment consists of automated app openings on a Wear OS smartwatch.
For this type of traffic, we observe that all three defenses perform similarly, reducing the attacker's accuracy by $31$--$36$ percentage points down to $\approx 30\%$ mean accuracy (Table~\ref{table:app-id-huaweiwatch-def}).
This highlights that on a specific class of traffic, \ie with more homogeneous traffic, the defenses are more efficient.
In this case, \defense{add_dummies} is the least expensive defense, requiring $46.8$KB of dummy traffic/sample, which is in the range of what the heaviest applications naturally use (Table~\ref{table:high-volume-apps-details}).
A deeper analysis of the results also shows that all defenses successfully confuse the attacker for ``medium-volume'' apps (Figure~\ref{fig:app-id-huaweiwatch-def-front-cm}).
``High-volume" applications still stand out, but the gradual hiding visible on Figure~\ref{fig:app-id-huaweiwatch-def-front-cm} suggests that increasing the parameters of the defense, \eg injecting more dummies in \defense{add_dummies}, could potentially protect better such applications.
However, this protection would come at an even higher cost for the applications that transmit smaller amount of traffic.
As in the ``wide'' experiment, we observe that \defense{pad} has a highly variable cost ranging from $6$KB to $1.5$MB.
We observe the latter on the opening of the \app{Camera}, which suggests that \defense{pad} is not adapted for constant-traffic.
As before, \defense{delay_group} and \defense{add_dummies} have a consistent cost across labels.

\para{Fine-Grained Action Fingerprinting on \app{DiabetesM}.}
In this case, we observe that all $3$ defenses \defense{pad}, \defense{delay_group}, \defense{add_dummies} are only moderately effective, reducing the attacker's accuracy by $\approx 10$ percentage points.
One possible explanation is the increased number of samples ($150$/label) compared to the previous experiments ($25$/label) that enable the adversarial classifier to adapt better to the defenses.
In §\ref{subsubsec:diabetesm}, we highlighted that timings were of importance to classify fine-grained actions in the application~\app{DiabetesM}.
However, in this case the attacker fingerprints a combination of the sizes and the timings, as the feature importance on \defense{delay_group} defended traces reveals (Figure~\ref{fig:action-id-diabetesm-def-delay_group-fi}).
In this experiment, we observe that all three defenses have a consistent cost across labels.

\subsection{Summary of the Defenses}
\label{subsec:defenses-take-aways}

Our experimental evaluation of defense approaches, such as regularization and randomization, against the traffic-analysis attacks presented in this work yields some interesting insights.
First, we find that these defenses achieve only a limited protection against our traffic-analysis attacks: although they do reduce the attacks' accuracy, the classifier trained by an eavesdropping adversary still performs significantly better than a random guess.
This indicates that even the defended Bluetooth traffic traces contain useful information for an adversary.
At the same time, the costs introduced by the defenses are high: to achieve their small levels of protection they introduce additional traffic and/or delays reaching an overhead in the range of 1$\times$ to 23$\times$ and delays up to $1$s per packet.
This raises a question about the applicability of such defenses for Bluetooth applications running on current wearable devices.
Furthermore, we find that the various defenses behave differently, depending on the adversarial task.
In particular, our results show that defending against application or action identification is somewhat easier than device identification, thus indicating that global traffic patterns are the hardest to hide.
Additionally, our evaluation shows that the defenses are not fair: we find that they do not provide the same level of protection across applications or actions (for instance, apps that communicate a lot are not well protected) and their costs are variable across applications or actions (we observe that applications that stream information have high padding costs, \eg \app{Camera} and fitness applications for workout monitoring).
Finally, our empirical evaluation of these defenses confirms the robustness of our attacking methodology as depending on the adversarial task and the defense, our classifier adapts its important features.
For instance, Bluetooth Low Energy device identification relied mostly on packet sizes (Figure~\ref{fig:dev-id-fi-le}), unlike Bluetooth Classic that used mostly timings (Figure~\ref{fig:dev-id-fi-cla}).
However, when we apply per-packet padding to the Low Energy traces, the classifier adjusts and gives higher importance to timings (Figure~\ref{fig:device-id-def-ble-pad-fi}).
Overall, our experimental results highlight the need for the design and evaluation of novel approaches for defending against traffic-analysis attacks on Bluetooth communications.

\section{Discussion}
\label{sec:discussion}

\para{Ethical Considerations.}
We provided every device manufacturer and app developer mentioned in this paper with our findings prior to the publication of this document.
%We make the datasets (and the code) available for review purposes~\citep{wearable-dataset-lbarman,wearable-tools}.
To minimize the risk of misuse, we make the dataset available only for research purposes upon request~\citep{wearable-tools}.
We discarded the traffic from other devices in the dataset.

\para{Impact of the Attacks.}
The traffic-analysis attacks presented in this work can be used to infer information from Bluetooth communications, despite the use of encryption.
Device identification enables tracking users only by observing encrypted communications, thus defeating MAC address randomization.
This does not require observing any pairings/paging events or plaintext identifiers.
Device identification can also facilitate active attacks by revealing the model and version of a communicating device.
We note that advertisers already use Bluetooth and Wi-Fi signals to passively and actively locate users (\eg consumers in a store)~\citep{google_track_bluetooth, tlf_tracking}.
Similarly, application and action fingerprinting leak sensitive actions performed by the wearer, \eg the recording of an insulin injection or a heartbeat measurement.
On a side note, Apple Watch has an ``arrhythmia alert'' feature that continuously measures the heart beat on the watch and sends notifications to the phone in case of irregular patterns that could indicate a stroke.
We could not simulate arrhythmia events, but all the evidence we have from our experiments suggests that such an action could be fingerprintable without an appropriate defense mechanism.
Therefore, a passive observer could identify users' susceptibility to heart attacks over the Bluetooth network, despite the encryption.
Finally, application-opening identification and action identification can be exploited to build user profiles and to serve targeted advertisements, as it is already the case with Bluetooth-based ``proximity advertising''~\citep{the_drum_london_cabs, proxbook, stores_secret_surveillance}.

\para{Cost of the Attack.}
The overall cost of the attack consists of purchasing a set of devices of interest (including both wearable devices and smartphones).
These devices are consumer-grade hardware that have accessible prices.
We also showed in §\ref{subsubsec:transferability} that the adversary does not have to train on every combination of devices and applications.
We suspect that after collecting data from enough devices, actions on new hardware can become classified without the overhead of training.
A counter-argument is the \textit{aging} of the dataset, which could force the adversary to re-train often.
We briefly demonstrate in Section~\ref{subsubsec:aging} how a dataset can be used over at least a month, but further study is needed to understand how quickly the usefulness of a dataset degrades.
We note that in other domains such as website fingerprinting, attacks have been successful with datasets that were several years old~\citep{sirinam2019triplet}.
We expect wearable devices' firmware, OSes and applications to change at a slower rate than websites.

\para{Bluetooth Sniffing Technologies.}
The adversary also needs a reliable Bluetooth sniffer.
The most accurate models are so-called ``wide-band'' scanners (\eg the Ellisys Vanguard~\citep{ellisys_vanguard} or the Frontline Sodera~\citep{fte_sodera}).
These models ignore the Bluetooth frequency hopping and concurrently capture the traffic of all channels.
The complexity and broad functionality of these devices comes at a high price ($\approx50$K USD).
However, recent research has demonstrated that similar results can be achieved using less expensive Software-Defined Radios (SDRs)~\citep{tabassam2008bluetooth, cominelli2020even, cominelli2020one}.
For instance, Cominelli \etal{} built an SDR sniffer that works on a single Ettus N310 board ($\approx 10$K USD) or two Ettus B210 boards ($2\times 2{,}000$ USD)~\citep{cominelli2020even}.
Finally, there also exists a consumer-grade class of cheaper, less accurate Bluetooth sniffers (\eg Ubertooth, $\approx 100$ USD).
They only listen on one channel at a time.
These low-end scanners attempt to \emph{follow} an active connection by brute-forcing the hopping pattern parameters~\citep{ryan2013bluetooth}.
When successful, this enables an inexpensive scanner to accurately capture all traffic simply by ``hopping along'' with the pair of communicating devices.
In practice today, this process is still imprecise and many packets are missed.
Nonetheless, researchers have shown that using two synchronized Ubertooth scanners leads to improved Bluetooth traffic capture rate~\citep{albazrqaoe2016practical, albazrqaoe2018practical}.
Although this work uses a commercial Bluetooth sniffer, there is an ongoing research trend focusing on less expensive, accurate Bluetooth sniffing.

\para{Impact of Packet Loss on the Attacks.}
In our experiments, we use a high-end sniffer that is co-located with the target devices.
In this configuration, the sniffer has close to $100\%$ packet capture rate.
In practice, lower-end sniffers will suffer packet loss.
Collisions from other devices and the distance between the target and the eavesdropper also increase loss.
We briefly investigate how the attack accuracy varies with degraded capture conditions.

First, we decouple the attacker accuracy, the packet loss and the capture conditions, and we explore the attack accuracy versus the loss rate only.
This loss can stem from many real-life parameters (distance, noise, multipath interference, the quality of the eavesdropping device, etc.).
As a generic approach, we study the effect of uniform packet loss on our dataset.
We simulate packet loss on the captured traces and re-run the application identification (``deep'') experiment (§\ref{subsubsec:app-id-wearos}).
We apply a uniform packet loss by dropping individual packets with a given probability.
We then use the methodology already presented (we split the traces into train/test and compute the average classification accuracy).
The experiment is repeated $10$ times per loss rate.
We observe (Appendix, Figure~\ref{fig:deep-loss}) that even with $50\%$ packet loss, the loss in accuracy is only $10$ percentage points, with the mean accuracy dropping from $64\%$ to $54\%$.
For high-volume apps, the mean accuracy drops from $90\%$ to $77\%$.
This experiment indicates that the approach is robust to packet losses: even when missing every other packet, the attacker is able to classify with significant accuracy.

Due to the difficulty of relating the various capture conditions to the loss rate, we only discuss experimental results generated by Albazrqaoe \etal{} with the most common inexpensive Bluetooth sniffers:
The authors describe how a single Ubertooth, when placed $10$m away from the target device and in the presence of significant 802.11 interference~\citep{albazrqaoe2016practical}, has between $25\%$ and $50\%$ packet loss.
In similar conditions, a BlueEar sniffer (composed of two synchronized Ubertooth) maintains packet losses below $10\%$~\citep{albazrqaoe2016practical}.
This observation suggests that the attack could also be performed with inexpensive Bluetooth sniffers.

\para{Other Attacks.}
There exist a number of \emph{active} attacks that can break the confidentiality of Bluetooth communications~\citep{antonioli2019knob,antonioli2020bias,wu2020blesa,wang2020bluedoor,zhang2020breaking}.
There are also attacks against the MAC randomization mechanism employed by the Bluetooth protocol~\citep{becker2019tracking,zuo2019automatic,becker2019tracking,celosia2019saving,martin2019handoff}.
Currently, these attacks are more economical to run than our traffic-analysis attacks that require a significant effort for training the adversarial machine-learning classifiers.
However, these attacks have already received significant visibility, and we expect that the Bluetooth Special Interest Group and device manufacturers will soon take them into account and apply the necessary countermeasures.
We remark that our approach is complementary to these attacks and will be applicable even when the above attacks are patched.
Finally, we remind the reader that unlike these works, ours considers a weaker adversary who passively observes ongoing communications.

\para{Implementation of Defenses.}
A defense against traffic analysis could be implemented at different layers of the stack involved with Bluetooth communications: the Bluetooth protocol, the OS, or the applications.
An implementation in a lower layer, \eg the Bluetooth stack, would provide application transparency, but specifying a single defense strategy that works across devices and applications without a prohibitive cost seems challenging.
On the contrary, application developers could protect against in-application actions fingerprinting by enumerating the data exchanges and ensuring that the traffic from sensitive actions ``blends-in''.
In this case, all sensitive actions would need to have the same patterns as some other non-sensitive actions (or ideally, all other actions).
Nonetheless, such an approach would only provide local (\ie in-application) protection.
To make the fingerprinting of applications and cross-application actions more difficult, the various developers would need to coordinate with each other.
Otherwise, a defense can easily become itself a fingerprint if only one or a few applications implement it.
Therefore, another interesting possibility is to create ``anti traffic-analysis policies'' in the OS of wearable devices.
Apps could request a particular defense strategy that matches their requirements in terms of latency, bandwidth, and battery usage.
Meanwhile, the operating system could standardize and maintain defense strategies transparently, making their deployment easy for developers.%\todo{discuss the use of Bluetooth link-layer encryption here, if there is smth to say?}

\section{Limitations}
\label{sec:limitations}

\para{Moderate Testbed Size.}
Our testbed consists of only $13$ devices.
Although this number is modest, our experiments incurred significant human costs in operating these non-automatable devices.
Our primary dataset consists of $98$ hours of raw recording.
Each of the $2{,}215$ $30$-second samples was recorded by a human and required sometimes minutes of resetting the devices, not to mention performing the physical activities corresponding to the action captured.
We made a best-effort to cover a comprehensive and diverse set of wearable devices from popular vendors.
We are hopeful that future work will further generalize the attack to more devices, which we could only hint towards with our transferability experiment (§\ref{subsubsec:transferability}).
We open-source the automation framework built to collect traces from Wear OS devices to facilitate further research on the topic~\citep{wearable-tools}.

\para{Closed-world Scenario.}
Our setting corresponds to a ``closed-world'' scenario where the adversary can model all possible actions/applications of the devices available in our testbed.
This could be justified due to the small number of applications currently available on wearable devices.
In future work, we plan to explore the performance of the attacks ``in the wild'', \eg by collecting real Bluetooth traffic traces around a campus or a gym and attempting to classify them.

\para{Method Scalability.}
We demontrated an example in which a model trained on a pair of devices can be used for classifying actions of other devices (§\ref{subsubsec:transferability}).
However, this experiment has been limited to two pairs of devices, and it is unclear whether a unique model could be successfully trained to classify actions originating from many classes of devices.
While we expect it to be the case for similar devices, our preliminary results show that a model trained to recognize applications on an Apple watch does not perform well when used to recognize applications on an Android device, highlighting that an adversary should take into account heterogeneous devices when training her classifier.
%Therefore, to ensure accurate classification, the best strategy for an attacker is to train a model on the target devices.
%Several devices can also be attacked with a single model (§\ref{subsubsec:transferability}), lowering the cost of the attack, but further study is needed to understand what classes of devices can be combined.

\para{Environment of the Capture.}
In this paper, we did not experiment with the range of capture and kept the sniffer close-by to the devices (max $ \approx2$m).
Bluetooth Classic and Low Energy have a maximum theoretical range that greatly varies depending on the Bluetooth flavor, the encoding, the sender/receiver's antenna gains and the transmitted power~\citep{bluetooth_range_calculator}.
For consumer devices, the range under optimal conditions is between $50$ and $100$m and, we estimate, from meters to tens of meters under realistic conditions.
Furthermore, our attacks were conducted in a single environment that is fairly noisy (with tens of active Wi-Fi and Bluetooth devices in vicinity).
We suspect that a less noisy environment (\eg a home, in the case of a nosy neighbor) would produce cleaner traces that are easier to train upon, facilitating the attacks, but further study is needed to understand the impact of noise and collisions on traffic analysis.
\color{black}

\section{Related Work}
\label{sec:related-work}

\para{Bluetooth Eavesdropping.}
The ability to sniff Bluetooth communications is essential for performing traffic-analysis attacks.
The first open-source Bluetooth Sniffer was BlueSniff~\citep{spill2007bluesniff}:
This work demonstrates how to retrieve the MAC address of communicating Bluetooth Classic devices and how to recover the hopping sequence.
Similar results are later shown on Bluetooth Low Energy~\citep{ryan2013bluetooth}: using an Ubertooth device, Mike Ryan showed how to recover the hopping sequence and eavesdrop on a single BLE connection.
The author also demonstrates that a pairing done with JustWorks or a 6-digit PIN can be decrypted.
Subsequently, Albazrqaoe \etal{} used two synchronized Ubertooth devices to obtain a capture accuracy greater than a single Ubertooth~\citep{albazrqaoe2016practical,albazrqaoe2018practical}.
Finally, Cominelli \etal{} rely on software-defined radio (SDR) to concurrently capture Bluetooth Classic traffic on all channels~\citep{cominelli2020even}, at a lower cost than full-band commercial sniffers.
Their most recent work uses a GPU to process BLE traffic in real-time~\citep{cominelli2020one}.

\para{Bluetooth Traffic Analysis.}
There exist few related works that perform traffic-analysis attacks on Bluetooth communications.
This is possibly due to the non-existence of reliable, inexpensive Bluetooth sniffing tools in the past.
Closest to ours is the work by Das \etal~\citep{das2016uncovering}.
They focus on $6$ Bluetooth Low Energy fitness trackers in a gym.
First, they demonstrate that BLE traffic is correlated with the wearer's movements, thus making it possible to infer if the wearer is idle, walking, or running.
Second, they show how the traffic is linked to the gait of the wearer, and that the encrypted traffic is enough to recognize a person with $97.6\%$ accuracy across $10$ users.
Acar \etal{} infer user actions in a smart home using a layered traffic-analysis attack~\citep{acar2018peek}.
Their methodology is similar to ours: they first perform device identification and then use it as a stepping stone to further infer device states and user activities.
However, their IoT testbed consists of only one device that communicates using Bluetooth (a smart BLE light bulb).
To the best of our knowledge, there is no work performing traffic-analysis attacks on Bluetooth Classic communications.

\para{Bluetooth Device Fingerprinting/Tracking.}
Several works focus on Bluetooth device fingerprinting, \ie device identification and tracking.
Their goal is either to propose an authentication mechanism, \eg to identify MAC spoofing, or to demonstrate an attack, \eg BLE device tracking despite the MAC address rotation.
Our device-identification attack (§\ref{sec:dev_id}) falls into the second category; however, we only use it as a first step towards the rest of our contributions (application and user-action identification) that are orthogonal to this category of related works.

On the defense side, Aksu \etal{} create a testbed composed of $6$ Bluetooth Classic smartwatches connected to a single smartphone, and they demonstrate that the smartwatches can be identified via their communications' timings~\citep{aksu2018identification}.
However, their model is different and they do not sniff Bluetooth traces in the air, rather collect them using the Bluetooth Host Controller Interface (HCI) log on the smartphone.
Huang \etal{} propose BlueID~\citep{huang2014blueid}, a system that prevents identity spoofing by fingerprinting the clock of the master device.

Concerning the attacks, Zuo \etal{} demonstrate that Bluetooth Low Energy devices can be recognized by plaintext identifiers found in their communications~\citep{zuo2019automatic}.
They suggest application- or protocol-level solutions to better rotate static identifiers.
We note that their solutions would not thwart our device identification attack that works on encrypted traffic (§\ref{sec:dev_id}).
Becker \etal{} use static identifiers differently~\citep{becker2019tracking}: they demonstrate that the rotation of MAC address and other static identifiers are not synchronized, which enables defeating the MAC randomization.
More generally, Celosia and Cunche examine BLE devices in the wild and show that they fail to properly rotate their MAC addresses, thus enabling tracking~\citep{celosia2019saving}.
Then, Korolova and Sharma show that the ``nearby devices'' list that Bluetooth devices maintain and which most applications can obtain, can be used to track users across applications~\citep{korolova2018cross}.
Targeting more specifically Apple's Continuity protocol, Martin \etal{} reverse-engineer the protocol and find flaws that defeat MAC address randomization and that leak information about the device types and user activities~\citep{martin2019handoff}.
Similarly, Celosia and Cunche also reverse-engineer the protocol and demonstrate how it reveals information about human activities in a smart home~\citep{celosia2020discontinued}.

\para{Other Bluetooth Attacks and Tools.}
There exist other attacks on the Bluetooth protocol that are orthogonal to our work; for instance, active attacks, or protocol-specific attacks on wearable devices.
We note that fixing these will likely not affect our higher-level attacks that rely on Bluetooth communication metadata.

We first list attacks on the Bluetooth protocol and implementations.
Antonioli \etal{} demonstrate how to break the key negotiation protocol of Bluetooth Classic~\citep{antonioli2019knob}, forcing it to a $1$-byte entropy encryption key.
The same authors reverse-engineer and identify vulnerabilities in Google Nearby Connections~\citep{antonioli2019nearby}, a protocol that uses a combination of Bluetooth Classic, Low Energy, and Wi-Fi for short-distance transfers.
Then, they also exploit role-switching (slave/master) and legacy pairings to perform a Man-in-the-Middle on Bluetooth Classic~\citep{antonioli2020bias}.
Wu \etal{} present an active spoofing attack on Bluetooth Low Energy by exploiting the reconnection procedure~\citep{wu2020blesa}, whereas Wang \etal{} demonstrate an active attack to bypass Bluetooth Low Energy authentication and encryption~\citep{wang2020bluedoor}.
Finally, Zhang \etal{} show a downgrade attack based on Bluetooth Low Energy's Secure Connection Only (SCO) mode~\citep{zhang2020breaking}.

In the category of Bluetooth tools, Mantz \etal{} present InternalBlue~\citep{mantz2019internalblue}, a Bluetooth experimentation framework that enables patching the Bluetooth firmware of Broadcom chips.
Similarly, Classen and Hollick show how to analyze Bluetooth communications using consumer devices~\citep{classen2019inside}, while Ruge \etal{} design an emulation framework and perform fuzzing to uncover vulnerabilities~\citep{ruge2020frankenstein}.
In particular, the authors find unattended Remote Code Executions (RCE) on some Bluetooth chips.

Focusing on wearable devices, Classen \etal{} perform an in-depth security analysis of the Fitbit ecosystem~\citep{classen2018anatomy}, analyzing the firmware and the application of a fitness tracker.
They find vulnerabilities that enable flashing malware, disabling encryption, and extracting private information about the users.
Hilts \etal{} perform a comparative analysis of the security and privacy of fitness trackers~\citep{hilts2016every}; they highlight security vulnerabilities and issues with their data policies.

\para{Other Traffic-Analysis Attacks.}
Danezis and Clayton first introduced traffic analysis in modern digital communications~\citep{danezis07introducing}.
Since then, Tor traffic has been the primary target of such attacks due to its popularity and its strong threat model~\citep{wang13improved,wang16realistically,wang2017walkie,panchenko2011website,hayes2016k,cai2014systematic,sirinam2018deep,sirinam2019triplet}.
Among the most well-known attacks, are the ``k-Nearest Neighbors'' by Wang \etal~\citep{wang2014effective}, CUMUL by Panchenko \etal, which relies on cumulative sums of bytes to create fingerprints~\citep{panchenko2016website}, and k-Fingerprinting, by Hayes and Danezis, which uses a combination of random forests and nearest neighbors for classification~\citep{hayes2016k}.
The most recent attacks rely on deep-learning approaches~\citep{sirinam2018deep} and $N$-shot learning~\citep{sirinam2019triplet} to achieve higher accuracy.

Traffic-analysis attacks have also been applied on TLS traffic, for instance, to recognize a streamed video~\cite{reed16leaky, schuster17beauty, reed17identifying} or to identify the operating system and the applications~\citep{muehlstein2017analyzing}.
Many recent related works focus on IoT traffic and smart homes~\citep{trimananda2020packet,srinivasan2008protecting,acar2018peek,apthorpe2017spying,apthorpe2019keeping,alshehri2020attacking};
while these works often employ a similar attack methodology, they do not consider the same ecosystem of devices, applications and actions.
We focus on devices and applications that have access to live fitness- or health-related information of the wearer, both in a fine-grained manner and over a long period of time.
Our contribution is the first work to perform an in-depth analysis of the fingerprintability of devices, applications and user actions in this ecosystem.
The different nature of the traffic also has a direct impact on the design of defenses.
%, due to their close relationship with human activities.
Similarly, traffic-analysis attacks have been studied in the context of smartphones to identify applications~\citep{zhang11inferring,taylor2016appscanner,taylor2017robust} or user activities (\eg sending an e-mail or browsing a web page)~\citep{conti2015analyzing,saltaformaggio2016eavesdropping}.
However, these attacks are not all equal:
The majority of them use network- or transport-layer headers~\citep{conti2015analyzing,saltaformaggio2016eavesdropping} or application layer headers~\citep{taylor2017robust, taylor2016appscanner, apthorpe2017spying, apthorpe2019keeping} and can use IP-based flow separation to facilitate the attack.
Therefore, they assume an adversary that is already on the target network.
On the contrary, some works consider a weaker adversary and perform the attack using 802.11 Wi-Fi frames~\citep{zhang11inferring,srinivasan2008protecting} or the metadata of tunneled traffic~\citep{alshehri2020attacking}, which cannot be easily de-multiplexed per device, application or action.
Due to the Bluetooth operation, our attacks fall in the latter category. %\todo{double-check the categories}

% Grundy Q, Chiu K, Held F, Continella A, Bero L, Holz R. Data sharing practices of medicines related apps and the mobile ecosystem: traffic, content, and network analysis. bmj. 2019 \citep{grundy2019data}

\para{Defenses Against Traffic-Analysis Attacks.}
A straightforward but costly defense against traffic analysis is to enforce identical communication patterns across the classes that the adversary aims to identify.
This task becomes increasingly complex and costly with a growing number of classes, hence this technique  applies only on a small scale, for example in one particular application.
Same as for the attacks, most defenses target Tor traffic~\citep{panchenko2011website,juarez2016toward,gong2020zero,wang2014effective,cai2014systematic,wang2017walkie,wright2009traffic,luo2011httpos} or the traffic from IoT devices and smart homes~\citep{apthorpe2017spying,apthorpe2019keeping,alshehri2020attacking,dyer2012peek}.
We distinguish three defense categories:
Regularization defenses make packet traces harder to distinguish by removing their differences, \eg by enforcing constant bit-rates and packet sizes~\citep{dyer2012peek,cai2014cs,cai2014systematic,apthorpe2017spying}, by altering the traces into the common closest ``super-sequence'' of packets~\citep{wang2014effective}, or by forbidding duplex communications~\citep{wang2017walkie}.
Obfuscation approaches aim at confusing the adversary by tweaking the setting, \eg by hiding traffic into another protocol~\citep{wright2009traffic,luo2011httpos}, or by loading two web pages at the same time in the case of website fingerprinting~\citep{panchenko2011website}.
Randomization defenses confuse the adversary by adding randomized dummy traffic~\citep{juarez2016toward,gong2020zero,apthorpe2019keeping,alshehri2020attacking, dyer2012peek}.
To the best of our knowledge, our work is the first to investigate the performance of regularization and randomization defenses against Bluetooth traffic-analysis attacks.

Traffic analysis is also a concern for anonymous communications systems that aim at enforcing similar traffic patterns across \emph{participants}.
The adversary considered in this case is a stronger global adversary and anonymous communication systems provide protection at the cost of a high bandwidth~\citep{barman2020prifi,chen2018taranet} or a high latency~\citep{corrigan2015riposte,angel2016unobservable,kwon2016riffle}.
It has been shown that spending either latency or bandwidth is a fundamental trade-off for traffic-analysis resistance~\citep{das2018anonymity}.
In the setting of the attacks presented in this paper, this trade-off does not apply; our goal is not to hide the source of the communication.

\section{Conclusion}
\label{sec:conclusion}
In this work, we have shown that encrypted Bluetooth communications between a wearable device and its connected smartphone leak information about their contents: a passive adversary can infer sensitive information by exploiting their metadata via traffic-analysis attacks.
Our empirical evaluation on a Bluetooth (Classic and Low Energy) traffic dataset generated by a diverse set of wearable devices demonstrates that an eavesdropper can accurately identify communicating devices to their model number, recognize user activities (\eg health monitoring or exercising), the opening of specific applications on smartwatches and fine-grained user actions (\eg recording an insulin injection), and extract the profile and habits of the wearer.
Our experimental analysis of common defense strategies against our traffic-analysis attacks, such as padding or delaying packets, and injecting dummy traffic, show that these do not provide sufficient protection, and that they introduce significant costs for Bluetooth communications.
Overall, this research highlights an open problem regarding the confidentiality of Bluetooth communications and the need for designing novel efficient defenses to address it.

\begin{acks}
This work has been made possible by the help of Friederike Groschupp and Stéphanie Lebrun.
We also wish to thank Daniele Antonioli, Sylvain Chatel, Jiska Classen, Ricard Delgado, and David Lazar for the constructive discussions and feedbacks on the drafts.
We are grateful to the ``Centre Suisse d'Electronique et Microtechnique'' (CSEM) for providing us with the Ellisys Bluetooth sniffer.
This work was supported in part by grant 200021\_178978/1 (PrivateLife) of the Swiss National Science Foundation (SNF).
Some illustrations in the figures have been made by the artists freepik, eucalyp and smashicons (flaticon.com).
\end{acks}

\bibliographystyle{ACM-Reference-Format}
\bibliography{bibliography.bib}

\appendix
\renewcommand{\thetable}{A\arabic{table}}
\setcounter{table}{0}
\section{Additional Figures and Tables}

\begin{figure}[h]

\subfigure[Number of trees, device identification (§\ref{sec:dev_id}).]{\includegraphics[width=0.32\linewidth]{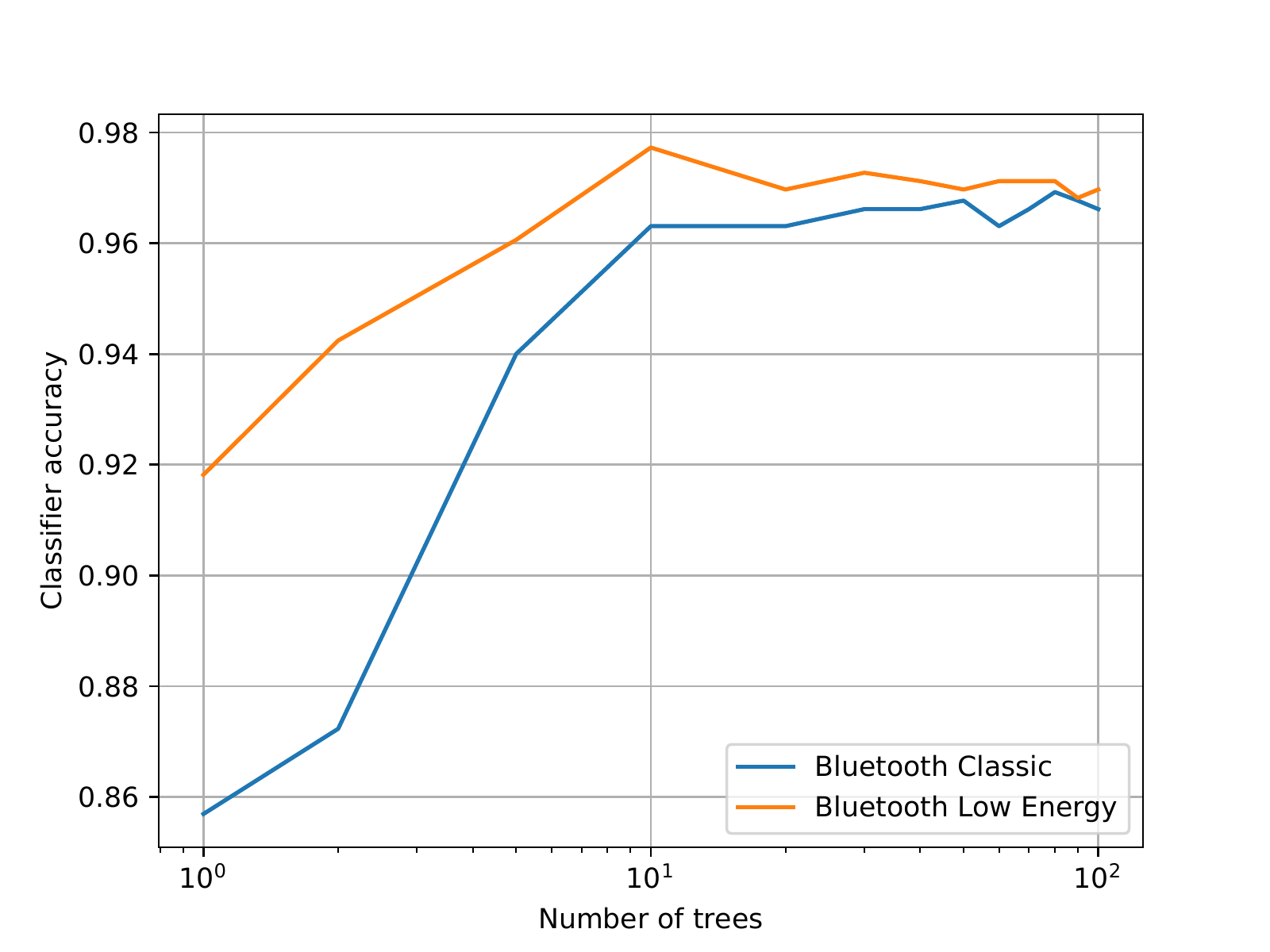}
\label{fig:number_of_features}
}\hspace{0.1cm} %
\subfigure[Number of features kept by the Recursive Feature Elimination (RFE), device identification (§\ref{sec:dev_id}).]{\includegraphics[width=0.32\linewidth]{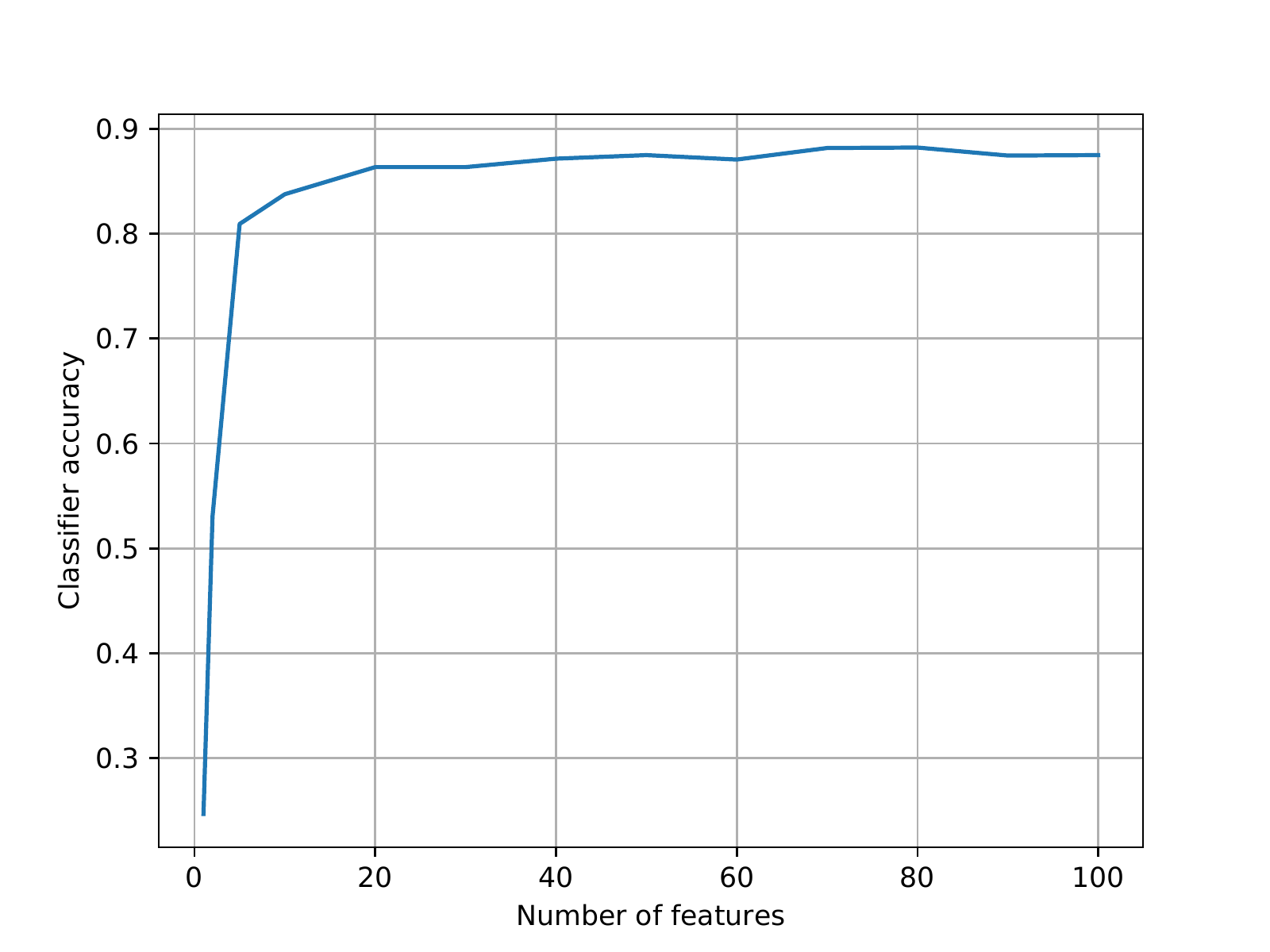}
\label{fig:action-id-wearable-nfeatures}
}\hspace{0.1cm} %
\subfigure[Number of samples per class, application identification (``deep'' experiment, §\ref{subsec:app-id-wearos}).]{\includegraphics[width=0.32\linewidth]{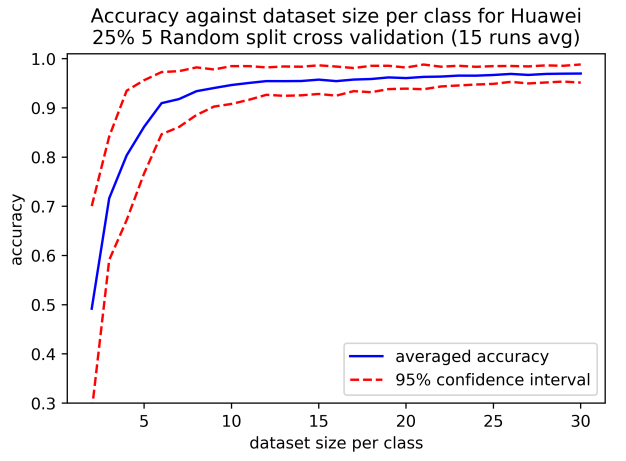}
\label{fig:number_of_samples}}
\caption{Choice of parameters versus mean classifier accuracy.}
\end{figure}
\vspace{0.1cm}
\begin{figure}[h]
\centering
\subfigure[Normalized confusion matrix per true label.]{\includegraphics[width=0.49\linewidth]{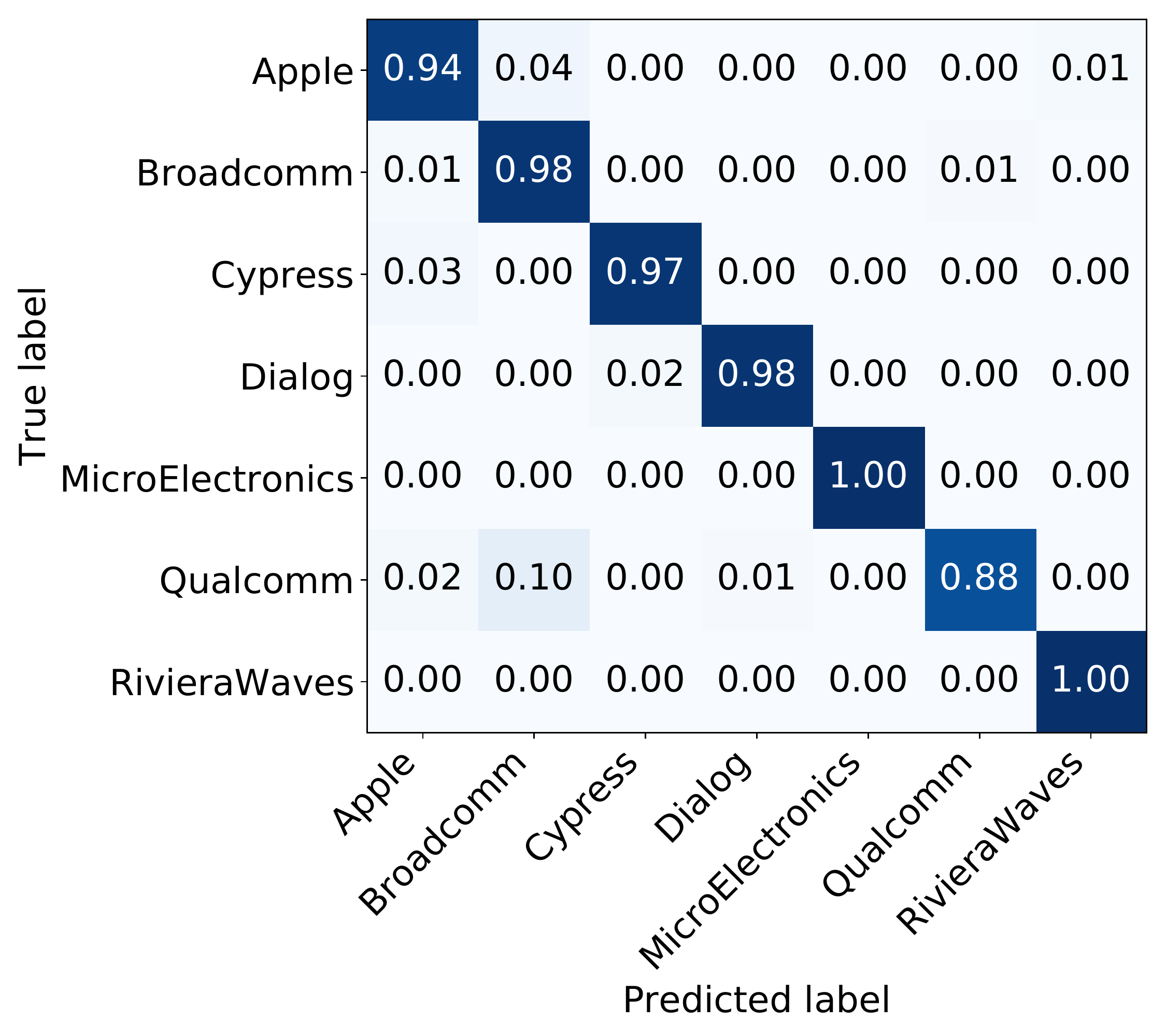}
\label{fig:chipset-id-cm}
} %
\subfigure[Feature importance.]{\includegraphics[width=0.49\linewidth]{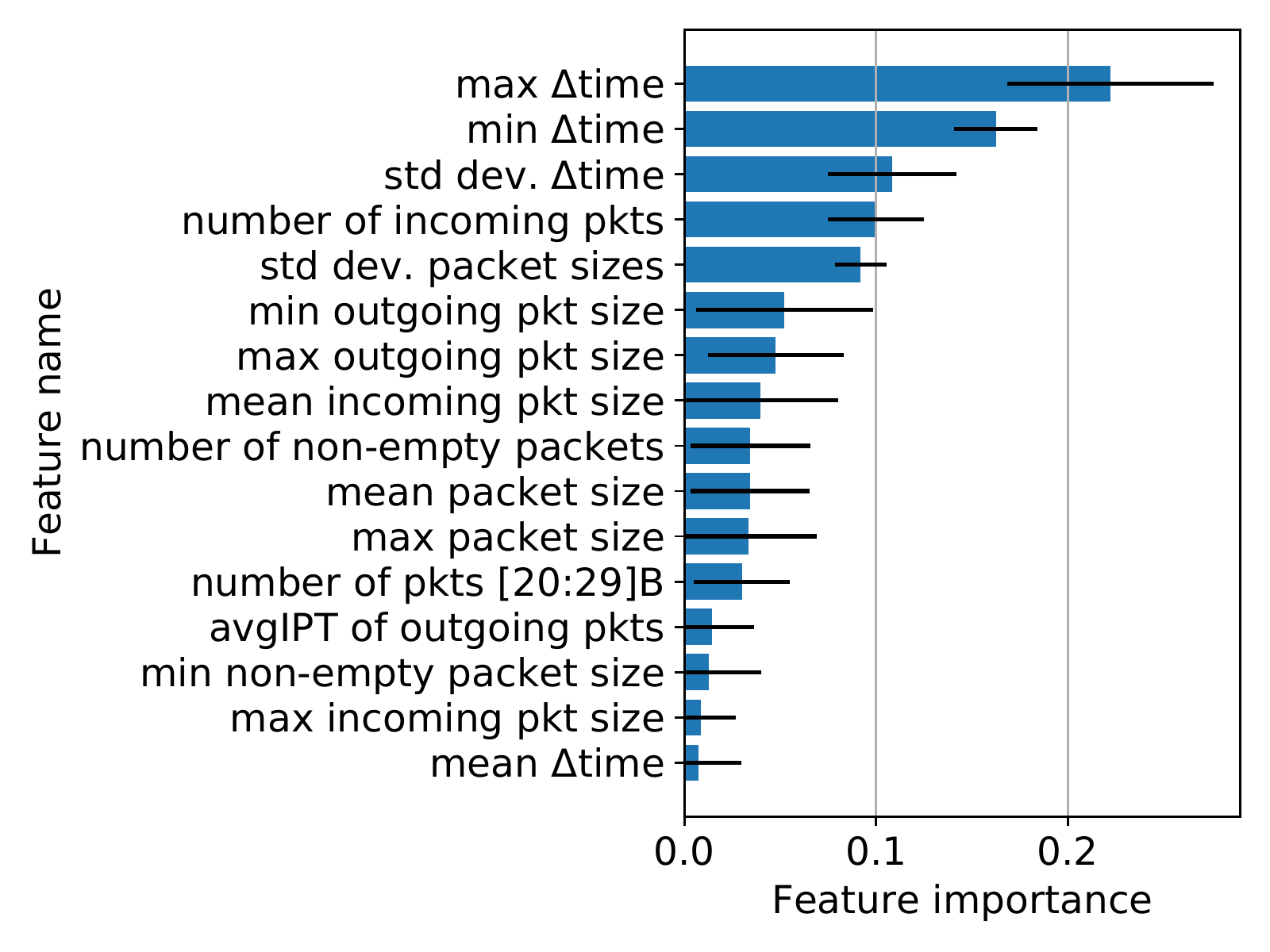}
\label{chipset-id-fi}}
\label{chipset-id}
\caption{Chipset identification (Classic \& LE)}
\end{figure}

\begin{figure}[h]
\centering
\includegraphics[width=\textwidth]{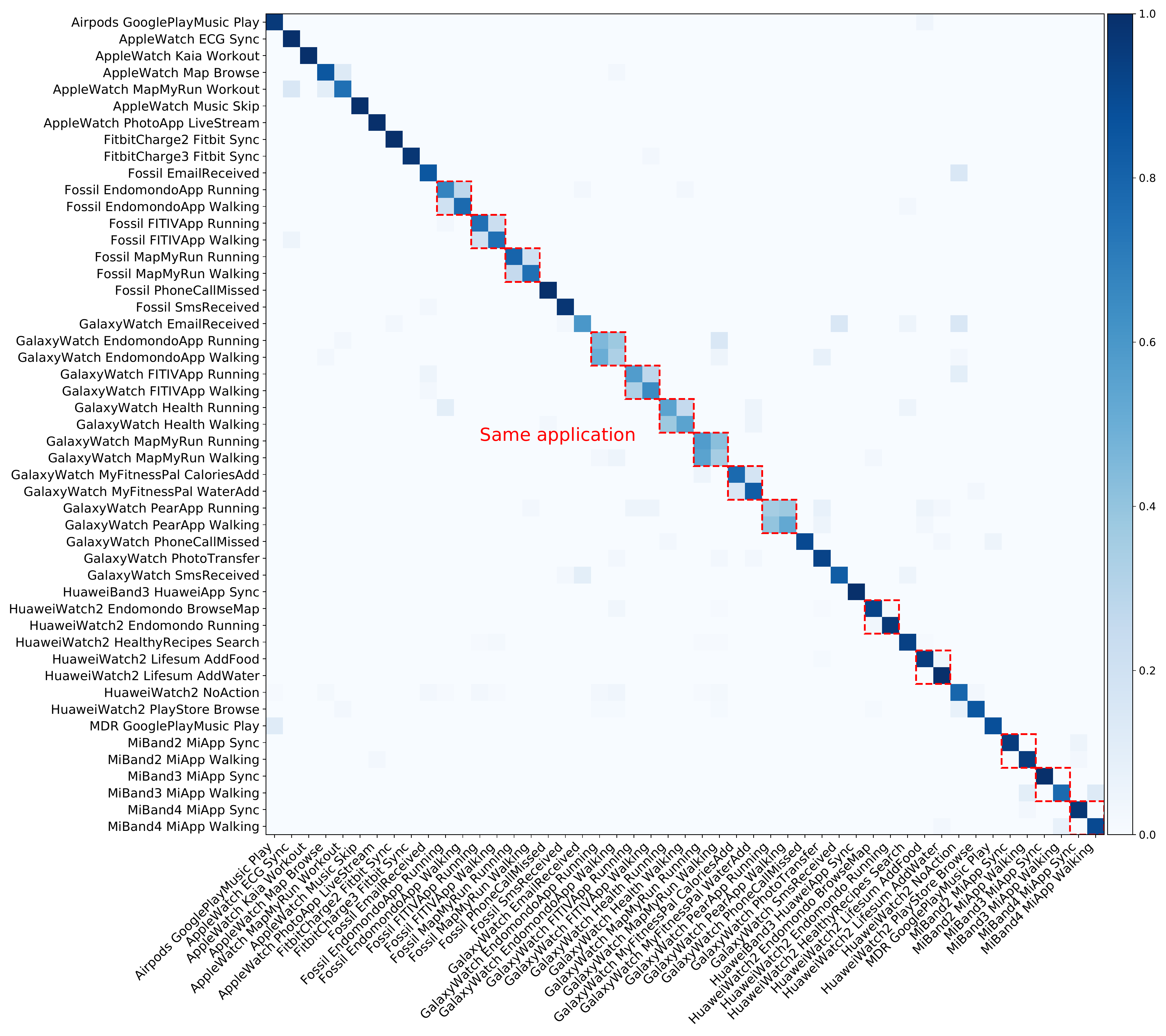}
\caption{Confusion matrix for human-triggered actions. Red dashed squares regroup the two actions ``start a walking workout'' and ``start a running workout'' within the same application.}\label{fig:application-identification-wearables-full}
\end{figure}

\begin{figure}[h]
\centering
\subfigure[Normalized confusion matrix for $56$ apps, Wear OS\@. Apps are sorted by increasing median transmitted volume.]{\includegraphics[width=0.32\linewidth]{app-id-huaweiwatch-cm.pdf}
\label{fig:wearos-app-id-confusion-matrix-full}
}\hspace{0.1cm} %
\subfigure[Normalized confusion matrix for the $18$ ``low-volume'' apps, Wear OS\@. The mean accuracy is $17\%$.]{\includegraphics[width=0.32\linewidth]{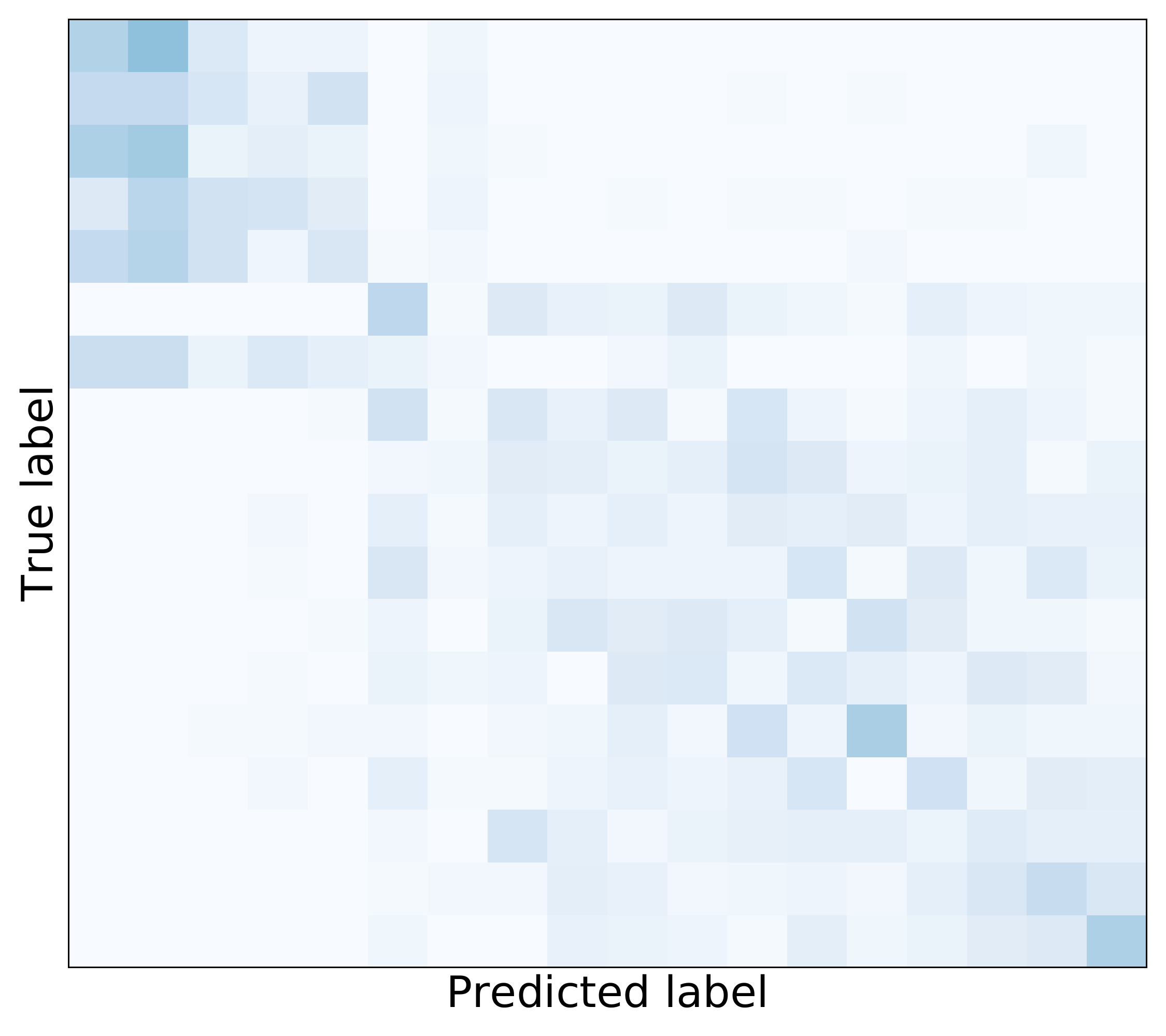}
\label{fig:wearos-app-id-confusion-matrix-filtered2}
}\hspace{0.1cm} %
\subfigure[Normalized confusion matrix for $38$ high-volume apps, Wear OS\@. The mean accuracy is $90\%$.]{\includegraphics[width=0.32\linewidth]{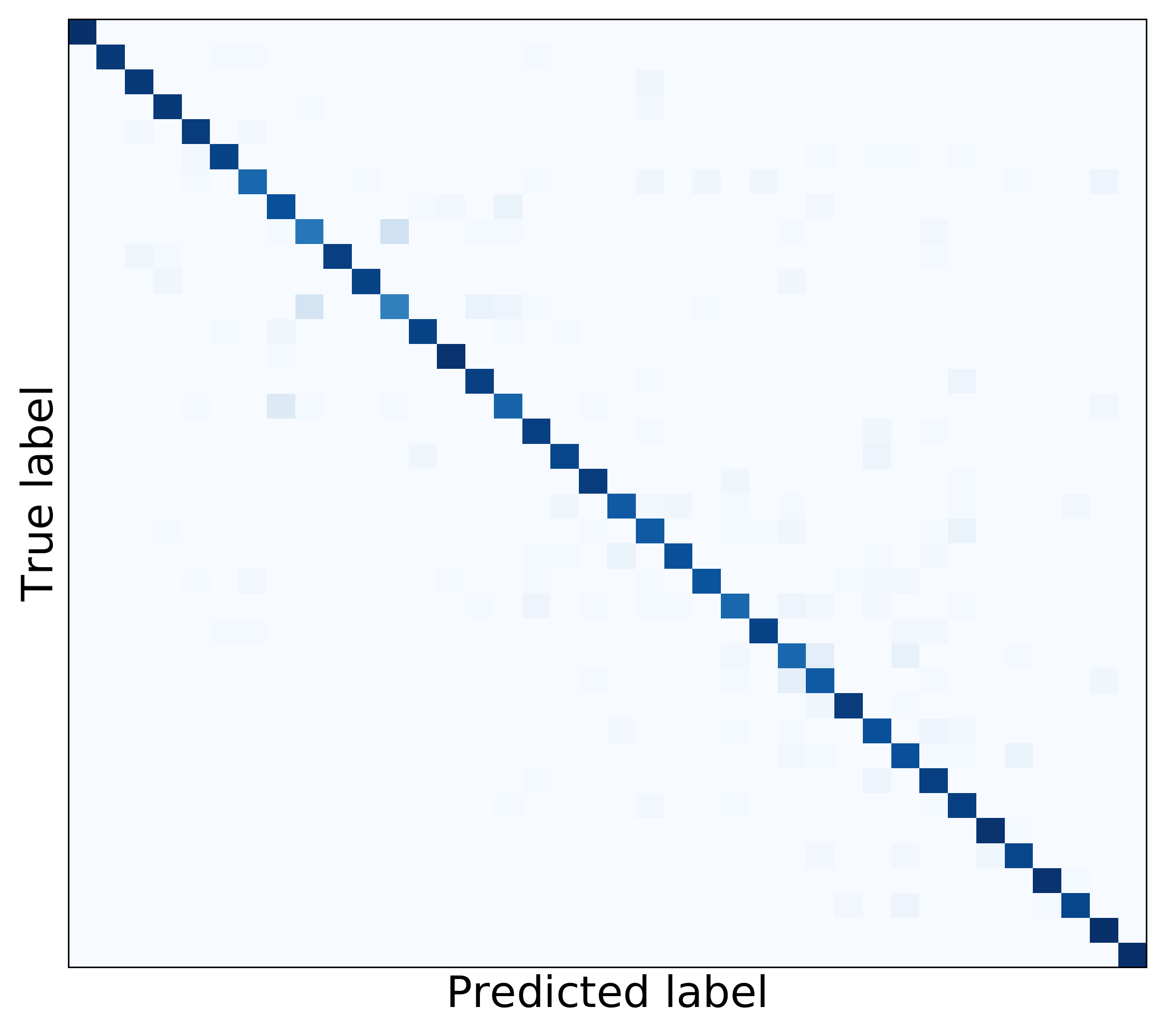}
\label{fig:wearos-app-id-confusion-matrix-filtered}
}
\label{fig:classifier_accuracy_apps_openings-full}
\caption{Normalized confusion matrices per true label for recognizing smartwatch application openings.}
\end{figure}

\vspace{0.2cm}

\begin{figure}[h]
\centering
\subfigure[Train on Huawei-Pixel. Test on Fossil-Nexus.]{\includegraphics[width=0.48\linewidth]{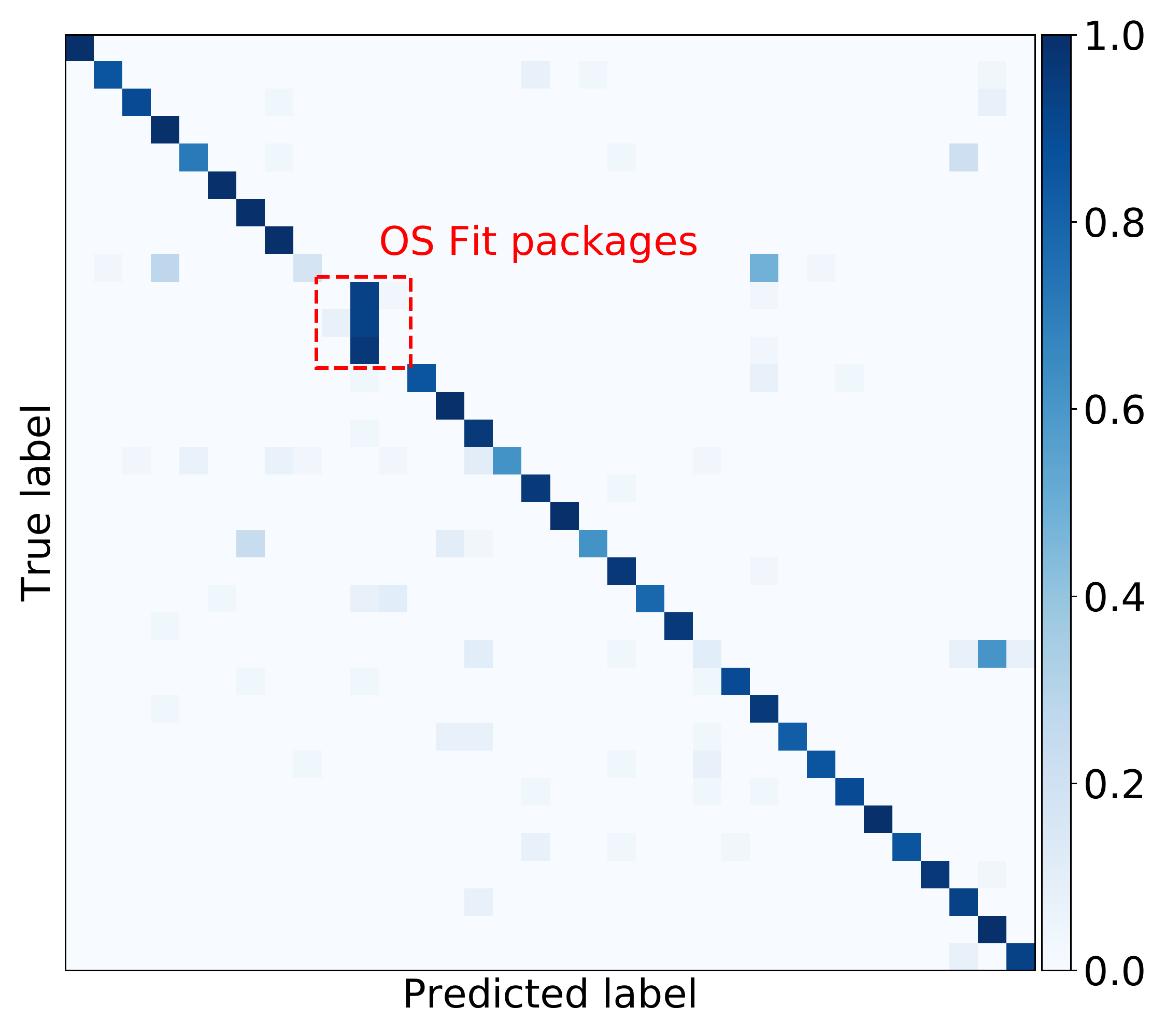}
\label{fig:transfer1}
} %
\subfigure[Train on Fossil-Nexus. Test on Huawei-Pixel.]{\includegraphics[width=0.48\linewidth]{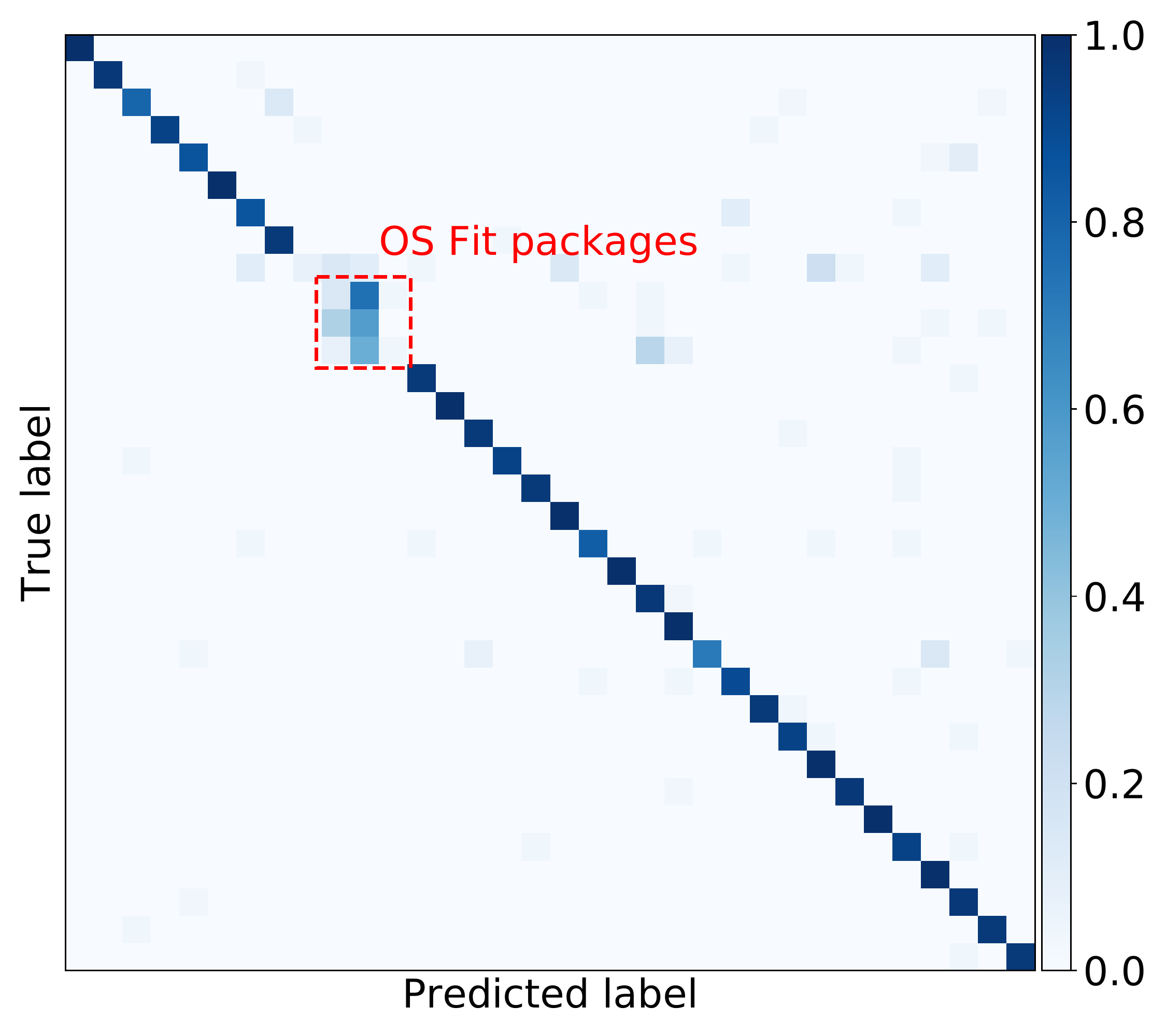}
\label{fig:transfer2}}
\caption{Transferability experiment for the application identification (``deep'' experiment).}
\label{fig:transfer}
\end{figure}

\begin{figure}[h]
\begin{minipage}{0.45\linewidth}
\centering
\includegraphics[width=\textwidth]{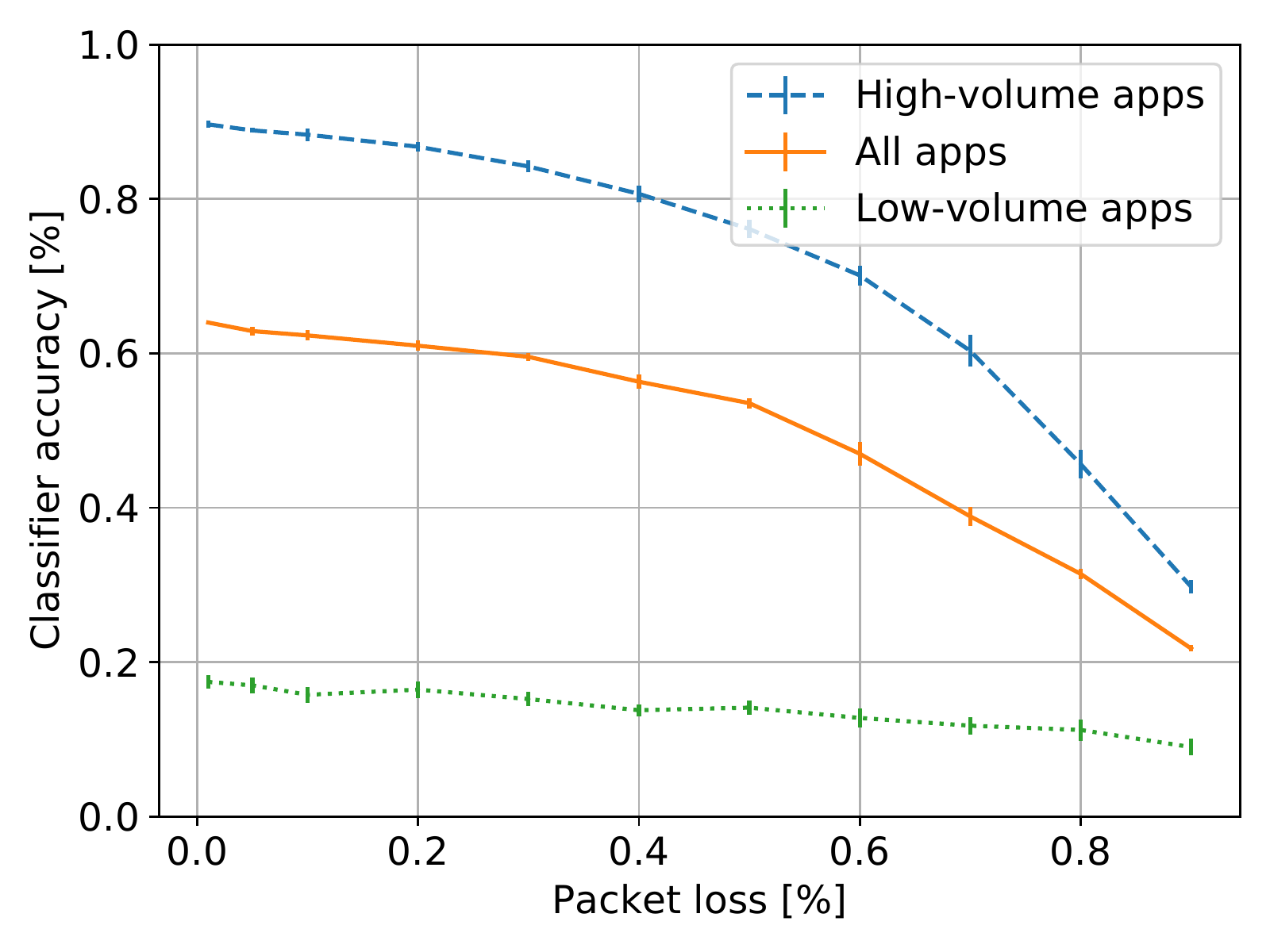}
\caption{Packet loss rate versus classifier accuracy for application identification (``deep'' experiment).}\label{fig:deep-loss}
\end{minipage}\qquad
\begin{minipage}{0.45\linewidth}
\centering
\includegraphics[width=\textwidth]{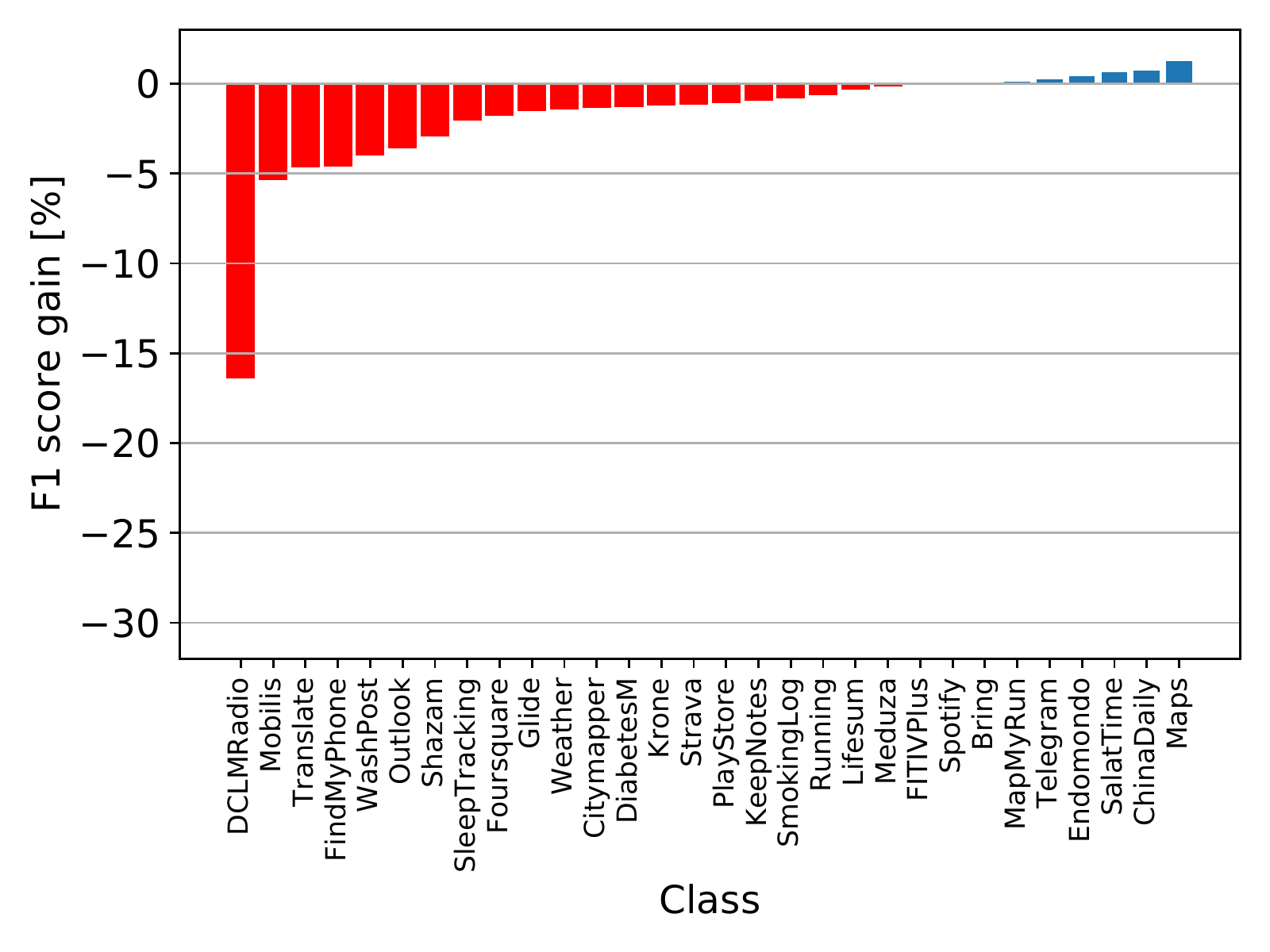}
\caption{Evolution of the F1 score per class, averaged over $29$ days, training over the initial three days. See Figure~\ref{fig:aging2} for the case of training on day 0 only.}\label{fig:aging-3-days}
\end{minipage}\qquad
\end{figure}

\begin{figure}[h]
\centering
\subfigure[Mean $W$ of the Rayleigth distribution with respect to the attacker accuracy (lower is better).]{\includegraphics[width=0.49\linewidth]{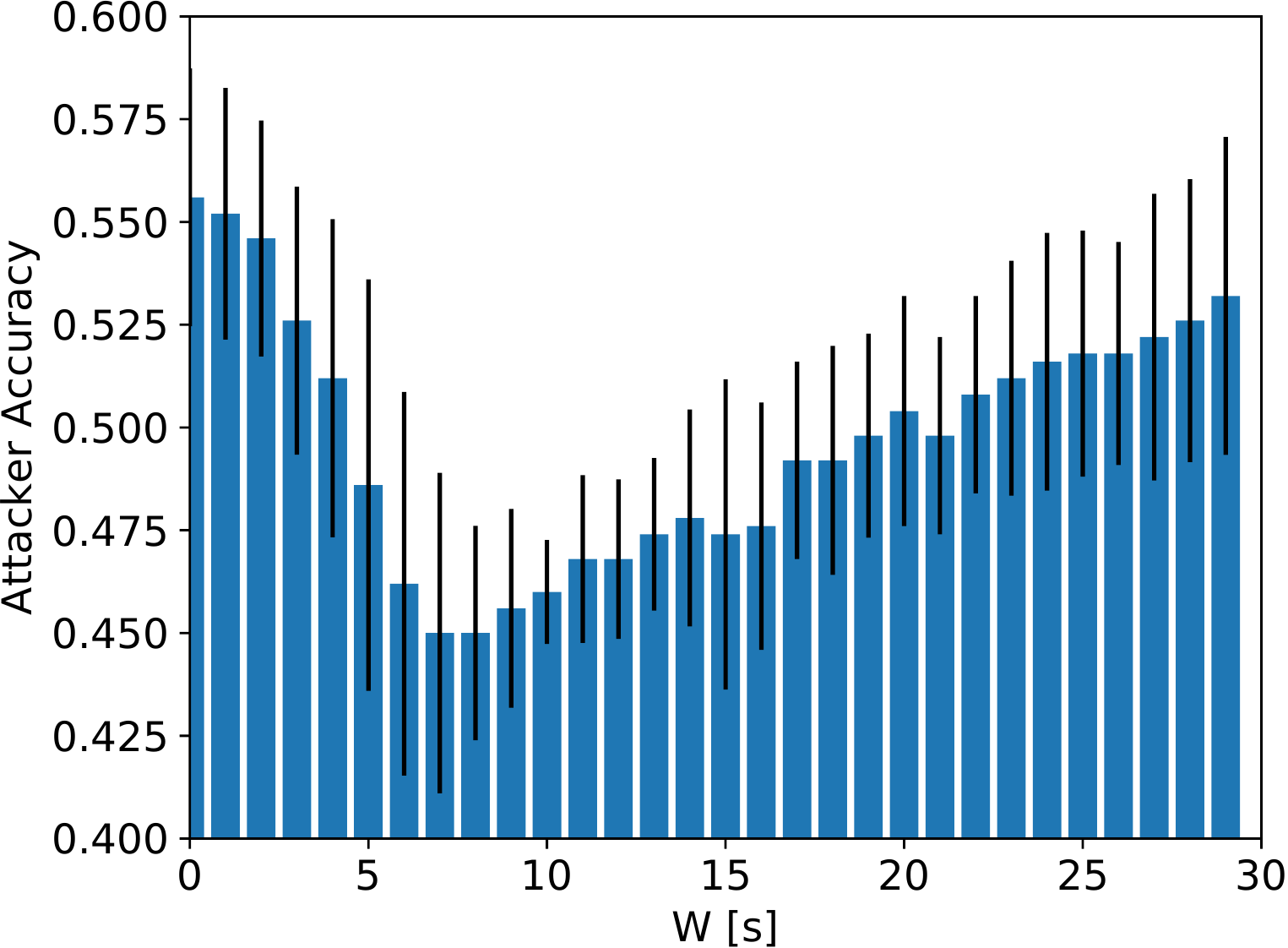}}\hspace{0.1cm} %
\subfigure[Number of dummies $N$ with respect to the cost of the defense.
The cost is averaged per sample.]{\includegraphics[width=0.49\linewidth]{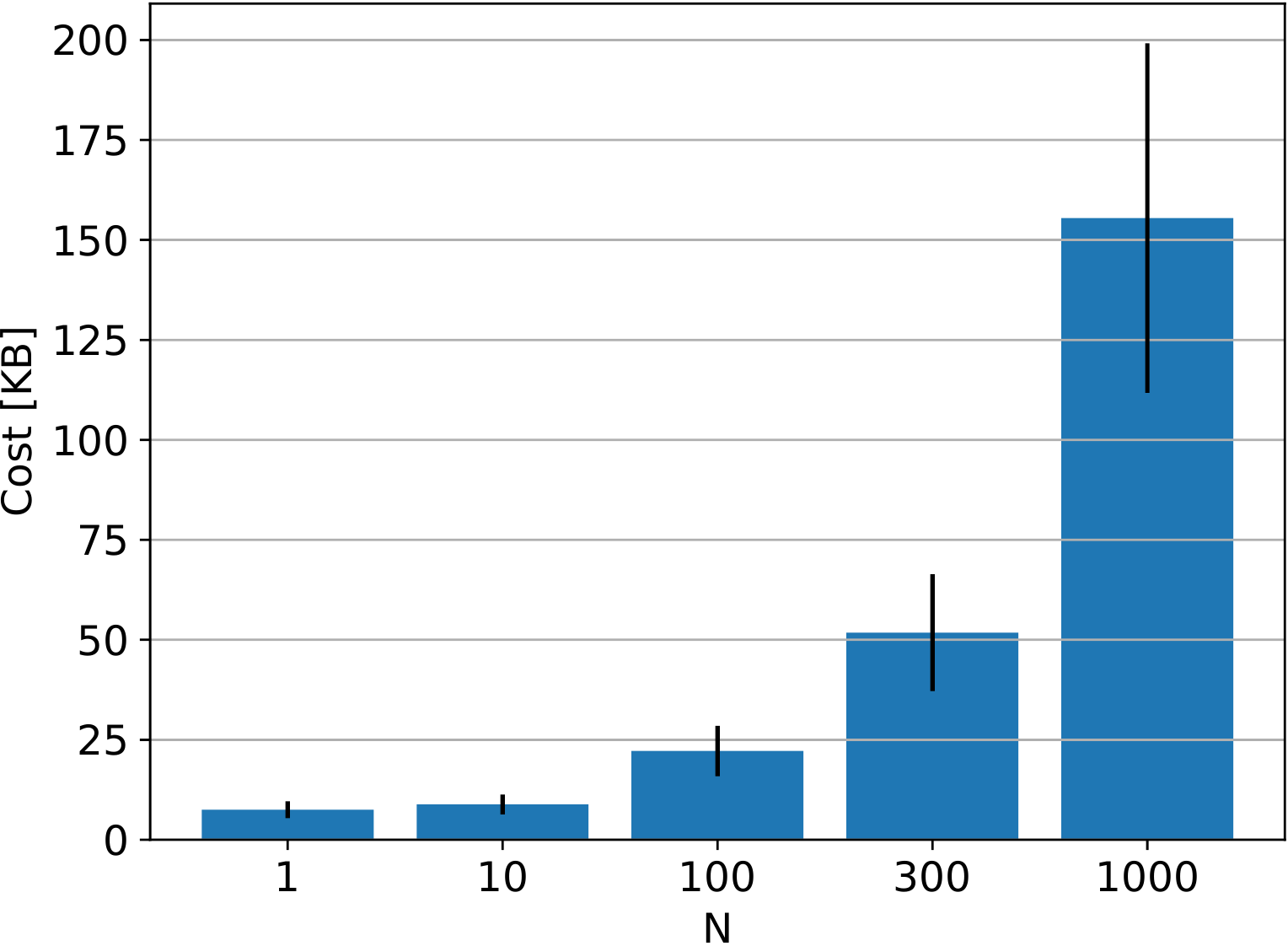}}
\caption{Parameter choices for \defense{add_dummies}, illustrated here with the samples for application identification (``deep'' experiment).
We select $W=6$s and $N=300$.}
\label{fig:add-dummies-params}
\end{figure}

\begin{figure}[h]
\centering
\subfigure[Feature importance when attacking \defense{pad}-defended traces, device identification, Bluetooth LE.]{\includegraphics[width=0.49\linewidth]{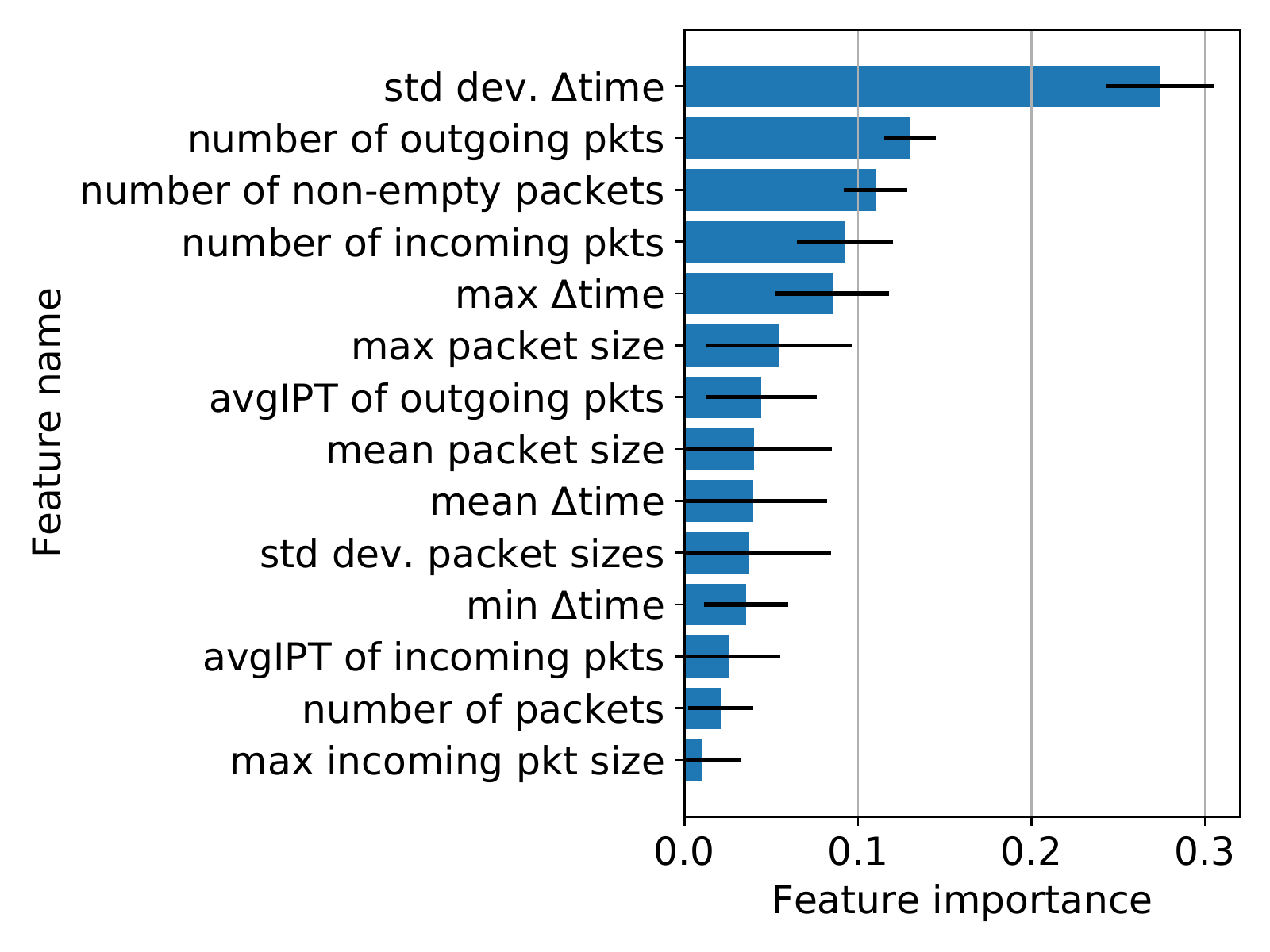}
\label{fig:device-id-def-ble-pad-fi}
}\hspace{0.05cm} %
\subfigure[Feature importance for action identification against \defense{delay_group}-defended traces, DiabetesM.]{\includegraphics[width=0.49\linewidth]{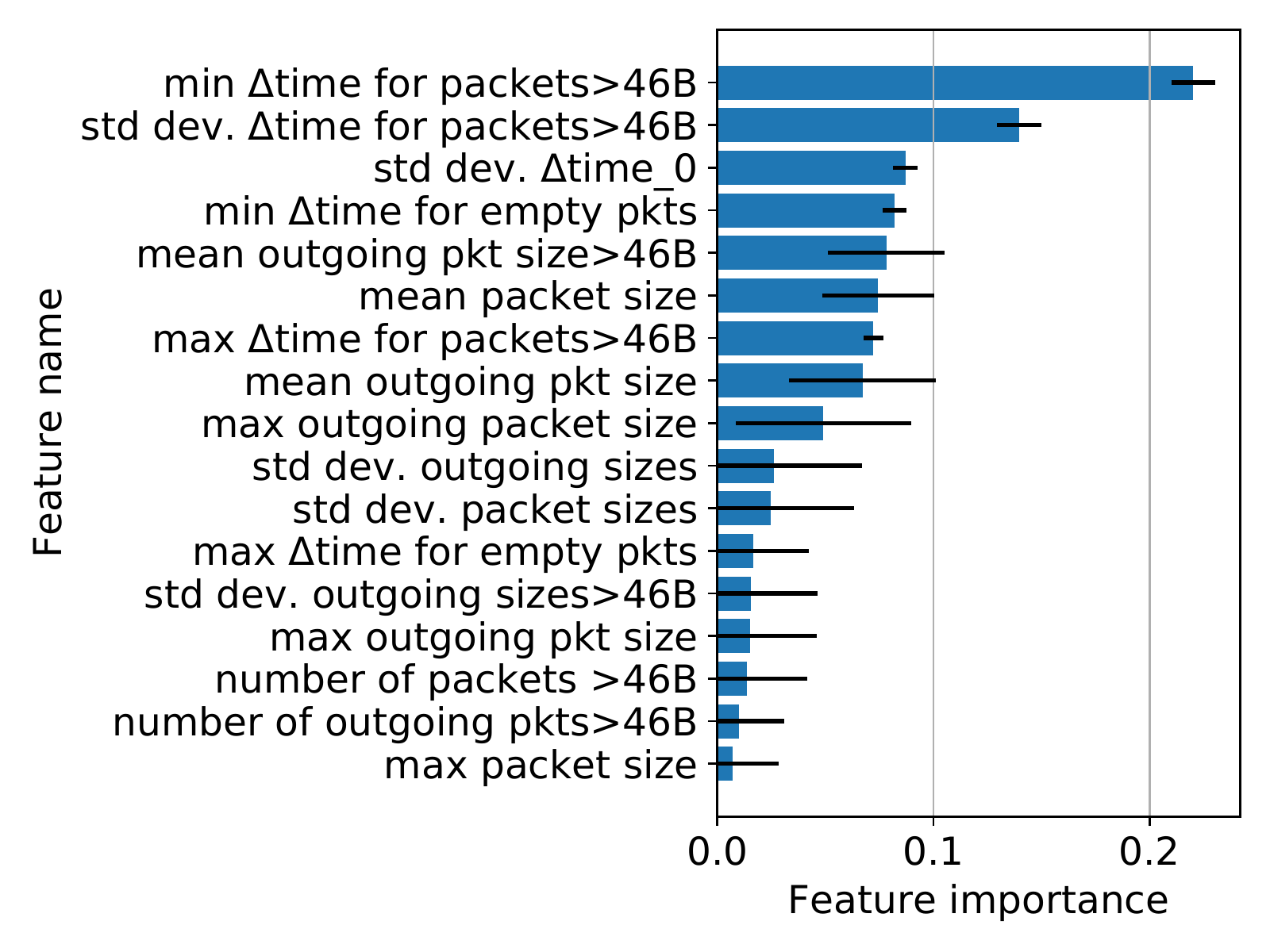}
\label{fig:action-id-diabetesm-def-delay_group-fi}}
\caption{Feature importance when attacking defended traces, two scenarios.}
\end{figure}

\begin{figure}[h]
\centering
\subfigure[Bluetooth Classic]{\includegraphics[width=0.49\linewidth]{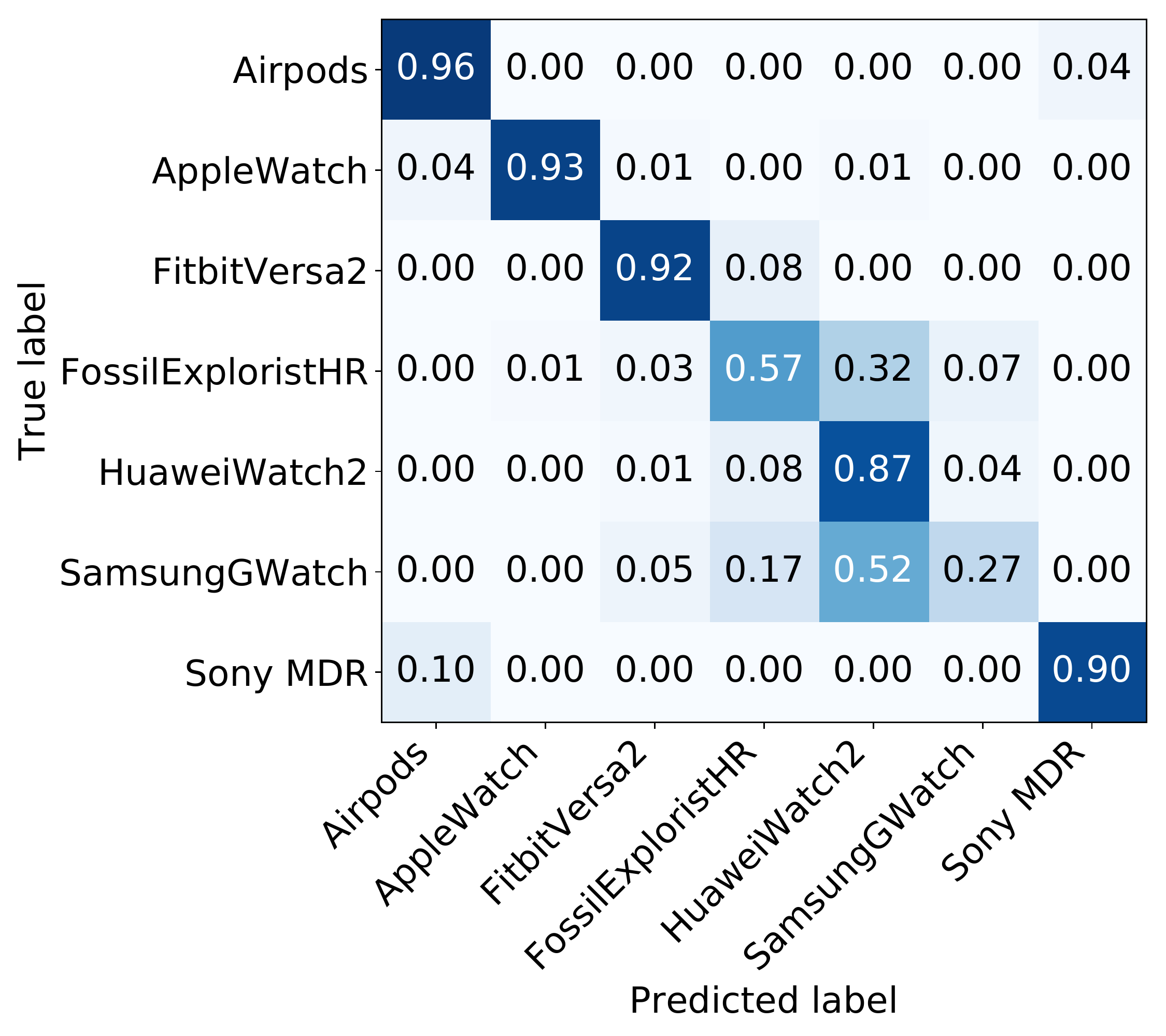}}\hspace{0.05cm} %
\subfigure[Bluetooth Low Energy]{\includegraphics[width=0.49\linewidth]{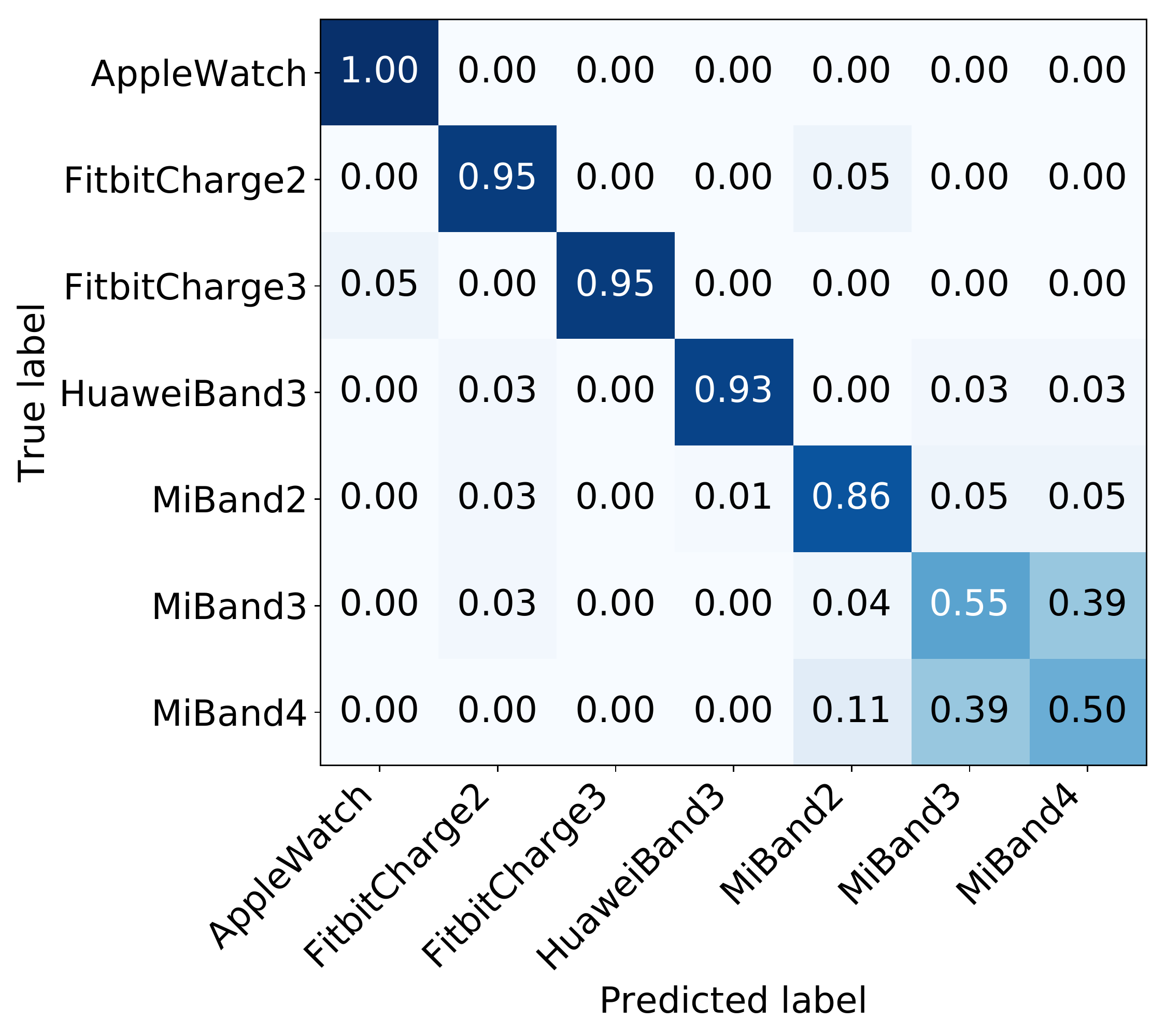}}
\caption{Normalized confusion matrices per true label, device identification against \defense{add_dummies}-defended traces.}
\label{fig:device-id-front-cm}
\end{figure}

\begin{figure}[h]
\centering
\subfigure[Action identification (``wide'' experiment on all wearables).]{\includegraphics[width=0.49\linewidth]{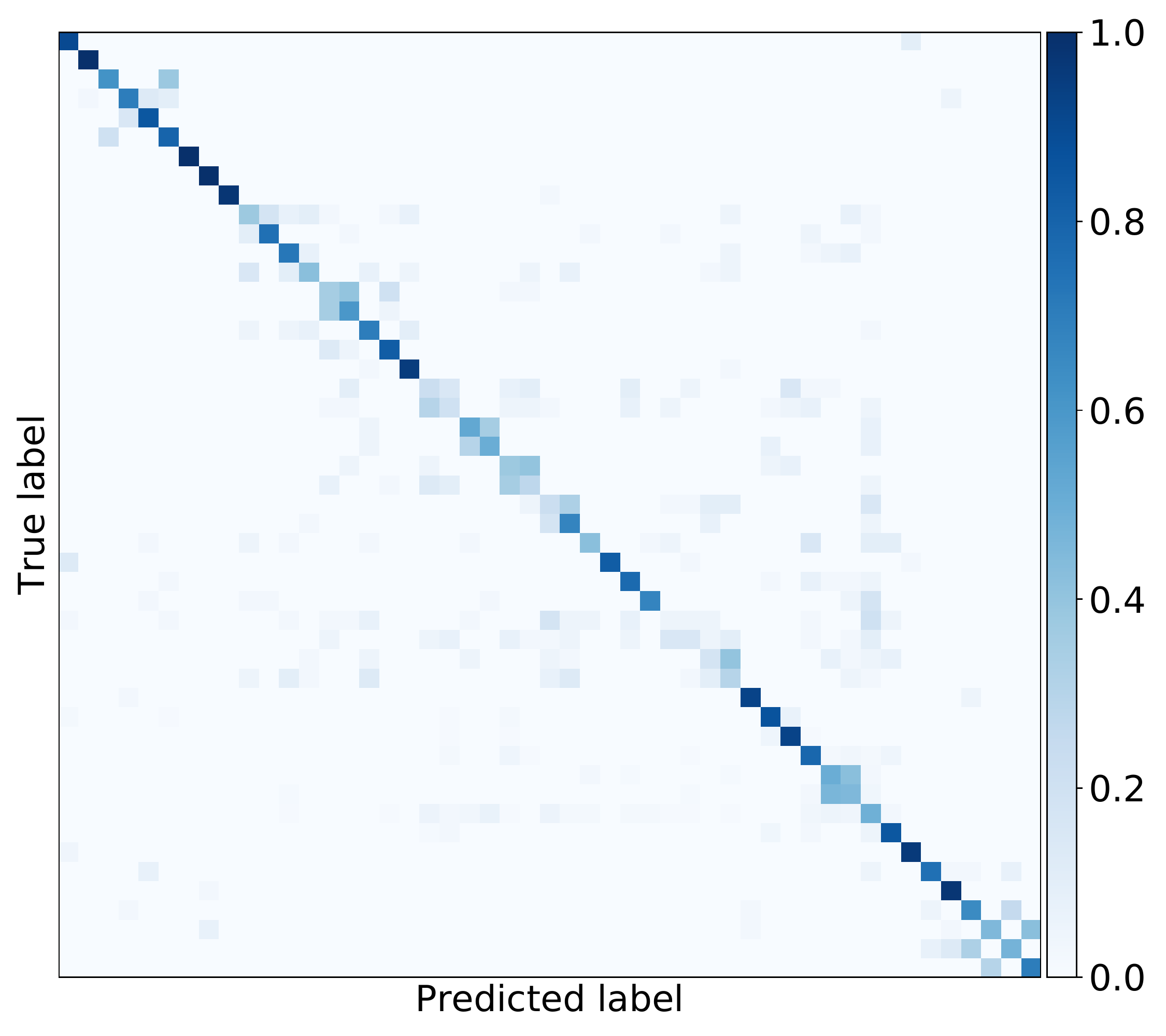}
\label{fig:action-id-wearables-def-front-cm}}\hspace{0.05cm} %
\subfigure[Application identification  (``deep'' experiment).]{\includegraphics[width=0.49\linewidth]{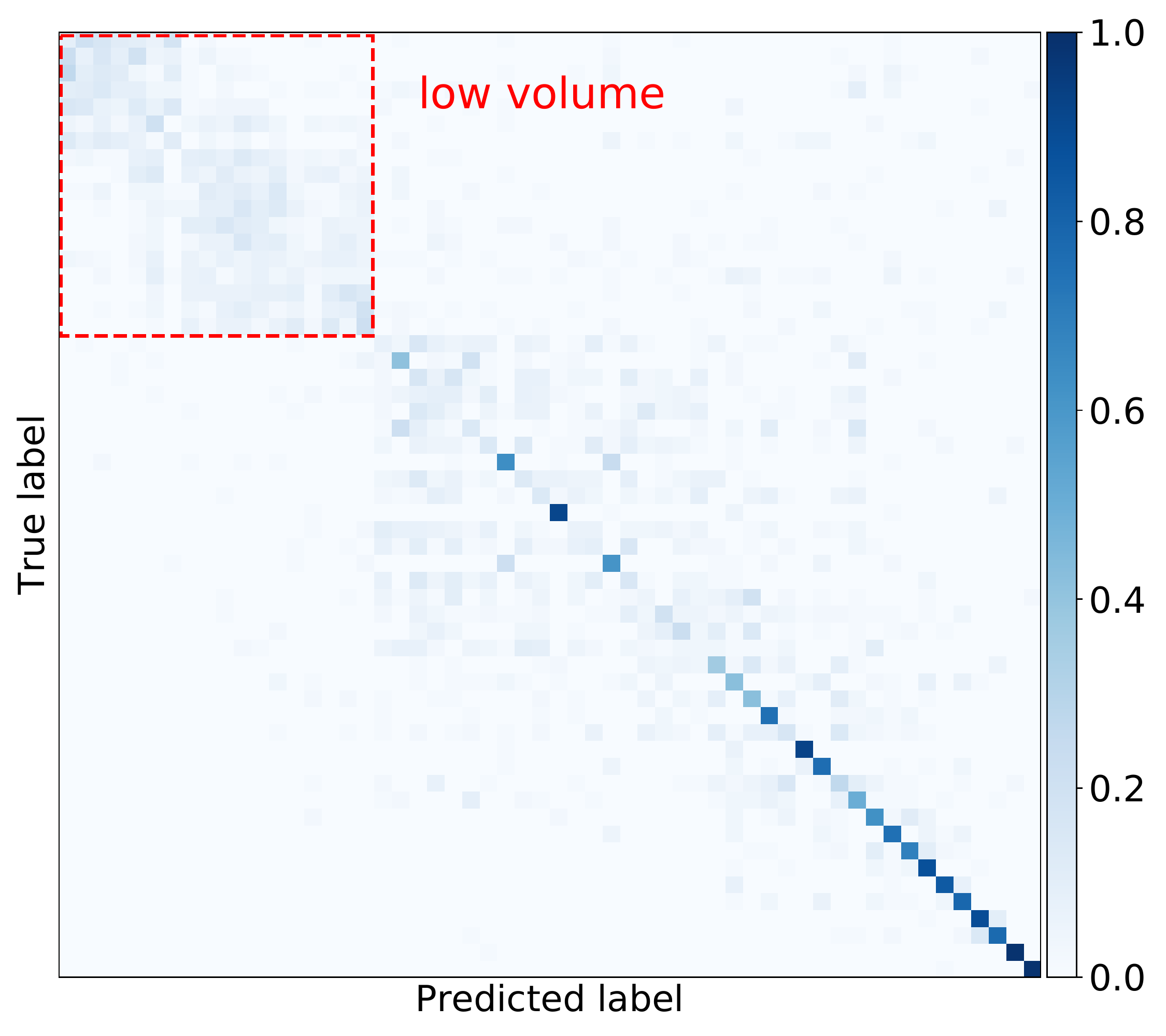}
\label{fig:app-id-huaweiwatch-def-front-cm}}
\caption{Normalized confusion matrices per true label.
Action and application identification against \defense{add_dummies}-defended traces.
The matrices are sorted by increasing median transmitted volume.}
\end{figure}
\begin{table}[h]
\caption*{Classifier performance for Device identification.}
\vspace{-0.1cm}
\small
\centering
\begin{minipage}[t]{0.49\textwidth}
\caption{Bluetooth Classic.}
\vspace{-0.2cm}
\begin{tabular}{lrrrr}
\textbf{Label} & \textbf{Precision} & \textbf{Recall} & \textbf{F1-score} \\
\dev{Airpods} & 0.83 & 0.9 & 0.87 \\
\dev{AppleWatch} & 0.98 & 0.93 & 0.96 \\
\dev{FitbitVersa2} & 1.0 & 1.0 & 1.0 \\
\dev{FossilExploristHR} & 0.96 & 0.98 & 0.97 \\
\dev{HuaweiWatch2} & 0.98 & 0.98 & 0.98 \\
\dev{SamsungGWatch} & 1.0 & 0.98 & 0.99 \\
\dev{Sony MDR} & 0.9 & 0.86 & 0.88 \\
\emph{Average} & 0.96 & 0.96 & 0.96 \\
\end{tabular}
\label{table:device-id-cm-cla}
\end{minipage}\hfill
\begin{minipage}[t]{0.49\textwidth}
\caption{Bluetooth LE.}
\begin{tabular}{lrrrr}
\textbf{Label} & \textbf{Precision} & \textbf{Recall} & \textbf{F1-score} \\
\dev{AppleWatch} & 0.99 & 1.0 & 1.0 \\
\dev{FitbitCharge2} & 1.0 & 1.0 & 1.0 \\
\dev{FitbitCharge3} & 1.0 & 0.95 & 0.97 \\
\dev{HuaweiBand3} & 1.0 & 1.0 & 1.0 \\
\dev{MiBand2} & 0.93 & 0.94 & 0.93 \\
\dev{MiBand3} & 0.88 & 0.98 & 0.92 \\
\dev{MiBand4} & 0.99 & 0.86 & 0.92 \\
\emph{Average} & 0.97 & 0.97 & 0.97 \\
\end{tabular}
\label{table:device-id-cm-ble}
\end{minipage}
\end{table}

\begin{table}[h]
\caption*{Classifier performances.}
\vspace{-0.1cm}
\small
\centering
\begin{minipage}[t]{0.49\textwidth}
\caption{Chipset identification.}
\vspace{-0.2cm}
\begin{tabular}{lrrrr}
\textbf{Label} & \textbf{Precision} & \textbf{Recall} & \textbf{F1-score} \\
\dev{Apple} & 0.9 & 0.94 & 0.92 \\
\dev{Broadcomm} & 0.95 & 0.98 & 0.96 \\
\dev{Cypress} & 0.98 & 0.98 & 0.98 \\
\dev{Dialog} & 0.98 & 0.98 & 0.98 \\
\dev{MicroElectronics} & 1.0 & 1.0 & 1.0 \\
\dev{Qualcomm} & 0.97 & 0.88 & 0.92 \\
\dev{RivieraWaves} & 0.98 & 1.0 & 0.99 \\
\emph{Average} & 0.96 & 0.96 & 0.96 \\
\end{tabular}
\label{table:chipset-identification-accuracy}
\end{minipage}\hfill
\begin{minipage}[t]{0.49\textwidth}
\caption{Action identification within \app{DiabetesM}.}
\vspace{-0.2cm}
\begin{tabular}{lrrrr}
\textbf{Label} & \textbf{Precision} & \textbf{Recall} & \textbf{F1-score} \\
\act{Add Calorie} & 0.44 & 0.46 & 0.45 \\
\act{Add Carbs} & 0.87 & 0.9 & 0.89 \\
\act{Add Fat} & 0.77 & 0.81 & 0.79 \\
\act{Add Glucose} & 0.85 & 0.91 & 0.88 \\
\act{Add Insulin} & 0.9 & 0.95 & 0.92 \\
\act{Add Proteins} & 0.37 & 0.28 & 0.32 \\
\emph{Average} & 0.7 & 0.7 & 0.7 \\
\end{tabular}
\label{table:action-id-diabetesm-cm}
\end{minipage}
\vspace{-0.3cm}
\end{table}

\begin{table}[h]
\caption{$34$ ``high-volume'' applications common between the two pairs of device \dev{Huawei Watch 2 - Pixel 2} and \dev{Fossil Q Explorist HR - Nexus 5}.}
\small
\begin{tabularx}{\linewidth}{ X }

\app{Bring}, \app{Calm}, \app{ChinaDaily}, \app{Citymapper}, \app{DCLMRadio}, \app{DiabetesM}, \app{Endomondo}, \app{FITIVPlus}, \app{FindMyPhone}, \app{Fit}, \app{FitBreathe}, \app{FitWorkout}, \app{FoursquareCityGuide}, \app{Glide}, \app{KeepNotes}, \app{Krone}, \app{Lifesum}, \app{MapMyRun}, \app{Maps}, \app{Meduza}, \app{Mobills}, \app{Outlook}, \app{PlayStore}, \app{Running}, \app{SalatTime}, \app{Shazam}, \app{SleepTracking}, \app{SmokingLog}, \app{Spotify}, \app{Strava}, \app{Telegram}, \app{Translate}, \app{WashPost}, \app{Weather} \\
\end{tabularx}
\label{table:transfer-apps}
\end{table}

\begin{table}[h]
\caption{Applications and Actions used for the long-run captures.}
\small
\begin{tabularx}{\linewidth}{ r X }
\textbf{Applications} & \app{AppInTheAir}, \app{Bring}, \app{Calm}, \app{ChinaDaily}, \app{Citymapper}, \app{DCLMRadio}, \app{DiabetesM}, \app{Endomondo}, \app{FITIVPlus}, \app{FindMyPhone}, \app{FoursquareCityGuide}, \app{Glide}, \app{KeepNotes}, \app{Krone}, \app{Lifesum}, \app{MapMyRun}, \app{Maps}, \app{Meduza}, \app{Mobills}, \app{Outlook}, \app{PlayStore}, \app{Qardio}, \app{Running}, \app{SalatTime}, \app{Shazam}, \app{SleepTracking}, \app{SmokingLog}, \app{Spotify}, \app{Strava}, \app{Telegram}, \app{Translate}, \app{WashPost}, \app{Weather}. \\

\textbf{Actions}      & \act{DiabetesM_AddCalorie}, \act{DiabetesM_AddCarbs}, \act{DiabetesM_AddFat}, \act{DiabetesM_AddGlucose}, \act{DiabetesM_AddInsulin}, \act{DiabetesM_AddProteins}, \act{Endomondo_BrowseMap}, \act{Endomondo_Running}, \act{FoursquareCityGuide_Coffees}, \act{FoursquareCityGuide_Leisure}, \act{FoursquareCityGuide_NightLife}, \act{FoursquareCityGuide_Restaurants}, \act{FoursquareCityGuide_Shopping}, \act{HealthyRecipes_SearchRecipe}, \act{Lifesum_AddFood}, \act{Lifesum_AddWater}, \act{PlayStore_Browse}.                               \\
\end{tabularx}
\label{table:applications-actions-longrun}
\end{table}

\begin{table}[h]
\vspace{-0.4cm}
\caption{Classifier performance for Action identification, ``Wide'' experiment on all wearables.}
\small
\centering
\vspace{-0.3cm}
\begin{tabular}{lrrrr}
\textbf{Label} & \textbf{Precision} & \textbf{Recall} & \textbf{F1-score} \\
\app{Airpods GooglePlayMusic Play} & 0.87 & 0.96 & 0.91 \\
\app{AppleWatch ECG Sync} & 0.83 & 1.0 & 0.91 \\
\app{AppleWatch Kaia Workout} & 1.0 & 1.0 & 1.0 \\
\app{AppleWatch Map Browse} & 0.83 & 0.85 & 0.84 \\
\app{AppleWatch MapMyRun Workout} & 0.79 & 0.75 & 0.77 \\
\app{AppleWatch Music Skip} & 1.0 & 1.0 & 1.0 \\
\app{AppleWatch PhotoApp LiveStream} & 0.98 & 1.0 & 0.99 \\
\app{FitbitCharge2 Fitbit Sync} & 0.98 & 1.0 & 0.99 \\
\app{FitbitCharge3 Fitbit Sync} & 1.0 & 0.98 & 0.99 \\
\app{FossilExploristHR EndomondoApp Running} & 0.66 & 0.68 & 0.67 \\
\app{FossilExploristHR EndomondoApp Walking} & 0.74 & 0.78 & 0.76 \\
\app{FossilExploristHR FITIVApp Running} & 0.71 & 0.75 & 0.73 \\
\app{FossilExploristHR FITIVApp Walking} & 0.71 & 0.75 & 0.73 \\
\app{FossilExploristHR MapMyRun Running} & 0.76 & 0.8 & 0.78 \\
\app{FossilExploristHR MapMyRun Walking} & 0.77 & 0.75 & 0.76 \\
\app{FossilExploristHR NoApp EmailReceived} & 0.83 & 0.85 & 0.84 \\
\app{FossilExploristHR NoApp PhoneCallMissed} & 0.98 & 1.0 & 0.99 \\
\app{FossilExploristHR NoApp SmsReceived} & 0.95 & 0.98 & 0.96 \\
\app{GalaxyWatch EndomondoApp Running} & 0.42 & 0.45 & 0.43 \\
\app{GalaxyWatch EndomondoApp Walking} & 0.32 & 0.32 & 0.32 \\
\app{GalaxyWatch FITIVApp Running} & 0.61 & 0.57 & 0.59 \\
\app{GalaxyWatch FITIVApp Walking} & 0.65 & 0.65 & 0.65 \\
\app{GalaxyWatch MapMyRun Running} & 0.47 & 0.57 & 0.52 \\
\app{GalaxyWatch MapMyRun Walking} & 0.31 & 0.35 & 0.33 \\
\app{GalaxyWatch MyFitnessPalApp CaloriesAdd} & 0.84 & 0.78 & 0.81 \\
\app{GalaxyWatch MyFitnessPalApp WaterAdd} & 0.73 & 0.82 & 0.78 \\
\app{GalaxyWatch NoApp EmailReceived} & 0.83 & 0.6 & 0.7 \\
\app{GalaxyWatch NoApp PhoneCallMissed} & 1.0 & 0.9 & 0.95 \\
\app{GalaxyWatch NoApp PhotoTransfer} & 0.79 & 0.92 & 0.85 \\
\app{GalaxyWatch NoApp SmsReceived} & 0.85 & 0.82 & 0.84 \\
\app{GalaxyWatch PearApp Running} & 0.47 & 0.35 & 0.4 \\
\app{GalaxyWatch PearApp Walking} & 0.58 & 0.52 & 0.55 \\
\app{GalaxyWatch SamsungHealthApp Running} & 0.58 & 0.55 & 0.56 \\
\app{GalaxyWatch SamsungHealthApp Walking} & 0.67 & 0.55 & 0.6 \\
\app{HuaweiBand3 HuaweiApp Sync} & 1.0 & 1.0 & 1.0 \\
\app{HuaweiWatch2 Endomondo BrowseMap} & 0.94 & 0.92 & 0.93 \\
\app{HuaweiWatch2 Endomondo Running} & 0.98 & 0.97 & 0.97 \\
\app{HuaweiWatch2 HealthyRecipes SearchRecipe} & 0.92 & 0.93 & 0.93 \\
\app{HuaweiWatch2 Lifesum AddFood} & 0.93 & 0.95 & 0.94 \\
\app{HuaweiWatch2 Lifesum AddWater} & 0.93 & 1.0 & 0.96 \\
\app{HuaweiWatch2 NoApp NoAction} & 0.81 & 0.79 & 0.8 \\
\app{HuaweiWatch2 PlayStore Browse} & 0.94 & 0.85 & 0.89 \\
\app{MDR GooglePlayMusic Play} & 0.96 & 0.88 & 0.92 \\
\app{MiBand2 MiApp Sync} & 1.0 & 0.95 & 0.97 \\
\app{MiBand2 MiApp Walking} & 0.88 & 0.95 & 0.92 \\
\app{MiBand3 MiApp Sync} & 1.0 & 1.0 & 1.0 \\
\app{MiBand3 MiApp Walking} & 0.91 & 0.78 & 0.84 \\
\app{MiBand4 MiApp Sync} & 0.93 & 0.98 & 0.95 \\
\app{MiBand4 MiApp Walking} & 0.88 & 0.9 & 0.89 \\
\emph{Average} & 0.82 & 0.82 & 0.82 \\
\end{tabular}
\label{table:action-id-wearables-cm}
\end{table}

\begin{table}[h]
\small
\centering
\begin{minipage}[t]{0.49\textwidth}
\caption{Details of transmitted volumes for the $18$ ``low-volume'' apps over $40$ recorded samples. \app{NoApp} corresponds to OS communications.}
\begin{tabular}{lrr}
\textbf{App}           & \textbf{Median [B]} & \textbf{Std dev. [B]} \\
\app{Reminders}        & 0.0                 & 23660.5               \\
\app{Battery}          & 0.0                 & 239.6                 \\
\app{DuaKhatqmAlQuran} & 11.0                & 43168.3               \\
\app{WearCasts}        & 37.0                & 398149.0              \\
\app{DailyTracking}    & 44.0                & 326.9                 \\
\app{ASB}              & 75.0                & 3511.6                \\
\app{NoApp}            & 96.5                & 3949.5                \\
\app{HeartRate}        & 104.0               & 21824.4               \\
\app{Workout}          & 119.0               & 5108.4                \\
\app{AthkarOfPrayer}   & 120.0               & 4835.5                \\
\app{Alarm}            & 122.5               & 8891.4                \\
\app{GooglePay}        & 127.0               & 2762.1                \\
\app{Flashlight}       & 152.5               & 4176.7                \\
\app{Phone}            & 154.5               & 2319.2                \\
\app{PlayMusic}        & 156.0               & 16294.5               \\
\app{HealthyRecipes}   & 167.5               & 3104.7                \\
\app{Sleep}            & 171.5               & 14460.4               \\
\app{Medisafe}         & 194.0               & 8802.6                \\
\end{tabular}
\label{table:low-volume-apps-details}
\end{minipage}\hfill
\begin{minipage}[t]{0.49\textwidth}
\caption{Details of transmitted volumes for the $38$ ``high-volume'' apps over $40$ samples.}
\begin{tabular}{lrr}
\textbf{App}              & \textbf{Median [KB]} & \textbf{Std dev. [KB]} \\
\app{SalatTime}           & 0.6                  & 5.5                    \\
\app{MapMyFitness}            & 0.6                  & 2.6                    \\
\app{Citymapper}          & 1.1                  & 3.4                    \\
\app{Calm}                & 1.2                  & 8.3                    \\
\app{Outlook}             & 1.4                  & 3.4                    \\
\app{DiabetesM}           & 1.4                  & 2.5                    \\
\app{SmokingLog}          & 1.5                  & 46.7                   \\
\app{MapMyRun}            & 1.8                  & 8.1                    \\
\app{SleepTracking}       & 1.9                  & 4.4                    \\
\app{Mobills}             & 2.1                  & 1.5                    \\
\app{Fit}                 & 2.1                  & 8.4                    \\
\app{Weather}             & 2.6                  & 9.4                    \\
\app{Running}             & 2.9                  & 8.0                    \\
\app{FitWorkout}          & 3.3                  & 144.6                  \\
\app{FitBreathe}          & 3.3                  & 3.4                    \\
\app{FoursquareCityGuide} & 3.8                  & 7.0                    \\
\app{Glide}               & 4.8                  & 5.2                    \\
\app{Translate}           & 5.8                  & 5.9                    \\
\app{Shazam}              & 6.2                  & 41.9                   \\
\app{Qardio}              & 6.5                  & 7.1                    \\
\app{Krone}               & 6.8                  & 13.2                   \\
\app{KeepNotes}           & 7.0                  & 4.8                    \\
\app{FindMyPhone}         & 8.0                  & 12.2                   \\
\app{Telegram}            & 8.7                  & 3.7                    \\
\app{Strava}              & 10.6                 & 4.7                    \\
\app{DCLMRadio}           & 16.6                 & 15.6                   \\
\app{Lifesum}             & 17.8                 & 11.0                   \\
\app{Endomondo}           & 17.9                 & 9.7                    \\
\app{PlayStore}           & 18.5                 & 24.3                   \\
\app{Maps}                & 18.7                 & 62.8                   \\
\app{AppInTheAir}         & 26.4                 & 14.4                   \\
\app{Bring}               & 34.5                 & 7.9                    \\
\app{Spotify}             & 38.8                 & 9.8                    \\
\app{Meduza}              & 40.9                 & 13.9                   \\
\app{FITIVPlus}           & 48.9                 & 11.7                   \\
\app{ChinaDaily}          & 55.1                 & 8.3                    \\
\app{WashPost}            & 120.1                & 14.2                   \\
\app{Camera}              & 598.0                & 141.5                  \\
\end{tabular}
\label{table:high-volume-apps-details}
\end{minipage}
\end{table}

\begin{table}[h]
\caption*{Classifier performance for App identification, Huawei Watch.}
\small
\centering
\begin{minipage}[t]{0.49\textwidth}
\caption{$18$ ``low-volume'' applications.}
\begin{tabular}{lrrrr}
\textbf{Label} & \textbf{Precision} & \textbf{Recall} & \textbf{F1-score} \\
\app{Battery} & 0.21 & 0.31 & 0.25 \\
\app{Reminders} & 0.14 & 0.25 & 0.18 \\
\app{DuaKhatqmAlQuran} & 0.08 & 0.06 & 0.07 \\
\app{WearCasts} & 0.26 & 0.18 & 0.21 \\
\app{DailyTracking} & 0.2 & 0.15 & 0.17 \\
\app{ASB} & 0.25 & 0.28 & 0.26 \\
\app{NoApp} & 0.06 & 0.02 & 0.04 \\
\app{HeartRate} & 0.17 & 0.15 & 0.16 \\
\app{Workout} & 0.11 & 0.1 & 0.11 \\
\app{AthkarOfPrayer} & 0.09 & 0.09 & 0.09 \\
\app{Alarm} & 0.06 & 0.05 & 0.05 \\
\app{GooglePay} & 0.08 & 0.09 & 0.08 \\
\app{Flashlight} & 0.13 & 0.14 & 0.13 \\
\app{Phone} & 0.34 & 0.34 & 0.34 \\
\app{PlayMusic} & 0.2 & 0.2 & 0.2 \\
\app{HealthyRecipes} & 0.11 & 0.12 & 0.12 \\
\app{Sleep} & 0.21 & 0.24 & 0.22 \\
\app{Medisafe} & 0.32 & 0.32 & 0.32 \\
\emph{Average} & 0.17 & 0.17 & 0.17 \\
\end{tabular}
\label{table:app-id-huaweiwatch-cm-low}
\end{minipage}\hfill
\begin{minipage}[t]{0.49\textwidth}
\caption{$38$ ``high-volume'' applications}
\begin{tabular}{lrrrr}
\textbf{Label} & \textbf{Precision} & \textbf{Recall} & \textbf{F1-score} \\
\app{SalatTime} & 1.0 & 1.0 & 1.0 \\
\app{MapMyFitness} & 1.0 & 0.96 & 0.98 \\
\app{Calm} & 0.94 & 0.96 & 0.95 \\
\app{Citymapper} & 0.94 & 0.96 & 0.95 \\
\app{DiabetesM} & 0.94 & 0.95 & 0.94 \\
\app{Outlook} & 0.96 & 0.92 & 0.94 \\
\app{SmokingLog} & 0.91 & 0.79 & 0.85 \\
\app{Fit} & 0.82 & 0.88 & 0.85 \\
\app{Running} & 0.78 & 0.72 & 0.75 \\
\app{MapMyRun} & 1.0 & 0.94 & 0.97 \\
\app{SleepTracking} & 0.99 & 0.92 & 0.95 \\
\app{Weather} & 0.76 & 0.69 & 0.72 \\
\app{Mobills} & 0.95 & 0.92 & 0.94 \\
\app{FitBreathe} & 0.96 & 0.99 & 0.98 \\
\app{FoursquareCityGuide} & 0.91 & 0.94 & 0.93 \\
\app{FitWorkout} & 0.84 & 0.8 & 0.82 \\
\app{Glide} & 0.88 & 0.94 & 0.91 \\
\app{Translate} & 0.94 & 0.91 & 0.92 \\
\app{Qardio} & 0.95 & 0.95 & 0.95 \\
\app{Krone} & 0.91 & 0.84 & 0.87 \\
\app{FindMyPhone} & 0.81 & 0.84 & 0.82 \\
\app{KeepNotes} & 0.95 & 0.88 & 0.91 \\
\app{Shazam} & 0.95 & 0.86 & 0.9 \\
\app{Strava} & 0.86 & 0.79 & 0.82 \\
\app{Telegram} & 0.95 & 0.92 & 0.94 \\
\app{Maps} & 0.74 & 0.79 & 0.76 \\
\app{Endomondo} & 0.78 & 0.84 & 0.81 \\
\app{DCLMRadio} & 0.96 & 0.95 & 0.96 \\
\app{Lifesum} & 0.8 & 0.88 & 0.84 \\
\app{PlayStore} & 0.8 & 0.88 & 0.83 \\
\app{AppInTheAir} & 0.82 & 0.94 & 0.88 \\
\app{Bring} & 0.82 & 0.94 & 0.88 \\
\app{Spotify} & 0.96 & 0.99 & 0.98 \\
\app{Meduza} & 0.9 & 0.91 & 0.91 \\
\app{FITIVPlus} & 0.99 & 0.99 & 0.99 \\
\app{ChinaDaily} & 0.96 & 0.91 & 0.94 \\
\app{WashPost} & 0.9 & 1.0 & 0.95 \\
\app{Camera} & 1.0 & 1.0 & 1.0 \\
\emph{Average} & 0.9 & 0.9 & 0.9 \\
\end{tabular}
\label{table:app-id-huaweiwatch-cm-high}
\end{minipage}
\end{table}

\begin{table}[h]
\caption*{Classifier performance when transferring the model between pairs of devices.}
\small
\centering
\begin{minipage}[t]{0.49\textwidth}
\caption{Train on Huawei-Pixel. Test on Fossil-Nexus.}
\begin{tabular}{lrrrr}
\textbf{Label} & \textbf{Precision} & \textbf{Recall} & \textbf{F1-score} \\
\app{Bring} & 1.0 & 1.0 & 1.0 \\
\app{Calm} & 0.96 & 0.86 & 0.91 \\
\app{ChinaDaily} & 0.96 & 0.89 & 0.93 \\
\app{Citymapper} & 0.74 & 1.0 & 0.85 \\
\app{DCLMRadio} & 0.91 & 0.71 & 0.8 \\
\app{DiabetesM} & 0.97 & 1.0 & 0.98 \\
\app{Endomondo} & 0.78 & 1.0 & 0.88 \\
\app{FITIVPlus} & 0.88 & 1.0 & 0.93 \\
\app{FindMyPhone} & 0.71 & 0.17 & 0.28 \\
\app{Fit} & 0.0 & 0.0 & 0.0 \\
\app{FitBreathe} & 0.3 & 0.93 & 0.46 \\
\app{FitWorkout} & 0.0 & 0.0 & 0.0 \\
\app{FoursquareCityGuide} & 1.0 & 0.86 & 0.92 \\
\app{Glide} & 0.85 & 1.0 & 0.92 \\
\app{KeepNotes} & 0.71 & 0.96 & 0.82 \\
\app{Krone} & 1.0 & 0.62 & 0.77 \\
\app{Lifesum} & 0.84 & 0.96 & 0.9 \\
\app{MapMyRun} & 1.0 & 1.0 & 1.0 \\
\app{Maps} & 0.95 & 0.62 & 0.75 \\
\app{Meduza} & 0.85 & 0.97 & 0.9 \\
\app{Mobills} & 1.0 & 0.79 & 0.88 \\
\app{Outlook} & 1.0 & 0.96 & 0.98 \\
\app{PlayStore} & 0.33 & 0.11 & 0.16 \\
\app{Running} & 0.96 & 0.89 & 0.93 \\
\app{SalatTime} & 0.57 & 0.96 & 0.72 \\
\app{Shazam} & 1.0 & 0.82 & 0.9 \\
\app{SleepTracking} & 0.96 & 0.86 & 0.91 \\
\app{SmokingLog} & 0.96 & 0.89 & 0.93 \\
\app{Spotify} & 1.0 & 1.0 & 1.0 \\
\app{Strava} & 1.0 & 0.86 & 0.92 \\
\app{Telegram} & 1.0 & 0.97 & 0.98 \\
\app{Translate} & 0.72 & 0.93 & 0.81 \\
\app{WashPost} & 0.57 & 1.0 & 0.73 \\
\app{Weather} & 0.93 & 0.93 & 0.93 \\
\emph{Average} & 0.81 & 0.81 & 0.81 \\
\end{tabular}
\label{table:app-id-transfer1-cm}
\end{minipage}\hfill
\begin{minipage}[t]{0.49\textwidth}
\caption{Train on Fossil-Nexus. Test on Huawei-Pixel.}
\begin{tabular}{lrrrr}
\textbf{Label} & \textbf{Precision} & \textbf{Recall} & \textbf{F1-score} \\
\app{Bring} & 1.0 & 1.0 & 1.0 \\
\app{Calm} & 1.0 & 0.97 & 0.98 \\
\app{ChinaDaily} & 0.92 & 0.79 & 0.85 \\
\app{Citymapper} & 1.0 & 0.93 & 0.96 \\
\app{DCLMRadio} & 0.93 & 0.86 & 0.89 \\
\app{DiabetesM} & 1.0 & 1.0 & 1.0 \\
\app{Endomondo} & 0.83 & 0.86 & 0.84 \\
\app{FITIVPlus} & 0.87 & 0.96 & 0.92 \\
\app{FindMyPhone} & 0.67 & 0.07 & 0.13 \\
\app{Fit} & 0.21 & 0.14 & 0.17 \\
\app{FitBreathe} & 0.3 & 0.57 & 0.39 \\
\app{FitWorkout} & 0.5 & 0.04 & 0.07 \\
\app{FoursquareCityGuide} & 0.93 & 0.96 & 0.95 \\
\app{Glide} & 1.0 & 1.0 & 1.0 \\
\app{KeepNotes} & 0.93 & 0.96 & 0.95 \\
\app{Krone} & 0.96 & 0.93 & 0.95 \\
\app{Lifesum} & 0.96 & 0.96 & 0.96 \\
\app{MapMyRun} & 0.88 & 1.0 & 0.93 \\
\app{Maps} & 0.92 & 0.82 & 0.87 \\
\app{Meduza} & 1.0 & 1.0 & 1.0 \\
\app{Mobills} & 0.74 & 0.97 & 0.84 \\
\app{Outlook} & 0.85 & 1.0 & 0.92 \\
\app{PlayStore} & 0.95 & 0.71 & 0.82 \\
\app{Running} & 0.86 & 0.89 & 0.88 \\
\app{SalatTime} & 0.96 & 0.96 & 0.96 \\
\app{Shazam} & 0.9 & 0.93 & 0.91 \\
\app{SleepTracking} & 0.78 & 1.0 & 0.88 \\
\app{SmokingLog} & 0.97 & 0.97 & 0.97 \\
\app{Spotify} & 1.0 & 1.0 & 1.0 \\
\app{Strava} & 0.81 & 0.93 & 0.87 \\
\app{Telegram} & 0.76 & 1.0 & 0.86 \\
\app{Translate} & 0.8 & 0.97 & 0.88 \\
\app{WashPost} & 0.93 & 0.96 & 0.95 \\
\app{Weather} & 0.96 & 0.96 & 0.96 \\
\emph{Average} & 0.86 & 0.86 & 0.86 \\
\end{tabular}
\label{table:app-id-transfer2-cm}
\end{minipage}
\end{table}

\end{document}